\documentclass[aps,
               prb,
               nofootinbib,
               eprint,]{revtex4-2}
\usepackage{amsmath}  
\usepackage{amsfonts} 
\usepackage{graphicx} 
\usepackage[scale=1.0]{OldStandard}
\usepackage{libertinust1math}
\usepackage{ifthen}
\usepackage{color}
\usepackage[section]{placeins}
\usepackage[caption=false]{subfig}  
\usepackage[makeroom]{cancel}
\usepackage[hidelinks]{hyperref}

\usepackage{dg-latex}
\usepackage{dg-qm}
\usepackage{dg-color}
\usepackage{dg-jarev}
\usepackage{submath}

\usepackage[justification=justified]{caption}
\newcommand{\dgcap}[2]{\captionsetup{justification=centerlast,margin={#1\textwidth,#2\textwidth}}}
\newcommand{\llpush}{\hfill \ }

\newcommand{\gdir}{./}  

\newcommand {\gl}     {g_{1'}}
\newcommand {\gal}    {\gamma_{1'}}
\newcommand {\gr}     {g_{2'}}
\newcommand {\gar}    {\gamma_{2'}}
\newcommand {\jp} [2] {\mbox{#1--#2}}
\newcommand {\imdb}   {\mbox{Im}\,[D(\beta)]}
\newcommand {\lgz}    {\lambda \rightarrow 0}
\newcommand {\lgzt}   {\itm{\,\lgz}}
\newcommand {\llgz}   {\lim_{\lgz}}

\begin{document}


\title{A didactically motivated reexamination of a particle's quantum mechanics \\ with square-well potentials}

\author{Domenico Giordano}
\email{Electronic mail: dg.esa.retired@gmail.com}
\affiliation{European Space Agency (retired), The Netherlands} 

\author{Pierluigi Amodio}
\email{Electronic mail: pierluigi.amodio@uniba.it}
\affiliation{Dipartimento di Matematica, Universit\`a di Bari, Italy}

\author{Felice Iavernaro}
\email{Electronic mail: felice.iavernaro@uniba.it}
\affiliation{Dipartimento di Matematica, Universit\`a di Bari, Italy}


\date{\today}

\begin{abstract}  We address two questions regarding square-well potentials from a didactic perspective.
The first question concerns whether or not the justification of the standard \textit{a priori} omission of the potential's vertical segments in the analysis of the eigenvalue problem is licit.
The detour we follow to find out the answer considers a trapezoidal potential, includes the solution, analytical and numerical, of the corresponding eigenvalue problem and then analyzes the behavior of that solution in the limit when the slope of the trapezoidal potential's ramps becomes vertical.
The second question, obviously linked to the first one, pertains whether or not eigenfunction's and its first derivative's continuity at the potential's jump points is justified as \textit{a priori assumption} to kick-off the solution process, as it is standardly accepted in textbook approaches to the potential's eigenvalue problem.
  \end{abstract}


\maketitle 


\tableofcontents

\section{Introduction \label{intro}}

A lot of quantum-mechanics textbooks \cite{ep1936,ep1950,am11961,dth1964,ls1968,cct1977,ll1977,db1989,sf1999g,sf1999e,bb2000,df2001,rg2004,dg2005,pa2005,pt2012}\footnote{Complete literature surveys are unattainable asymptotic ideals. The list we provide contains only the textbooks we consulted but we trust they constitute a sufficiently representative sample.} consider, discuss, and solve the eigenvalue problem related to the one-dimensional symmetrical/unsymmetrical finite and/or infinite square-well potential.
The subject has been seemingly analyzed in a variety of substantially similar manners, which we group together in and label as standard textbook approaches (\sta) for future reference, and the outcome of those analyses is looked upon as established body of knowledge to be taught routinely.
So, why would one wish to go through a reexamination?
The inspiration came from a student's interesting and subtle remark:
\begin{quote}
  We are taught about square-well potentials (\swp), such as, say, the one shown in \Rfi{ufp.sw}, as useful idealizations of practical cases;\footnote{A typical example can be found at page 246 of Bohm's textbook \cite{db1989} where the Ramsauer effect is described.} however, when we deal with the eigenvalue problem, we utilize for all intents and purposes the discontinuous potential (\dswp) shown in \Rfi{ufp.swh} which is a somewhat different representation of the original \swp\ because the Heaviside's functions ignore the presence of the vertical segments. How do we know beforehand that the omission of the vertical segments, which, after all, are legitimate portions of the potential required by idealizations, is irrelevant for the solution of the eigenvalue problem?
\end{quote}
A teacher's very probable reaction, naturally banking on the mature body of knowledge offered by the \sta, would be to reassure the student that even if a way could be thought of {absorbing into the analysis the \swp's vertical segments} then, in the end, the same results would be obtained and nothing new would be found;
a typical student would presumably be convinced by such a reassurance because it conveniently minimizes the learning process, obviously.
On the other hand, there exist a fraction, maybe small, of curious students to whom that reassurance would prove less effective.
In the back of their mind, the wisdom delivered by that witty master of physics that Feynman was in the last paragraph of his incisive article \cite{rf1969tpt} about science's meaning,
\begin{quote}
  It is necessary to teach both to accept and to reject the past with a kind of balance that takes considerable skill. Science alone of all the subjects contains within itself the lesson of the danger of belief in the infallibility of the greatest teachers of the preceding generations.
\end{quote}
would keep bouncing back and forth together with other tempting reflections such as,
  ``How do I know \textit{beforehand} that I am not going to find out anything new?
    And even if that would turn out to be the case, how do I know whether or not I will at least \textit{learn} something new by following other paths if I do not explore them?''
So, imagining such a state of mind, we gathered encouragement and thrust from Feynman's advice, ``So carry on. Thank you.'', concluding his cited article and went on with the reexamination described in the sequel.
Our effort is dedicated particularly to the students in the second camp.
%
\begin{figure}[h]
   \subfloat[Square well]                         {\label{ufp.sw}  \resizebox{.31\textwidth}{!}{\includegraphics*[trim=25 20 60 60]{\gdir/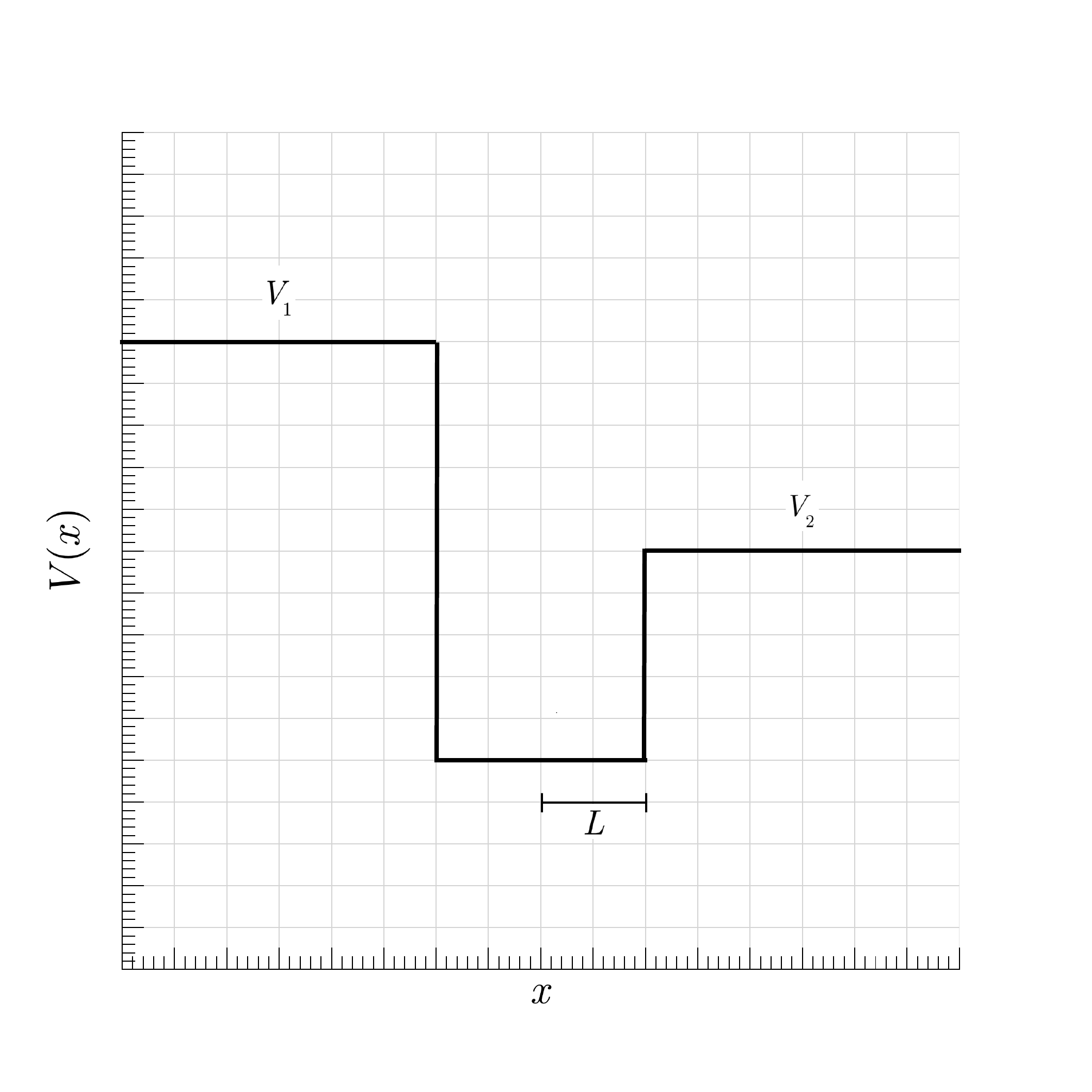}}} \hfill
   \subfloat[Square well with Heaviside functions]{\label{ufp.swh} \resizebox{.31\textwidth}{!}{\includegraphics*[trim=25 20 60 60]{\gdir/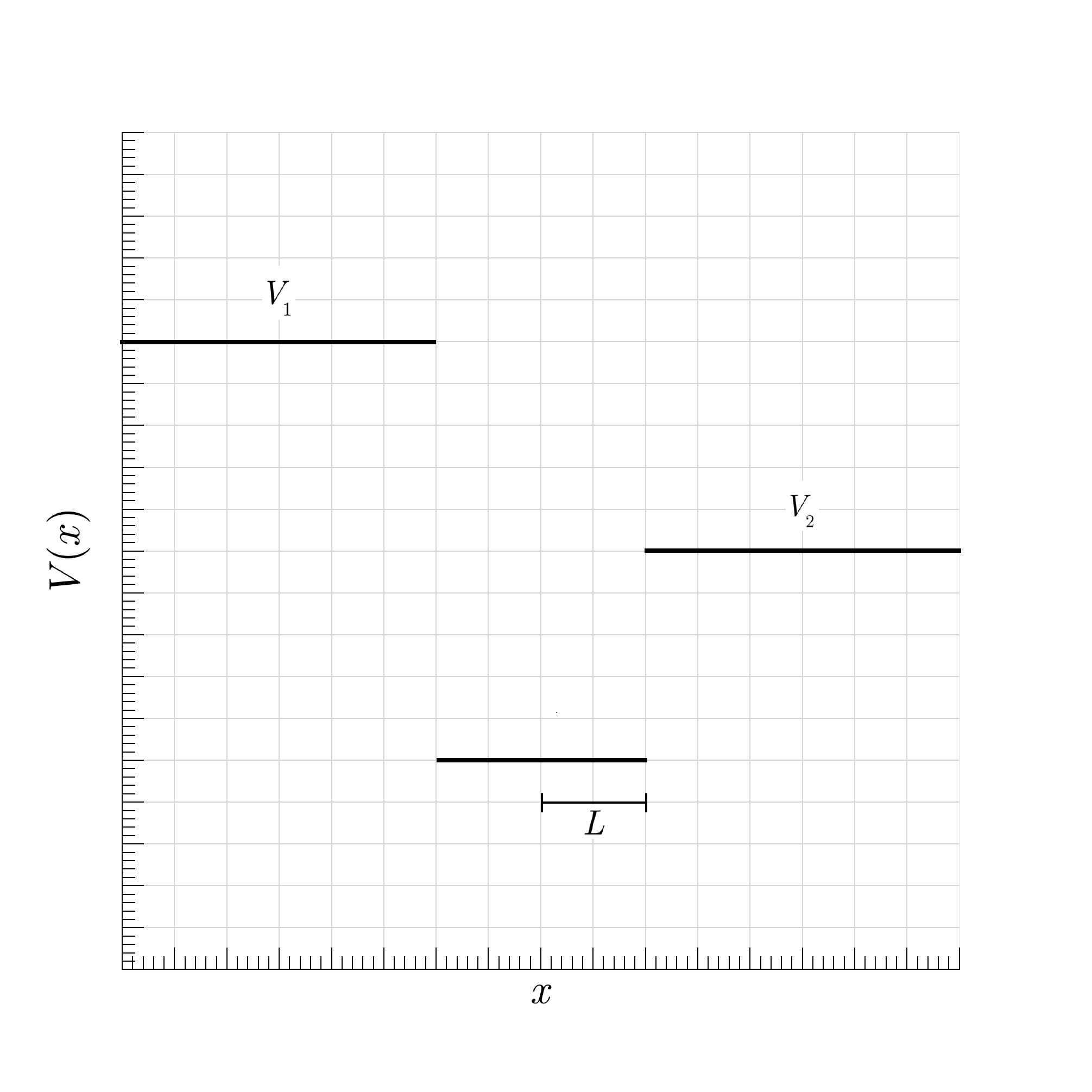}}} \hfill
   \subfloat[Trapezoidal well]                    {\label{ufp.tw}  \resizebox{.31\textwidth}{!}{\includegraphics*[trim=25 20 60 60]{\gdir/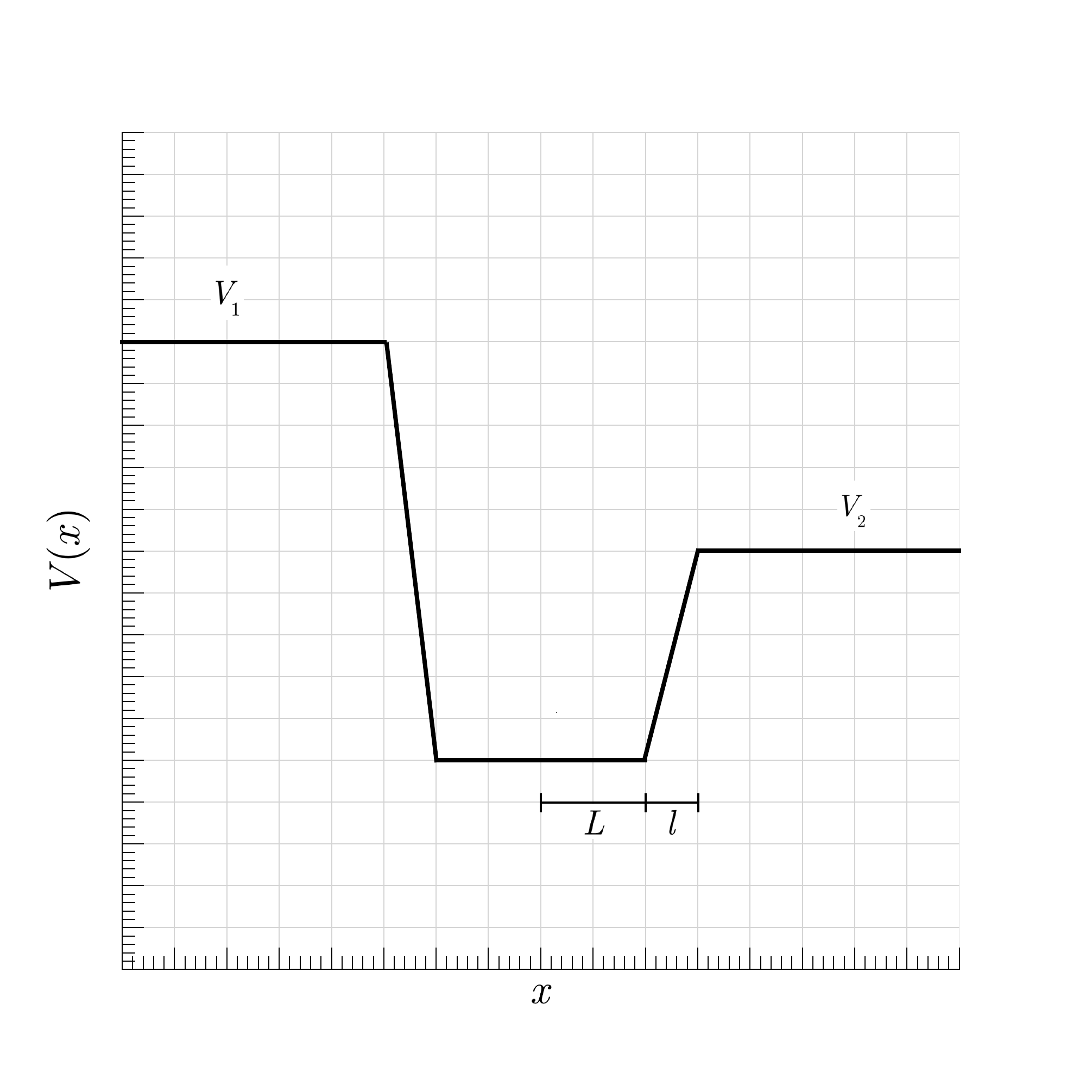}}}
   \caption{Finite unsymmetrical potentials considered in this study; units are arbitrary.\label{ufp}}
\end{figure}
%

Our approach confides in and complies with the natural philosophy's famous principle \textit{Saltus natura non facit}
(Nature does not make jumps)\footnote{The mentioning of this ``Loi de la Continuit\'{e}'' (\textit{Law of Continuity}), in Leibniz's words \cite{gl1898}, or ``old canon in natural history'', in Darwin's words \cite{cd1860}, may appear somewhat peculiar or, maybe, even disturbing to the eyes of quantum-mechanics purists that subscribe to the Copenhagen interpretation; of course, we intend no provocation to resume bygone fierce debates and simply invoke the principle only to justify our hesitation regarding the conceptual applicability of the discontinuous potential of \Rfi{ufp.swh}.} that so much inspired several scientific eminences of the past in different departments of science \cite{cvl1755,gl1896,gl1898,cd1860}.
Indeed, we relinquish the \dswp, take as starting point the trapezoidal-well potential (\twp) sketched in \Rfi{ufp.tw}, solve the corresponding eigenvalue problem analytically and numerically, and investigate the solution's behavior when the slope of the \twp's oblique segments becomes vertical \mbox{$(l \rightarrow 0)$}.
It seemed to us a rather straightforward conceptual pathway to follow in order to avoid the omission of the \swp's vertical segments.
We were delighted to discover, although only after having carried out almost completely our study, that the same idea had been proposed and probed by Branson \cite{db1979ajp}\footnote{We are grateful to S. De Vincenzo (Universidad Central de Venezuela, Caracas) for bringing Branson's article to our attention.} in 1979.
We keep in great regard Branson's article because it drew our attention towards another important issue connected with the \dswp\,: the presumed continuity of eigenfunctions and their derivatives at the potential's jump points ($x=\pm L$ in \Rfi{ufp.swh});\footnote{More precisely, it is not actually considered an issue in the \sta\ but Branson drew attention to the unsatisfactoriness of the mathematical explanations provided in the \sta\ to justify the continuity's assumption.} we will tell our point of view about this matter in \Rse{ccjp}.
We were also pleased to discover in \mbox{Fig. 6-1} at page 237 of Tipler and Llewellyn's textbook \cite{pt2012} that those authors used the \twp\ of \Rfi{ufp.tw} with $V_{1}=V_{2}$ to characterize the quantum dynamics of an electron between two electrodes in a vacuum tube, a fine schematization not so far from real-life applications.
\section{Quantum-mechanics problem with the trapezoidal-well potential \label{ev.twp}}
\subsection{Formulation and preliminary considerations regarding boundary conditions \label{pc}}
We consider a particle on the $x$ axis subjected to the \twp
\begin{equation}\label{twp}
   V(x) = \begin{cases}
             V_{1}                      & \quad                  x \le -\,( l + L )    \\[2ex]
             - V_{1}\,\dfrac{x + L}{l}  & \quad -\,( l + L ) \le x \le -   L           \\[2ex]
             0                                     & \quad -   L        \le x \le +   L           \\[2ex]
             + V_{2}\,\dfrac{x - L}{l}  & \quad +   L        \le x \le +\,( L + l )    \\[2ex]
             V_{2}                      & \quad +\,( L + l ) \le x                     \\
   \end{cases}
\end{equation}
shown in \Rfi{ufp.tw}.
The particle's hamiltonian is simply
\begin{equation}\label{Ham}
   \ham = - \hpib\pd{2}{}{x} + V(x)
\end{equation}
and its quantum mechanics is governed by the \SEq
\begin{equation}\label{seq}
  \ih \pd{}{\wf}{t} = - \hpib\pd{2}{\wf}{x} + V(x)\,\wf
\end{equation}
The praxis in quantum-mechanics textbooks is to introduce at this point the standard variable-separation technique and to launch onto the analysis of the eigenvalue problem governed by the time-independent \SEq ; as representative example, we mention Griffiths' didactically remarkable textbook \cite{dg2005}.
We believe that such a way of proceeding is somehow incomplete because it gives the student only a partial view inasmuch as it puts in evidence exclusively the suitableness of the mathematical operators intervening in the \SEq\ [\REq{seq}] for variable-separation techniques and totally disregards the equally important role of the boundary conditions which, we are convinced, deserve attention already at this stage of the problem formulation.
From a mathematical point of view, \REq{seq} is a second-order partial differential equation whose integration requires an initial condition
\begin{equation}\label{seq.ic}
  \wf(x,0) = F(x)
\end{equation}
and appropriate boundary conditions.
In the one-dimensional case we are considering, there are two boundaries ($x=\pm \infty$) and, therefore, we need two conditions involving wavefunction and its first derivative; we may write them formally as follows
\begin{subequations}\label{seq.bc} \seqn
  \begin{align}
     G_{1}[\wf(-\infty,t),\wfx(-\infty,t),\wf(+\infty,t),\wfx(+\infty,t)] = 0 \label{seq.bc.1}\\[.25\baselineskip]
     G_{2}[\wf(-\infty,t),\wfx(-\infty,t),\wf(+\infty,t),\wfx(+\infty,t)] = 0 \label{seq.bc.2}
  \end{align}
\end{subequations}
Explicit examples of mathematical nature embedded in \REqq{seq.bc} are: the prescription of the wavefunction
\begin{subequations}\label{seq.bc.i} \seqn
  \begin{align}
     \wf(-\infty,t) - \Theta_{1}(t) = 0 \label{seq.bc.i.1}\\[.25\baselineskip]
     \wf(+\infty,t) - \Theta_{2}(t) = 0 \label{seq.bc.i.2}
  \end{align}
\end{subequations}
by means of supposedly known functions, or the periodicity condition
\begin{subequations}\label{seq.bc.p} \seqn
  \begin{align}
     \wf (-\infty,t) - \wf (+\infty,t) = 0 \label{seq.bc.p.1}\\[.25\baselineskip]
     \wfx(-\infty,t) - \wfx(+\infty,t) = 0 \label{seq.bc.p.2}
  \end{align}
\end{subequations}
or conditions of the Sturm-Liouville type
\begin{subequations}\label{seq.bc.sl} \seqn
  \begin{align}
     \mu_{1} \wf(-\infty,t) + \mu_{2} \wfx(-\infty,t) = 0 \label{seq.bc.sl.1}\\[.25\baselineskip]
     \nu_{1} \wf(+\infty,t) + \nu_{2} \wfx(+\infty,t) = 0 \label{seq.bc.sl.2}
  \end{align}
in which $\mu, \nu$ are given constants.
\end{subequations}
Of course, \REqq{seq.bc} must encode in mathematical terms information about what is physically going on at the boundaries.
It may happen sometimes that an explicit and crystal clear grasp of the boundary conditions is not in our possession but that occurrence does not either entitle us to ignore or exempt us from keeping in mind their conceptual necessity, at least formally.
Now, within a mere mathematical context, there is really nothing particularly special about the above differential-equation problem [\REq{seq}, \REq{seq.ic}, \REqq{seq.bc}]: if initial $(F)$ and boundary ($G_{1},G_{2}$) conditions are explicitly specified then the set of the mentioned equations is a ready intake to feed numerical-solution machineries.
In this regard, an analogy comes quickly to mind:
heat-transfer engineers solve routinely a similar set either numerically or via variable separation when possible.
Their unknown is the temperature and, obviously, the terms in their \REq{seq} have different physical meanings; the imaginary unit does not appear but its appearance in our case is an almost irrelevant computational preoccupation because modern\footnote{And even not so modern such as the old good \textsc{fortran}.} programming languages handle complex numbers smoothly.
Within a physical context, quantum mechanics casts a peculiar nuance on the differential-equation problem we are considering.
From a quantum-mechanical point of view, the acceptable solutions to \REq{seq} must conform to a very strict requirement:
the wavefunction must be normalizable
\begin{equation}\label{wfn}
   \intinf{\wfc(x,t)\cdot\wf(x,t)} = 1
\end{equation}
otherwise the energy operator \itm{\ene=\eopt}
is not hermitean and the macroscopic observable energy does not turn out to be real \cite{dg2021ejp}
\begin{equation}\label{moe}
   \moE \neq \moEc
\end{equation}
Non-compliant solutions have, therefore, no physical significance.
Incisive statements emphasizing this aspect were expressed, for example, by Bohm \cite[pag. 178]{db1989},
\begin{quote}
  If this requirement [our \REq{wfn}] is not satisfied, then we cannot even normalize the probability, so that it is impossible to give the wave function a meaning in terms of physically observable averages.
\end{quote}
and Griffiths \cite[pag. 13 (his emphasis)]{dg2005},
\begin{quote}
  ... \textbf{non-normalizable} solutions cannot represent particles, and must be rejected. Physically realizable states correspond to the \textbf{square-integrable} solutions to Schr\"{o}dinger's equation.
\end{quote}
The normalization condition [\REq{wfn}] has twofold repercussions on the boundary conditions.
If the energy operator is hermitean then so must be the hamiltonian
\begin{equation}\label{moh}
  \moHc = \moEc = \moE = \moH
\end{equation}
or equivalently
\begin{equation}\label{moh.i}
  \intinf{[\cco{(\ham\wf)}\,\wf - \wfc\,\ham\wf]} = 0
\end{equation}
The submission of our hamiltonian [\REq{Ham}] to the hermiticity test represented by \REq{moh.i} yields the following constraint
\begin{equation}\label{bc.c1}
  \left(\wfc \pd{}{\wf}{x} - \wf \pd{}{\wfc}{x}\right)_{x=+\infty} - \left(\wfc \pd{}{\wf}{x} - \wf \pd{}{\wfc}{x}\right)_{x=-\infty} = 0
\end{equation}
on the boundary conditions.
Of the explicit examples listed after \REqq{seq.bc}, the periodicity conditions [\REqq{seq.bc.p}] are the only ones that comply with \REq{bc.c1}; the Sturm-Liouville conditions [\REqq{seq.bc.sl}] do only if the coefficients' ratios are real
\begin{equation}\label{seq.sl.c1}
   \cco{\left(\frac{\mu_{1}}{\mu_{2}}\right)} = \frac{\mu_{1}}{\mu_{2}}    \qquad\qquad    \cco{\left(\frac{\nu_{1}}{\nu_{2}}\right)} = \frac{\nu_{1}}{\nu_{2}}
\end{equation}
Nothing can be said \textit{a priori} about the wavefunction-prescription conditions [\REqq{seq.bc.i}] for arbitrary functions $\Theta_{k}(t)$.
A more severe constraint is levied by the boundaries' locations being situated at \mbox{$x=\pm\infty$}.
These locations are a bit hostile in view of normalization operations; they restrict the boundary conditions even more than \REq{bc.c1} because they require the vanishing of the wavefunction \cite{db1989,dg2005,pa2005}\footnote{On the necessity of wavefunction vanishing at $(\pm)$ infinity as required by the normalization condition, Griffiths wrote in footnote 12 at page 14 of his textbook \cite{dg2005}: ``A good mathematician can supply you with pathological counterexamples,...''. Hilariously, his prophecy came perfectly true when one of us (FI) engaged in such a mathematically refined, somehow even amusing, task.}
\begin{equation}\label{seq.bc.i.v}
   \wf(-\infty,t) = \wf(+\infty,t) = 0
\end{equation}
\REqqb{seq.bc.i.v} are a particular case of wavefunction-prescription condition [\REqq{seq.bc.i} with $\Theta_{k}(t)=0$], comply with \REq{bc.c1} and, in so doing, safeguard the hermiticity [\REq{moh.i}] of the hamiltonian [\REq{Ham}]; \textit{de facto}, they also imply the vanishing of the wavefunction's corresponding first derivatives. 

\subsection{Boundary conditions with variable separation \label{vsbc}}

We rejoin now the beaten path of the literature by applying the standard variable-separation technique
\begin{equation}\label{vst}
   \wf(x,t) = \Phi(t)\cdot\psi(x)
\end{equation}
which splits the \SEq\ [\REq{seq}] in two separated and independent differential-equation problems
\begin{subequations}\label{evp}  \seqn
   \begin{gather}
      \ih \pd{}{\Phi}{t} = \epsilon \Phi                  \label{evp.t} \\[.4\baselineskip]
      -\hpib \pd{2}{\psi}{x} + V(x)\,\psi = \epsilon \psi \label{evp.s}
   \end{gather}
\end{subequations}
The temporal one [\REq{evp.t}] is easily integrated
\begin{equation}\label{evp.t.i}
  \Phi(t) = \Phi(0)\cdot \exp\left(-i \frac{\epsilon t}{\hbar}\right)
\end{equation}
but we put its integral [\REq{evp.t.i}] on hold for the time being because the exploitation of the initial condition [\REq{seq.ic}] is premature at this moment.
The integration of the time-independent \SEq\ [\REq{evp.s}] involves more elaboration.
It definitely requires two boundary conditions that, obviously, we should derive from the general ones [\REqq{seq.bc}] by substituting in them the variable-separated wavefunction [\REq{vst}].
This move calls for due attention because it leads us to face a crucial conceptual filter that reveals the importance of giving the boundary conditions the deserved attention:
if \REqq{seq.bc} transform to
\begin{subequations}\label{evp.s.bc} \seqn
  \begin{align}
     G_{1}[\psi(-\infty),\efx(-\infty),\psi(+\infty),\efx(+\infty)] = 0 \label{evp.s.bc.1}   \\[.25\baselineskip]
     G_{2}[\psi(-\infty),\efx(-\infty),\psi(+\infty),\efx(+\infty)] = 0 \label{evp.s.bc.2}
  \end{align}
\end{subequations}
then we have green light to proceed with separated variables; otherwise this would be the end of the road because the boundary conditions do not permit the existence of variable-separated solutions, the receptive mathematical structure of the \SEq\ [\REq{seq}] towards variable separation [\REq{vst}] notwithstanding.
We are on safe ground with the wavefunction vanishing at ($\pm$) infinity because, after the substitution of \REq{vst}, \REqq{seq.bc.i.v} go smoothly into the eigenfunctions' vanishing
\begin{equation}\label{evp.bc.i.v}
   \psi(-\infty) = \psi(+\infty) = 0
\end{equation}
Although these considerations may appear a bit formal, they deliver, we believe, an important didactical message that was best expressed in a generalized manner by Tanner \cite{act1991ajp} in 1991:
\begin{quote}
   Although the \SEq\ might be separable in some coordinates, the boundary conditions can recouple the variables.
\end{quote}
The applicability extent of such a statement is really wide.
It is true, for example, in the case of spatially confined molecules whose time-independent \SEq\ cannot be separated in terms of center-of-mass and internal coordinates due to the variable recoupling imposed by the confinement boundary conditions.
Tanner also complained that:
\begin{quote}
  A representative sample of relevant sections (on the hydrogen atom, center of mass, etc.) of introductory textbooks on quantum mechanics revealed no discussion of this difficulty.
\end{quote}
We tend to side with him.
Textbooks invariably focus on the separation of the mathematical operators appearing in the \SEq, be it either time-dependent or \mbox{-independent}.
Exceptions paying due attention to boundary conditions are rare; among them, Persico's great textbook \cite{ep1936,ep1950} shines through.\footnote{The textbook in Italian \cite{ep1936} received a very positive review in Nature \textbf{139}, 394 (1937). The not better identified reviewer, who enigmatically signed as H. T. H. P., valued Persico's efforts as ``We owe a deep debt of gratitude to Dr. Persico for undertaking the useful task of presenting, in a single volume of reasonable size, a unified account of all aspects of the subjects.'' and concluded with ``... the only serious defect of the book is that it is in Italian. Will some publisher consider the possibility of an English translation?'' His exhortation was fulfilled 13 years later by Prentice-Hall which published the English translation \cite{ep1950} by G. Temmer; the English translation was then reviewed, again positively, by M. Lax in American Journal of Physics \textbf{19}, 478 (1951).}

The time-independent \SEq\ [\REq{evp.s}] with the \twp\ [\REq{twp}] and the eigenfunction-vanishing boundary conditions [\REqq{evp.bc.i.v}] constitute the eigenvalue problem we wish to solve.
Before launching onto the solution process, however, we wish to spend a few more words to emphasize further the importance of the boundary conditions.
In this regard, we ask: how do we know if an eigenfunction $\psi$ corresponding to a determined eigenvalue $\epsilon$ is a unique solution\footnote{Griffiths dedicated problem 2.45 at page 87 of his textbook \cite{dg2005} to this matter but his emphasis was more on the absence of non-degenerate states.}  to \REq{evp.s}?
Let us suppose that two eigenfunctions $\psi_{1},\psi_{2}$ exist for the same eigenvalue; if they are linearly independent then their Wronskian \cite{wb2017}
\begin{equation}\label{w}
  W\left[ \psi_{1},\psi_{2} \right] = \begin{vmatrix} \psi_{1} & \psi_{2} \\[2ex]  \pd{}{\psi_{1}}{x} & \pd{}{\psi_{2}}{x} \end{vmatrix}
  = \psi_{1} \pd{}{\psi_{2}}{x} -  \pd{}{\psi_{1}}{x} \psi_{2}
\end{equation}
never vanishes.
Both eigenfunctions must verify differential equation [\REq{evp.s}]
\begin{subequations}\label{evp.u}  \seqn
   \begin{gather}
      -\hpib \pd{2}{\psi_{1}}{x} + V(x)\,\psi_{1} = \epsilon \psi_{1}    \label{evp.u.1} \\[.4\baselineskip]
      -\hpib \pd{2}{\psi_{2}}{x} + V(x)\,\psi_{2} = \epsilon \psi_{2}    \label{evp.u.2}
   \end{gather}
and boundary conditions [\REqq{evp.bc.i.v}]
   \begin{gather}
      \psi_{1}(-\infty) = \psi_{1}(+\infty) = 0    \label{bc.u.1} \\[.4\baselineskip]
      \psi_{2}(-\infty) = \psi_{2}(+\infty) = 0    \label{bc.u.2}
   \end{gather}
simultaneously by definition.
The potential in \REqq{evp.u} can be any and needs not necessarily be the \twp\ of \REq{twp}.
We can multiply \REq{evp.u.1} by $\psi_{2}$, \REq{evp.u.2} by $\psi_{1}$, and subtract to obtain a vanishing expression
\begin{equation}\label{evp.u.3}
  \psi_{1} \pd{2}{\psi_{2}}{x} - \psi_{2} \pd{2}{\psi_{1}}{x} = 0
\end{equation}
which we can transform, through a simple game of derivative regrouping and expanding, into a form
\begin{equation}\label{evp.u.4}
  \pd{}{}{x}\left( \psi_{1} \pd{}{\psi_{2}}{x} -  \pd{}{\psi_{1}}{x} \psi_{2}  \right) = \pd{}{W}{x} = 0
\end{equation}
that proves the Wronskian's invariance; thus, if the Wronskian is continuous in $(-\infty,+\infty)$, and we plant here a flag to which we will need to return during the discussion of \Rse{ccjp}, then it is constant and we can conveniently evaluate it at the boundaries
\begin{equation}\label{w.b}
  W = \psi_{1}(-\infty) \pds{}{\psi_{2}}{x}{\!\!x=-\infty} - \pds{}{\psi_{1}}{x}{\!\!x=-\infty}\psi_{2}(-\infty)
    = \psi_{1}(+\infty) \pds{}{\psi_{2}}{x}{\!\!x=+\infty} - \pds{}{\psi_{1}}{x}{\!\!x=+\infty}\psi_{2}(+\infty)
\end{equation}
\end{subequations}
We understand at once from \REq{w.b} how the eigenfunctions' uniqueness is crucially hanging on the knowledge of the boundary conditions.
We can confide in those [\REqd{bc.u.1}{bc.u.2}] we have adopted in our eigenvalue problem because they reassuringly make the Wronskian vanish ($W=0$), imply the linear dependence of $\psi_{1},\psi_{2}$ and, in so doing, compel unambiguously the uniqueness.
So, the eigenstates are not degenerate: for a specified eigenvalue there is one and only one eigenfunction.
This conclusion goes hand in hand with two other important properties whose proofs are disseminated throughout the majority of the textbooks cited in the beginning of \Rse{intro}: the eigenvalues are real \itm{\cco{\epsilon} = \epsilon}
and the eigenfunctions are orthogonal
\begin{equation}\label{ef.ortho}
   \intmpi{\cco{\psi}_{\epsilon'}(x)\cdot\psi_{\epsilon''}(x)}{x} = 0
\end{equation}
with $\epsilon',\epsilon''$ being two distinct eigenvalues.
An interesting consequence of the eigenfunction-uniqueness proof is that we are given the freedom to choose the eigenfunctions to be either real or pure imaginary.
Indeed, if we break down the eigenfunction explicitly into its real and complex parts
\begin{equation}\label{ef.split}
  \psi = u + i v
\end{equation}
and substitute into differential equation [\REq{evp.s}] and boundary conditions [\REqq{evp.bc.i.v}] then we reach again the same structure of \REqs{evp.u.1}{bc.u.2} with $\psi_{1},\psi_{2}$ replaced by $u,v$ and the vanishing Wronskian
\begin{equation}\label{w.b.uv}
  W = u(-\infty) \pds{}{v}{x}{\!\!x=-\infty} - \pds{}{u}{x}{\!\!x=-\infty}v(-\infty)
    = u(+\infty) \pds{}{v}{x}{\!\!x=+\infty} - \pds{}{u}{x}{\!\!x=+\infty}v(+\infty)
    =0
\end{equation}
Thus, $u,v$ are linearly dependent and the eigenfunction $\psi$ is proportional to anyone of them through an inessential proportionality constant that we can choose either real or imaginary as it pleases us.

We imagine the reader to be sufficiently sensitized about the importance of the boundary conditions and, therefore, move on with the analysis of the eigenvalue problem.

\subsection{The eigenvalue problem\label{tevp}}

\subsubsection{Nondimensional formulation \label{ndf}}

We begin by formulating the eigenvalue problem in nondimensional form.
We scale the $x$ coordinate with the semi-extension of the potential well (\Rfi{ufp.tw})
\begin{subequations}\label{nd}\seqn
	\begin{equation}\label{nd.x}
	   x = \xi\cdot L
	\end{equation}
	and the eigenfunction with the inverse squared root of the total extension
	\begin{equation}\label{nd.ef}
	   \psi = \frac{\phi}{\sqrt{2L}}
	\end{equation}
\end{subequations}
In this way, the time-independent \SEq\ [\REq{evp.s}] and the wavefunction-vanishing boundary conditions [\REqq{evp.bc.i.v}] turn into the nondimensional forms
\begin{subequations}\label{ndevp}  \seqn
   \begin{gather}
      - \pd{2}{\phi}{\xi} + v(\xi)\,\phi = \beta\,\phi \label{ndevp.s}  \\[.4\baselineskip]
      \phi(-\infty) = \phi(+\infty)      = 0           \label{ndevp.bc}
   \end{gather}
The nondimensional eigenvalue in \REq{ndevp.s} is defined as
\begin{equation}\label{ndev}
   \beta = \frac{2mL^{2}\epsilon}{\hbar^{2}}
\end{equation}
while the nondimensional \twp\ 
\begin{equation}\label{ndevp.twp}
   v(\xi) = \frac{2mL^{2}}{\hbar^{2}} V(x) =
    \begin{cases}
             v_{1}                                        & \quad                        \xi \le -\,( \lambda + 1 )    \\[2ex]
             - v_{1}\,\dfrac{\xi + 1}{\lambda}            & \quad -\,( \lambda + 1 ) \le \xi \le -1                    \\[2ex]
             0                                            & \quad -1                 \le \xi \le +1                    \\[2ex]
             + v_{2}\,\dfrac{\xi - 1}{\lambda}            & \quad +1                 \le \xi \le +\,( 1 + \lambda )    \\[2ex]
             v_{2}                                        & \quad +\,( 1 + \lambda ) \le \xi                     \\
   \end{cases}
\end{equation}
descends from \REq{twp} and includes three solution-controlling characteristic numbers
\begin{gather}
   \subeqn{v_{s} = \frac{2mL^{2}V_{s}}{\hbar^{2}}}{s=1,2}{1} \label{ndv1.2} \\[.4\baselineskip]
   \lambda = \frac{l}{L}                                     \label{ndlL}
\end{gather}
\end{subequations}
\begin{figure}[h]
   \resizebox{.47\textwidth}{!}{\includegraphics*[trim=33 20 60 60]{\gdir/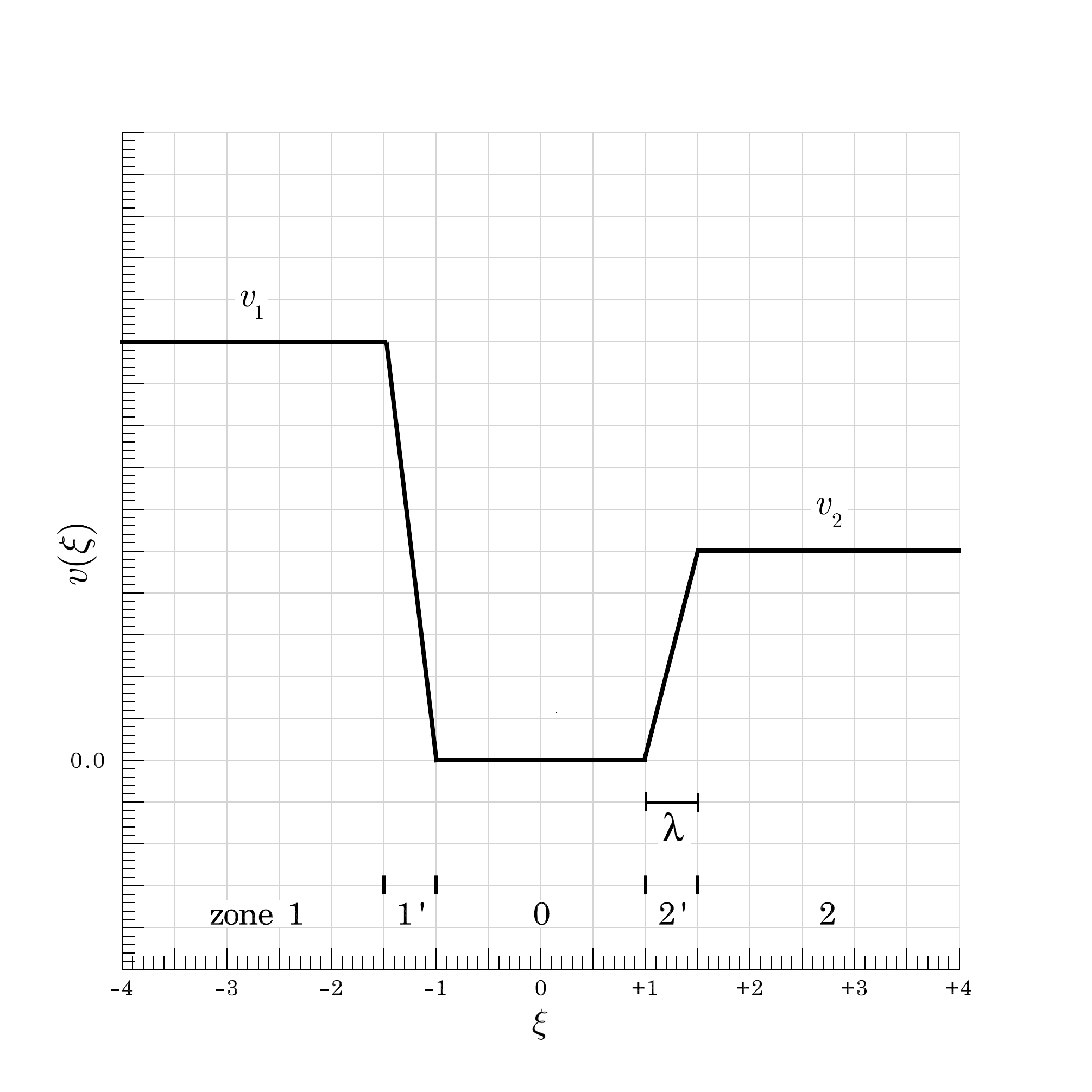}}
   \caption{Nondimensional \twp.\label{ndtwp}}
\end{figure}
%
We assume \itm{v_{2} \leq v_{1}} for a mere reason of convenience; obviously, the limitation does not restrict the results in any way.
We have graphically illustrated the nondimensional \twp\ [\REq{ndevp.twp}] in \Rfi{ndtwp} in view of the forthcoming analysis.
The potential subdivides the $x$ axis in five zones, in each of which the nondimensional time-independent \SEq\ [\REq{ndevp.s}] must be integrated separately.
The zonal solutions can be joined by imposing the continuity of the eigenfunction and of its first derivative at the junction points, that is, the points at which the \twp's slope is discontinuous; the unquestionable legitimacy of the claimed continuity conditions is guaranteed by the \twp's continuity. 
In turn, the continuity of the eigenfunction's second derivative at the junction points is guaranteed by \REq{ndevp.s}. 
The analytical integration is described in the following sections.
In parallel, we have carried out the integration also numerically by a method based on high-order finite differences \cite{pa2011jnaiam,pa2015cnsns,pa2020lncs-b} implemented in the code HOFiD\_MSP that can solve multiparameter spectral BV-ODE problems.
In our numerical calculations, we transform the integration interval $(-\infty,\infty)$ into a finite interval by means of a simple variable change and we utilize 6th-order formulae on a grid whose resolution consists of 2505 points, distributed in groups of 501 equispaced points in each zone.
\FloatBarrier\noindent

\subsubsection{Analytical integration in the zones with constant potential\label{z102}}

If we introduce the dummy parameter $v_{0}=0$ and set for brevity
\begin{equation}\label{betas}
   \subeqn{\beta_{s} = \beta - v_{s}}{s=1,0,2}{0}
\end{equation}
in the zones where the \twp\ is constant [\REq{ndevp.twp} top, central, bottom] then the nondimensional differential equation [\REq{ndevp.s}] becomes
\begin{equation}\label{ndevp.s.12}
    \pd{2}{\phi_{s}}{\xi} + \beta_{s}\,\phi_{s} = 0
\end{equation}
and, from that, we obtain the general integral
\begin{equation}\label{ndevp.s.12.i}
  \phi_{s}(\xi) = A_{s} \sin(\xi\sqrt{\beta_{s}}) + B_{s} \cos(\xi\sqrt{\beta_{s}})
\end{equation}
The imposition of the boundary conditions [\REq{ndevp.bc}] in the left- and right-most zones yields
\begin{subequations}\label{ndevp.bc.12} \seqn
  \begin{align}
     \phi_{1}(-\infty) = \lim_{\xi\rightarrow-\infty}\left[ A_{1} \sin(\xi\sqrt{\beta_{1}}) + B_{1} \cos(\xi\sqrt{\beta_{1}})  \right] = 0 \label{seq.bc.12.1} \\[.25\baselineskip]
     \phi_{2}(+\infty) = \lim_{\xi\rightarrow+\infty}\left[ A_{2} \sin(\xi\sqrt{\beta_{2}}) + B_{2} \cos(\xi\sqrt{\beta_{2}})  \right] = 0 \label{seq.bc.12.2}
  \end{align}
\end{subequations}
Physically meaningful solutions can be extracted from \REqq{ndevp.bc.12} only if the arguments of the trigonometric functions are complex; that, in turn, implies the negativity of the parameters $\beta_{s}$.
This occurrence produces the limitation [\REq{betas}]
\begin{equation}\label{betas.lt.vs}
   \subeqn{\beta \leq v_{s}}{s=1,2}{0}
\end{equation}
and bounds the eigenvalues to lie below the lowest potential level, $v_{2}$ in our case.
In accordance with \REq{betas.lt.vs}, we rearrange \REq{betas} as
\begin{equation}\label{negbetas}
  \beta_{s} = - ( v_{s} - \beta ) = - k_{s}
\end{equation}
and transform the trigonometric functions of \REqq{ndevp.bc.12} into exponential functions
\begin{subequations}\label{ndevp.bc.12.exp} \seqn
  \begin{align}
     \phi_{1}(-\infty) = \frac{1}{2i}\lim_{\xi\rightarrow-\infty}\left[ (A_{1}+i B_{1}) \exp(-\xi\sqrt{k_{1}}) - (A_{1}-i B_{1})\cancel{\exp(\xi\sqrt{k_{1}})}  \right] = 0 \label{seq.bc.12.exp.1} \\[.25\baselineskip]
     \phi_{2}(+\infty) = \frac{1}{2i}\lim_{\xi\rightarrow+\infty}\left[ (A_{2}+i B_{2}) \cancel{\exp(-\xi\sqrt{k_{2}})} - (A_{2}-i B_{2}) \exp(\xi\sqrt{k_{2}})  \right] = 0 \label{seq.bc.12.exp.2}
  \end{align}
\end{subequations}
The crossed terms in \REqq{ndevp.bc.12.exp} vanish in the limit; thus, boundary-condition compliance requires
\begin{subequations}\label{ndevp.coeffs.12} \seqn
  \begin{align}
     A_{1} = - i B_{1} \label{ndevp.coeffs.1} \\[.25\baselineskip]
     A_{2} = + i B_{2} \label{ndevp.coeffs.2}
  \end{align}
\end{subequations}
Taking into account \REq{negbetas} and \REq{ndevp.coeffs.1}, we obtain the solution
\begin{equation}\label{ndevp.ef.1}
  \phi_{1}(\xi) = B_{1} \exp(+\xi\sqrt{k_{1}})
\end{equation}
in zone 1; similarly but from \REq{ndevp.coeffs.2}, we deduce the solution
\begin{equation}\label{ndevp.ef.2}
  \phi_{2}(\xi) = B_{2} \exp(-\xi\sqrt{k_{2}})
\end{equation}
in zone 2.

It is interesting to wonder what happens if $\beta$ is forced by deliberate assignment to infringe the limitation in \REq{betas.lt.vs}.
Obviously, the general integral [\REq{ndevp.s.12.i}] stands valid and $\sqrt{\beta_{s}}$ becomes real but the imposition of the boundary conditions [\REqq{ndevp.bc.12}] remains idle because the limits for $\xi\rightarrow\pm\infty$ of the trigonometric functions are indeterminate. 
Thus, rewinding to \REqd{wfn}{moh.i} through the sequence \REq{nd.ef}, \REqq{evp.bc.i.v}, \REq{vst}, \REq{seq.bc.i.v}, \REq{bc.c1}, we reach an unavoidable impasse: the hamiltonian's hermiticity test fails and the wavefunction cannot be coerced into normalization.
There is nothing else left to do than to enforce Griffiths' verdict quoted just before \REq{moh}: non-normalizable solutions must be rejected because they correspond to physically irrealizable states.
Does this mean that we should throw those solutions away? No, they can still be of service as mathematical ingredients to compose physically acceptable solutions but this angle of the subject is somewhat tangential to our main theme focused on the bound states and, therefore, we refer the interested reader to the lucid explanations provided by Griffiths in Sec. 2.4 at page 59 of his textbook \cite{dg2005}.

In the central zone, the \twp\ vanishes [\REq{ndevp.twp} central] so that $\beta_{0} = \beta - v_{0} = \beta$ and the general integral [\REq{ndevp.s.12.i}] becomes
\begin{equation}\label{ndevp.s.0.i}
  \phi_{\,\!0}(\xi) = A_{0} \sin(\xi\sqrt{\beta}) + B_{0} \cos(\xi\sqrt{\beta})
\end{equation}
With regard to the argument of the trigonometric functions in \REq{ndevp.s.0.i}, it is worth noticing that, as far as the imposed boundary conditions [\REqq{ndevp.bc.12}] are concerned, there is really nothing in them preventing the existence of negative eigenvalues.
The latter occurrence should not be ruled out simply on the basis of the presence of $\sqrt{\beta}$ in \REq{ndevp.s.0.i}.
Indeed, assuming hypothetically $\beta<0$,  we could write
\begin{equation}\label{nbetas}
  \sqrt{\beta} = i \sqrt{-\beta}
\end{equation}
and transform the trigonometric functions with complex argument $i\xi\sqrt{-\beta}$ into exponential functions
\begin{equation}\label{ndevp.s.0.i.nev}
  \phi_{\,\!0}(\xi) = \frac{B_{0}+ iA_{0}}{2} \exp(\xi\sqrt{-\beta}) + \frac{B_{0}- iA_{0}}{2}  \exp(-\xi\sqrt{-\beta})
\end{equation}
with real argument $\xi\sqrt{-\beta}$.
The exponential functions in \REq{ndevp.s.0.i.nev} would be harmless and well behaved because the continuity conditions for the determination of the coefficients have to be imposed at the boundaries of the central zone located at $\xi=\mp 1$.
Yet, we will discover soon the reason for eigenvalue positivity; for now, we just have to wait patiently a little bit longer.

\subsubsection{Analytical integration in the zones with linear potential\label{z1p2p}}

The integration of the nondimensional differential equation [\REq{ndevp.s}] in the zones with linear potential requires familiarity with the Airy's differential equation and functions  \cite{ma1972,ov2004} but it is very straightforward.
Let us begin with the left zone.
The \twp\ decreases linearly [\REq{ndevp.twp}, 2nd line from top] from the level $v_{1}$ down to the bottom of the well and the differential equation [\REq{ndevp.s}] becomes
\begin{equation}\label{ndevp.s.1p}
    \pd{2}{\phi_{1'}}{\xi} + \left( \beta + v_{1}\,\dfrac{\xi + 1}{\lambda} \right)\,\phi_{1'} = 0
\end{equation}
The independent-variable linear transformation
\begin{equation}\label{ivt.1p}
   \xi = - \eta \left(\dfrac{\lambda}{v_{1}}\right)^{1/3} - \left( \dfrac{\lambda}{v_{1}}\beta + 1 \right)
\end{equation}
converts \REq{ndevp.s.1p} into the Airy differential equation
\begin{equation}\label{airyde.1p}
    \pd{2}{\phi_{1'}}{\eta} -\eta\,\phi_{1'} = 0
\end{equation}
whose general integral is a linear combination of the Airy functions
\begin{equation}\label{ndevp.s.1p.i}
  \phi_{1'}(\eta) = A_{1'}\,\Ai(\eta) + B_{1'}\,\Bi(\eta)
\end{equation}
Things are pretty much similar in the right zone.
The \twp\ increases linearly [\REq{ndevp.twp}, 2nd line from bottom] from the bottom of the well up to the level $v_{2}$ and the differential equation [\REq{ndevp.s}] becomes
\begin{equation}\label{ndevp.s.2p}
    \pd{2}{\phi_{2'}}{\xi} + \left( \beta - v_{2}\,\dfrac{\xi - 1}{\lambda} \right)\,\phi_{2'} = 0
\end{equation}
The independent-variable linear transformation
\begin{equation}\label{ivt.2p}
   \xi = +\zeta  \left(\dfrac{\lambda}{v_{2}}\right)^{1/3} + \left( \dfrac{\lambda}{v_{2}}\beta + 1 \right)
\end{equation}
converts \REq{ndevp.s.2p} into another Airy differential equation
\begin{equation}\label{airyde.2p}
    \pd{2}{\phi_{2'}}{\zeta} -\zeta\,\phi_{2'} = 0
\end{equation}
with general integral
\begin{equation}\label{ndevp.s.2p.i}
  \phi_{2'}(\zeta) = A_{2'}\,\Ai(\zeta) + B_{2'}\,\Bi(\zeta)
\end{equation}
With the obtainment of \REqd{ndevp.s.1p.i}{ndevp.s.2p.i}, our task is quickly completed.
However, it is useful to introduce here for future reference some characteristics and consequences of the indepen\-dent-variable transformations [\REqd{ivt.1p}{ivt.2p}], we took advantage of to carry out the integration, in view of their recurrent use in the forthcoming sections.

The differentials
\begin{subequations}\label{ivt.d} \seqn
  \begin{align}
     d\xi = - d\eta  \left(\dfrac{\lambda}{v_{1}}\right)^{1/3}   \label{ivt.d.1p} \\[.25\baselineskip]
     d\xi = + d\zeta \left(\dfrac{\lambda}{v_{2}}\right)^{1/3}   \label{ivt.d.2p}
  \end{align}
\end{subequations}
help to derive transformations between derivatives with respect to old and new variables
\begin{subequations}\label{ivt.der} \seqn
  \begin{align}
     \pd{}{}{\xi} = - \left(\dfrac{v_{1}}{\lambda}\right)^{1/3} \pd{}{}{\eta}    \label{ivt.der.1p} \\[.25\baselineskip]
     \pd{}{}{\xi} = + \left(\dfrac{v_{2}}{\lambda}\right)^{1/3} \pd{}{}{\zeta}   \label{ivt.der.2p}
  \end{align}
\end{subequations}
The inverse transformations
\begin{subequations}\label{i.ivt} \seqn
  \begin{align}
     \eta  = - \left(\dfrac{v_{1}}{\lambda}\right)^{1/3} \left( \xi + 1 + \dfrac{\lambda}{v_{1}}\beta \right)   \label{i.ivt.1p} \\[.25\baselineskip]
     \zeta = + \left(\dfrac{v_{2}}{\lambda}\right)^{1/3} \left( \xi - 1 - \dfrac{\lambda}{v_{2}}\beta \right)   \label{i.ivt.2p}
  \end{align}
\end{subequations}
are also useful because they allow to obtain the characteristic values of the new variables $\eta$ and $\zeta$ at the junction points \jp{1}{1'}, \jp{1'}{0,} located respectively at \itm{\xi=-(1+\lambda)} and  \itm{\xi=-1}, that belong to the left zone and \jp{0}{2'}, \jp{2'}{2}, located respectively at \itm{\xi=+1} and \itm{\xi=+(1+\lambda)}, that belong to the right zone; we find respectively
\begin{subequations}\label{mp.left} \seqn
  \begin{align}
     \etab & = + \left(\dfrac{\lambda}{v_{1}}\right)^{2/3} k_{1} > 0 \makebox[0em][l]{\hspace*{5.00em} \jp{1}{1'}}    \label{mp.left.11p} \\[.25\baselineskip]
     \etah & = - \left(\dfrac{\lambda}{v_{1}}\right)^{2/3} \beta     \makebox[0em][l]{\hspace*{6.90em} \jp{1'}{0}}    \label{mp.left.1p0}
  \end{align}
\end{subequations}
in the left zone and
\begin{subequations}\label{mp.right} \seqn
  \begin{align}
     \zetah & = - \left(\dfrac{\lambda}{v_{2}}\right)^{2/3} \beta      \makebox[0em][l]{\hspace*{6.90em} \jp{0}{2'}} \label{mp.right.02p} \\[.25\baselineskip]
     \zetab & = + \left(\dfrac{\lambda}{v_{2}}\right)^{2/3} k_{2} > 0  \makebox[0em][l]{\hspace*{5.00em} \jp{2'}{2}} \label{mp.right.2p2}
  \end{align}
\end{subequations}
in the right zone.
The overlined values are always positive; the circumflexed values' sign depends on that of the eigenvalue.
They conform to the following, easily demonstrable, limitations
\begin{subequations}\label{cv.lim}\seqn
    \begin{align}
       \etah  & \leq \etab  = \etah  + \left(\lambda\sqrt{v_{1}}\right)^{2/3}   \label{cv.lim.m} \\[.25\baselineskip]
       \zetah & \leq \zetab = \zetah + \left(\lambda\sqrt{v_{2}}\right)^{2/3}   \label{cv.lim.c}
    \end{align}
\end{subequations}
and, expectedly, they fix the ranges of the new variables
\begin{subequations}\label{ivt.range}\seqn
    \begin{align}
       \etah  & \leq \eta  \leq \etab   \label{ivt.range.m} \\[.25\baselineskip]
       \zetah & \leq \zeta \leq \zetab  \label{ivt.range.c}
    \end{align}
\end{subequations}

\subsubsection{Eigenfunction's and its first derivative's continuity at junction points \label{ecjp}}

The eigenfunction's components [\REqs{ndevp.ef.1}{ndevp.s.0.i}, \REqd{ndevp.s.1p.i}{ndevp.s.2p.i}] we obtained by analytical integration involve the presence and require the determination of the eight coefficients $B_{1}, A_{1'}, B_{1'}, A_{0}, B_{0}, A_{2'}, B_{2'}, B_{2}$ and of the eigenvalue $\beta$; nine unknowns in total.
They can be found by imposing the continuity of the eigenfunction and of its first derivative, two conditions therefore, in the four zone-junction points; accordingly, this imposition permits the formulation of eight equations.
The additional equation needed to balance the number of unknowns descends from the reformulation of the wavefunction's normalization condition [\REq{wfn}] in terms of the eigenfunctions; as well known, the most convenient choice is the normalization of the eigenfunctions
\begin{equation}\label{efn}
   \intmpi{\cco{\psi}(x)\,\psi(x)}{x} = 1
\end{equation}
which, according to the adopted variable scaling [\REqq{nd}], goes into the nondimensional form
\begin{equation}\label{efn.nd}
   \frac{1}{2}\intmpi{\cco{\phi}(\xi)\,\phi(\xi)}{\xi} = 1
\end{equation}

Let us begin with the junction point \jp{1}{1'} located at \itm{\xi=-(1+\lambda)} and in correspondence of which $\eta=\etab>0$ [\REq{mp.left.11p}].
The eigenfunction's components [\REqd{ndevp.ef.1}{ndevp.s.1p.i}] can be soldered mathematically with the continuity joint
\begin{subequations}\label{jp11p} \seqn
  \begin{gather}
     \phi_{1}\left(-(1+\lambda)\right)          = \phi_{1'}(\etab)                                                                  \label{jp11p.ef}  \\[.25\baselineskip]
     \pdatb{}{\phi_{1}}{\xi}{\xi=-(1+\lambda)}  = - \left(\dfrac{v_{1}}{\lambda}\right)^{1/3}\pdatb{}{\phi_{1'}}{\eta}{\eta=\etab}  \label{jp11p.efd}
  \end{gather}
\end{subequations}
The right-hand side of \REq{jp11p.efd} descends from the derivative transformation indicated in \REq{ivt.der.1p}.
After derivatives are done and all necessary substitutions are in place, \REqq{jp11p} evolve into the algebraic system
\begin{subequations}\label{jp11p.as} \seqn
  \begin{align}
     B_{1} \exp[-(1+\lambda)\sqrt{k_{1}}] & = A_{1'}\,\Ai(\etab) + B_{1'}\,\Bi(\etab)                                               \label{jp11p.as.ef}  \\[.25\baselineskip]
     B_{1} \exp[-(1+\lambda)\sqrt{k_{1}}] & = - \dfrac{1}{\sqrt{\etab}} \left[ A_{1'}\,\Ai'(\etab) + B_{1'}\,\Bi'(\etab) \right]    \label{jp11p.as.efd}
  \end{align}
\end{subequations}
On the right-hand side of \REq{jp11p.as.efd}, we have complied with the standard notation \cite{ma1972,ov2004} reserved for the first derivatives of the Airy functions.
\REqqb{jp11p.as} fix two coefficients in terms of a third one; to that aim, they can be subtracted and rearranged as
\begin{equation}\label{jp11p.Ap}
    A_{1'}\,\Ai(\etab) + B_{1'}\,\Bi(\etab) = - \dfrac{1}{\sqrt{\etab}} \left[ A_{1'}\,\Ai'(\etab) + B_{1'}\,\Bi'(\etab) \right]
\end{equation}
to extract, for example, the coefficient $A_{1'}$
\begin{equation}\label{A1p}
    A_{1'} = - B_{1'}\cdot f_{1'}
\end{equation}
The factor $f_{1'}$ in \REq{A1p} is conveniently set to
\begin{equation}\label{fbar}
    f_{1'} = \frac{\sqrt{\etab}\,\,\Bi(\etab) + \Bi'(\etab)}{\sqrt{\etab}\,\,\Ai(\etab) + \Ai'(\etab) }
\end{equation}
to simplify the notation; it is always real because $\etab>0$. 
The coefficient $B_{1}$ can then be obtained from \REqq{jp11p.as} in two different but, obviously, equivalent ways
\begin{equation}\label{B1}
    B_{1} =   B_{1'}                        \left[ \Bi(\etab)  - f_{1'}\,\Ai(\etab) \right] \exp[(1+\lambda)\sqrt{k_{1}}]
          = -   \dfrac{B_{1'}}{\sqrt{\etab}}\left[ \Bi'(\etab) - f_{1'}\,\Ai'(\etab) \right] \exp[(1+\lambda)\sqrt{k_{1}}]
\end{equation}
from which we also extract, as collateral result, a useful identity
\begin{equation}\label{fbar.i}
    \Bi(\etab) - f_{1'}\,\Ai(\etab) = - \dfrac{\Bi'(\etab) - f_{1'}\,\Ai'(\etab)}{\sqrt{\etab}}
\end{equation}
that permits to interchange Airy's functions with their first derivatives and viceversa; we can also deduce \REq{fbar.i} from appropriate rearrangement of \REq{fbar}; after proper generalization, it will prove useful in \Rse{mt}.

Basically,  we must apply repeatedly the procedure followed for the junction point \jp{1}{1'} to the other junction points.
Let us see where it leads to for junction point $\jp{1'}{0}$ located at $\xi=-1$ and for which $\eta=\etah$.
The continuity requirement
\begin{subequations}\label{jp1p0} \seqn
  \begin{gather}
     \phi_{1'}(\etah) = \phi_{0}(-1)                                                                                       \label{jp1p0.ef}  \\[.25\baselineskip]
     - \left(\dfrac{v_{1}}{\lambda}\right)^{1/3} \pdatb{}{\phi_{1'}}{\eta}{\eta=\etah} = \pdatb{}{\phi_{\mspace{1mu}0}}{\xi}{\xi=-1}   \label{jp1p0.efd}
  \end{gather}
\end{subequations}
generates the algebraic system
\begin{subequations}\label{jp1p0.as} \seqn
  \begin{align}
      B_{1'} \left[ \Bi(\etah)  - f_{1'}\,\Ai(\etah)  \right]                       & = -A_{0} \sin(\sqrt{\beta}) + B_{0} \cos(\sqrt{\beta})    \label{jp1p0.as.ef}  \\[.25\baselineskip]
    - \frac{B_{1'}}{\sqrt{-\etah}} \left[ \Bi'(\etah) - f_{1'}\,\Ai'(\etah) \right] & =  A_{0} \cos(\sqrt{\beta}) + B_{0} \sin(\sqrt{\beta})    \label{jp1p0.as.efd}
  \end{align}
\end{subequations}
in which only the coefficient $B_{1'}$ appears because we exclude the coefficient $A_{1'}$ with the aid of \REq{A1p}.
Member-to-member division of \REqq{jp1p0.as} eliminates the former coefficient 
\begin{equation}\label{A0B0.1}
   \frac{- A_{0} \sin(\sqrt{\beta}) + B_{0} \cos(\sqrt{\beta})}{A_{0} \cos(\sqrt{\beta}) + B_{0} \sin(\sqrt{\beta})} = \gl
\end{equation}
The factor $\gl$ in \REq{A0B0.1} is set to
\begin{equation}\label{gleft}
   \gl = - \sqrt{-\etah} \frac{\Bi(\etah)  - f_{1'}\,\Ai(\etah)}{\Bi'(\etah)  - f_{1'}\,\Ai'(\etah)}
\end{equation}
again to simplify the notation; it is either real or pure imaginary according to the sign of the eigenvalue [\REq{mp.left.1p0}].
We hold on \REq{A0B0.1} \label{holdA0B0} as it stands instead of proceeding to solve for one of the two coefficients appearing in it; 
the reason behind this decision will surface in \Rse{efc}.
The coefficient $B_{1'}$ follows from \REqq{jp1p0.as} in either of the two equivalent forms
\begin{equation}\label{B1p}
   B_{1'} = \frac{-A_{0} \sin(\sqrt{\beta}) + B_{0} \cos(\sqrt{\beta})}{\Bi(\etah)  - f_{1'}\,\Ai(\etah)}
          = - \sqrt{-\etah} \frac{ A_{0} \cos(\sqrt{\beta}) + B_{0} \sin(\sqrt{\beta})}{\Bi'(\etah) - f_{1'}\,\Ai'(\etah)}
\end{equation}

We trust the continuity-implementation recipe to be sufficiently clear by now.
Its application to the junction points \jp{0}{2'} and \jp{2'}{2} is nothing else than the conceptual mirroring of what we have done so far with the junction points \jp{1}{1'} and \jp{1'}{0}.
Therefore, we believe we can confidently skip the details and list only the final output.
The continuity requirement at the junction point \jp{2'}{2} located at \itm{\xi=1+\lambda} and for which $\zeta=\zetab>0$
\begin{subequations}\label{jp2p2} \seqn
  \begin{gather}
     \phi_{2'}(\zetab) = \phi_{2}(1+\lambda)                                                                   \label{jp2p2.ef}  \\[.25\baselineskip]
     \left(\dfrac{v_{2}}{\lambda}\right)^{1/3} \pdatb{}{\phi_{2'}}{\zeta}{\zeta=\zetab}
                                       = \pdatb{}{\phi_{2}}{\xi}{\xi=1+\lambda}   \label{jp2p2.efd}
  \end{gather}
\end{subequations}
leads to
\begin{equation}\label{A2p}
    A_{2'} = - B_{2'}\cdot f_{2'}
\end{equation}
with
\begin{equation}\label{f2p}
    f_{2'} = \frac{\sqrt{\zetab}\,\,\Bi(\zetab) + \Bi'(\zetab)}{\sqrt{\zetab}\,\,\Ai(\zetab) + \Ai'(\zetab) }
\end{equation}
and
\begin{equation}\label{B2}
    B_{2} =   B_{2'}                        \left[ \Bi(\zetab)  - f_{2'}\,\Ai(\zetab) \right] \exp[(1+\lambda)\sqrt{k_{2}}]
          = -  \dfrac{B_{2'}}{\sqrt{\zetab}}\left[ \Bi'(\zetab) - f_{2'}\,\Ai'(\zetab) \right] \exp[(1+\lambda)\sqrt{k_{2}}]
\end{equation}
The continuity requirement at the junction point \jp{0}{2'} located at $\xi=1$ and for which $\zeta=\zetah$
\begin{subequations}\label{jp02p} \seqn
  \begin{gather}
     \phi_{0}(1)                    = \phi_{2'}(\zetah)                                                                    \label{jp02p.ef}  \\[.25\baselineskip]
     \pdatb{}{\phi_{0}}{\xi}{\xi=1} = + \left(\dfrac{v_{2}}{\lambda}\right)^{1/3} \pdatb{}{\phi_{2'}}{\zeta}{\zeta=\zetah} \label{jp02p.efd}
  \end{gather}
\end{subequations}
produces a second equation involving the coefficients $A_{0}$ and $B_{0}$
\begin{equation}\label{A0B0.2}
   \frac{A_{0} \sin(\sqrt{\beta}) + B_{0} \cos(\sqrt{\beta})}{A_{0} \cos(\sqrt{\beta}) - B_{0} \sin(\sqrt{\beta})} = \gr
\end{equation}
with
\begin{equation}\label{gright}
   \gr = + \sqrt{-\zetah} \frac{\Bi(\zetah)  - f_{2'}\,\Ai(\zetah)}{\Bi'(\zetah)  - f_{2'}\,\Ai'(\zetah)}
\end{equation}
and fixes the coefficient $B_{2'}$
\begin{equation}\label{B2p}
   B_{2'} = \frac{A_{0} \sin(\sqrt{\beta}) + B_{0} \cos(\sqrt{\beta})}{\Bi(\zetah)  - f_{2'}\,\Ai(\zetah)}
          = \sqrt{-\zetah} \frac{ A_{0} \cos(\sqrt{\beta}) - B_{0} \sin(\sqrt{\beta})}{\Bi'(\zetah) - f_{2'}\,\Ai'(\zetah)}
\end{equation}
The factor $f_{2'}$ is always real because $\zetab>0$ [\REq{mp.right.2p2}]; the factor $\gr$ is either real or pure imaginary according to the sign of the eigenvalue [\REq{mp.right.02p}].

As anticipated in the beginning of this section, we have obtained eight equations [\REq{A1p}, \REq{B1}, \REq{A0B0.1}, \REq{B1p}, \REq{A2p}, \REq{B2}, \REq{A0B0.2}, \REq{B2p}] to determine the eight coefficients and the eigenvalue; we still have \REq{efn.nd} in reserve but, right now, its exploitation is not required yet.
\REqdb{A0B0.1}{A0B0.2} are those of utmost importance and deserve particular attention because they generate the eigenvalues.
We take up their study in next section.

We wish to conclude with a reassurance to the reader concerned with the listed equations' seeming mathematical cumbersomeness, perhaps particularly perceived from the presence of Airy functions and their first derivatives.
We did the coding in \textsc{octave}, a programming language within which Airy functions and derivatives are built-in intrinsic functions, and the calculations went smooth and flawless.

\subsubsection{Eigenvalues \label{ev}}

Let us rewrite \REqd{A0B0.1}{A0B0.2} in a slightly rearranged but more convenient form 
\newcommand{\sublAzBz}{$_{1}$}
\newcommand{\eqsubone}{$_{1}$}
  \begin{align}
      A_{0} \left[ \sin(\sqrt{\beta}) + \gl \cos(\sqrt{\beta}) \right] + B_{0} \left[ \gl \sin(\sqrt{\beta}) - \cos(\sqrt{\beta}) \right] & = 0 \taglabel{A0B0.1}{\eqsubone} \label{A0B0.1.1}  \\[.25\baselineskip]
      A_{0} \left[ \sin(\sqrt{\beta}) - \gr \cos(\sqrt{\beta}) \right] + B_{0} \left[ \gr \sin(\sqrt{\beta}) + \cos(\sqrt{\beta}) \right] & = 0 \taglabel{A0B0.2}{\eqsubone} \label{A0B0.2.1}
  \end{align}
and let us look at it as a homogeneous algebraic system
\begin{equation}\label{A0B0.m}
  \begin{bmatrix}
     \,\sin(\sqrt{\beta}) + \gl \cos(\sqrt{\beta}) \quad & \gl \sin(\sqrt{\beta}) - \cos(\sqrt{\beta})\, \\[1.5ex]
       \sin(\sqrt{\beta}) - \gr \cos(\sqrt{\beta}) \quad & \gr \sin(\sqrt{\beta}) + \cos(\sqrt{\beta})
  \end{bmatrix} \cdot
  \begin{bmatrix} A_0 \\[1.5ex] B_{0} \end{bmatrix} = 0
\end{equation}
for the coefficients $A_{0}, B_{0}$.
The vanishing of its determinant leads to the transcendental equation
\begin{equation}\label{eigenv}
   D(\beta) = \left( 1 + \gl\gr \right) \sin(2\sqrt{\beta}) + \left( \gl - \gr \right) \cos(2\sqrt{\beta}) = 0
\end{equation}
that generates the eigenvalues.
Basically, all eigenvalue-generating equations encountered in the literature we consulted, textbooks \cite{ep1936,ep1950,am11961,dth1964,ls1968,cct1977,ll1977,db1989,sf1999g,sf1999e,bb2000,df2001,rg2004,dg2005,pa2005,pt2012} as well as specialized papers  \cite{pp1955ajp,rm1976ajp,cs1978jomp,bcr1990ajp,ds1992ejp,da2000ajp,pp2000jomp,rb2005jopa,oda2006ajp,vb2021oam,sdv2013rmdf,vb2013pm,vb2015pm,krn2015.05arxiv}, are particular cases embedded in \REq{eigenv}, all with \mbox{$\lambda=0$} obviously, most of them with symmetrical potential \mbox{($v_{1}=v_{2}$)} and just a few \cite{am11961,dth1964,ll1977} with unsymmetrical potential \mbox{($v_{1}\neq v_{2}$)}.
Very ingenious analytical as well as graphical ways have been proposed and exploited to extract the roots of those transcendental equations; however, the exploitability of these options, although sometimes still rather elaborated mathematically, is possible only for relatively simplified situations, such as the one involving symmetrical potentials for example.
The mathematical transcendence of \REq{eigenv} with respect to the variable $\beta$ is extreme in our case with \mbox{$\lambda\neq 0$} because it concatenates the complexity of \REqd{mp.left}{mp.right}, \REqd{fbar}{f2p}, \REqd{gleft}{gright}.
Therefore, we had no other option than to follow a numerical approach based on the Newton-Raphson method, a fruitful idea proposed by Memory \cite{jm1977ajp} already in 1977.
Of course, Barsan's warning \cite[bottom of page 3023]{vb2015pm}:
\begin{quote}
   ... the eigenvalue equations ... are transcendental equations, whose analytical solutions are difficult to obtain.
   Of course, they can be calculated numerically, with high precision, but their dependence on the physical parameters of the problem is totally lost.
\end{quote}
did not escape our attention but we believe that the fear of the loss mentioned in his last sentence is unfounded if one works with nondimensional variables.

The first question we may wish to settle regarding \REq{eigenv} concerns whether or not it can produce negative eigenvalues.
Usually, approaches in the literature \cite{pp1955ajp,dth1964,ls1968,bb2000,dg2005,oda2006ajp,pt2012,vb2015pm,krn2015.05arxiv} deduce the answer \textit{a posteriori} within the search of the eigenvalues with graphical methods; but we feel more comfortable with an approach helped by analytical support.
The path to follow consists in assuming hypothetically \mbox{$\beta<0$}, applying the switch of \REq{nbetas} and working out the consequences on the function $D(\beta)$. 
The expression found at the end of the mathematical manipulations turns out to be a pure imaginary non-linear combination of hyperbolic functions
\begin{equation}\label{eigenv.n}
   D(\beta) = i \left[ \left( 1 + \gl\,\gr \right) \sinh(+2\sqrt{-\beta}) - i \left( \gl - \gr \right) \cosh(2\sqrt{-\beta})  \right]
\end{equation}
\begin{figure}[h]
\dgcap{.05}{.05}
   \subfloat[\twp]         {\label{nev.twp} \resizebox{.47\textwidth}{!}{\includegraphics*[trim=33 20 60 60]{\gdir/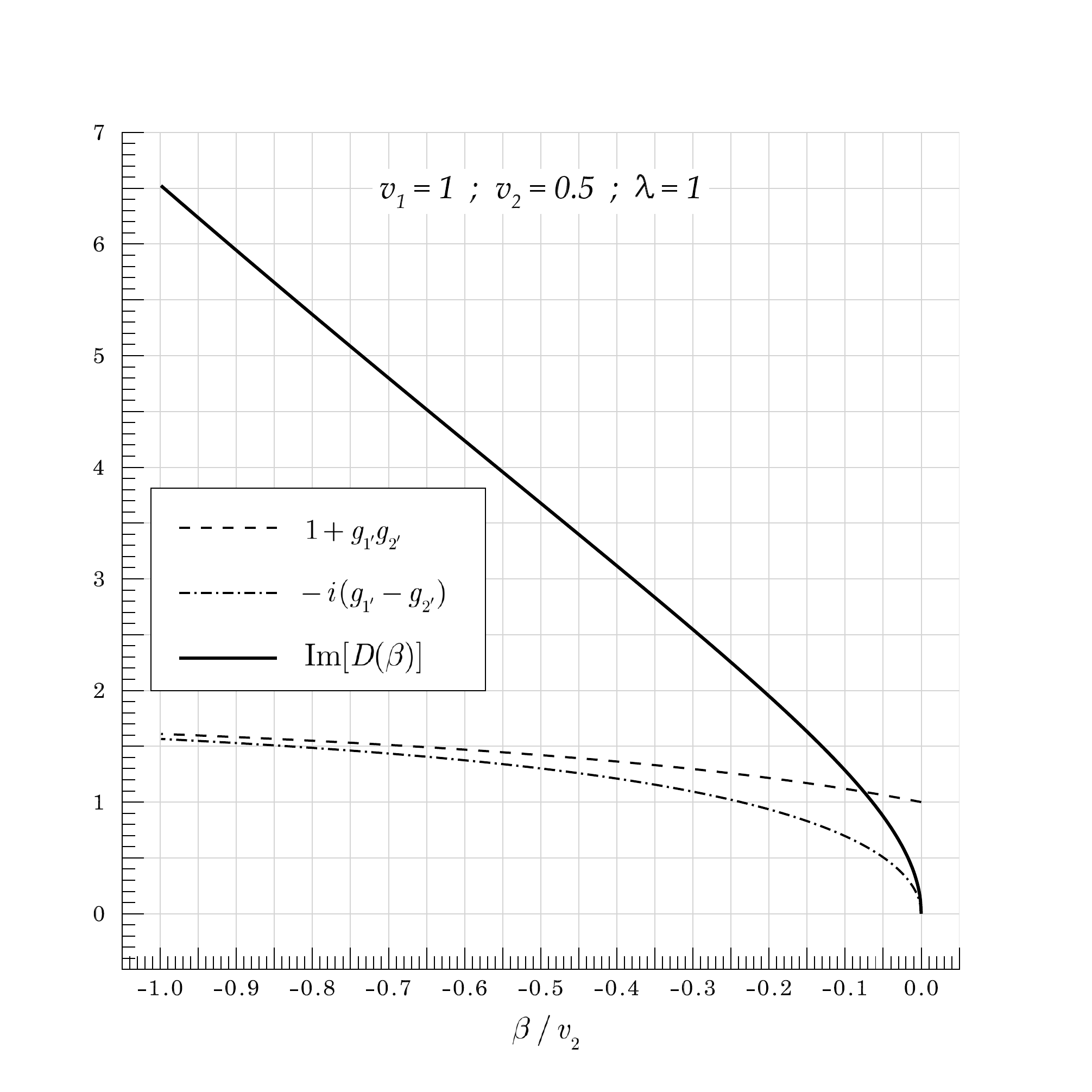}}}
   \subfloat[Virtual \swp] {\label{nev.swp} \resizebox{.47\textwidth}{!}{\includegraphics*[trim=33 20 60 60]{\gdir/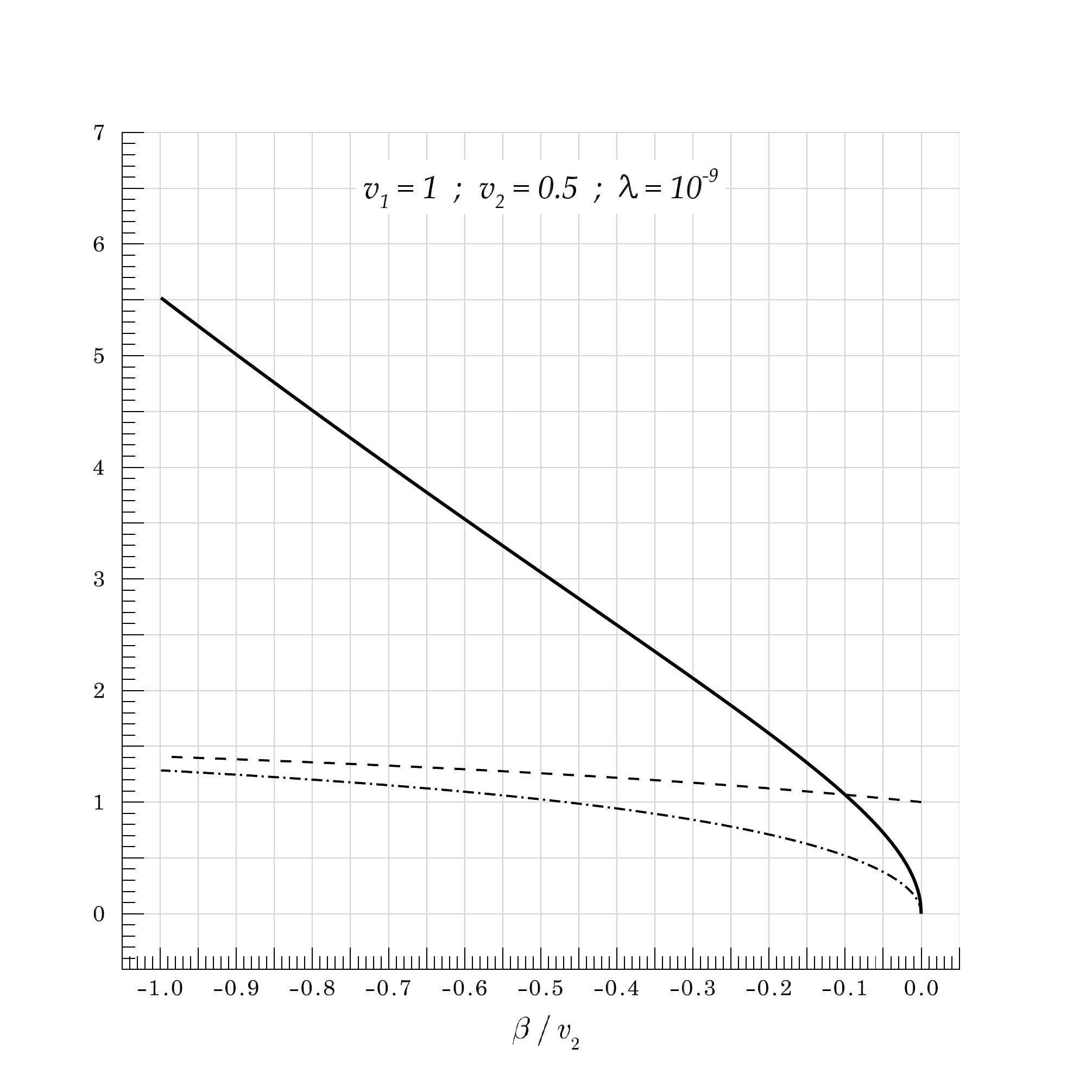}}}
   \caption{The function $\imdb$ and the coefficients of the hyperbolic functions in \REq{eigenv.n} versus $\beta/v_{2}$ in the interval [-1,0].\llpush\label{nev}}
\end{figure}
%
The quantity in square brackets is real and, therefore, represents $\imdb$ because the factors $\gl, \gr$ are pure imaginary [\REq{mp.left.1p0}, \REq{mp.right.02p}, \REqd{gleft}{gright}].
The hyperbolic functions are always positive; therefore, the responsibility for the sign of $\imdb$ falls on their coefficients, which, let us not forget, also depend on $\beta$.
Given the mathematical cumbersomeness of the coefficients, what we need to do is to draw their graphs versus $\beta$ to understand their behavior.
\Rfib{nev} provides two examples: a \twp\ with \itm{v_{1}=1, v_{2}=0.5, \lambda=1} in \Rfi{nev.twp} and a virtual\footnote{We trust the reader would agree with the assertion that a \twp\ with a steepness characterized by $\lambda=10^{-9}$ can be considered \textit{square} for all practical purposes.} \swp\  with \mbox{$v_{1}=1, v_{2}=0.5, \lambda=10^{-9}$} in \Rfi{nev.swp}.
They indicate that the coefficients of the hyperbolic functions are monotonic and positive and the function $\imdb$ never vanishes on the left of \itm{\beta=0}.
We have tested several combinations of the characteristic numbers \mbox{$v_{1}, v_{2}, \lambda$} and found out that the curves expectedly shift a bit but their monotonicity is never compromised and the general picture remains similar to those shown in \Rfi{nev}.
So, we can rest assured that negative eigenvalues do not exist and conclude that the eigenvalues are also bounded from below; then, \REq{betas.lt.vs} upgrades to the final form
\begin{subequations}\label{..betas.b}\seqn
	\begin{equation}\label{betas.b.a}
	   0 < \beta \leq v_{2}
	\end{equation}
	or even better
	\begin{equation}\label{betas.b.r}
	   0 < \frac{\beta}{v_{2}} \leq 1
	\end{equation}
\end{subequations}
Thus, all eigenvalues reside within the potential well; accordingly, we can forget \REq{ndevp.s.0.i.nev} and retain exclusively \REq{ndevp.s.0.i} as eigenfunction's component in the central zone.

\begin{figure}[h]
\dgcap{.05}{.05}
   \subfloat[Coefficients] {\label{c1-D.coef}\resizebox{.47\textwidth}{!}{\includegraphics*[trim=33 20 60 60]{\gdir/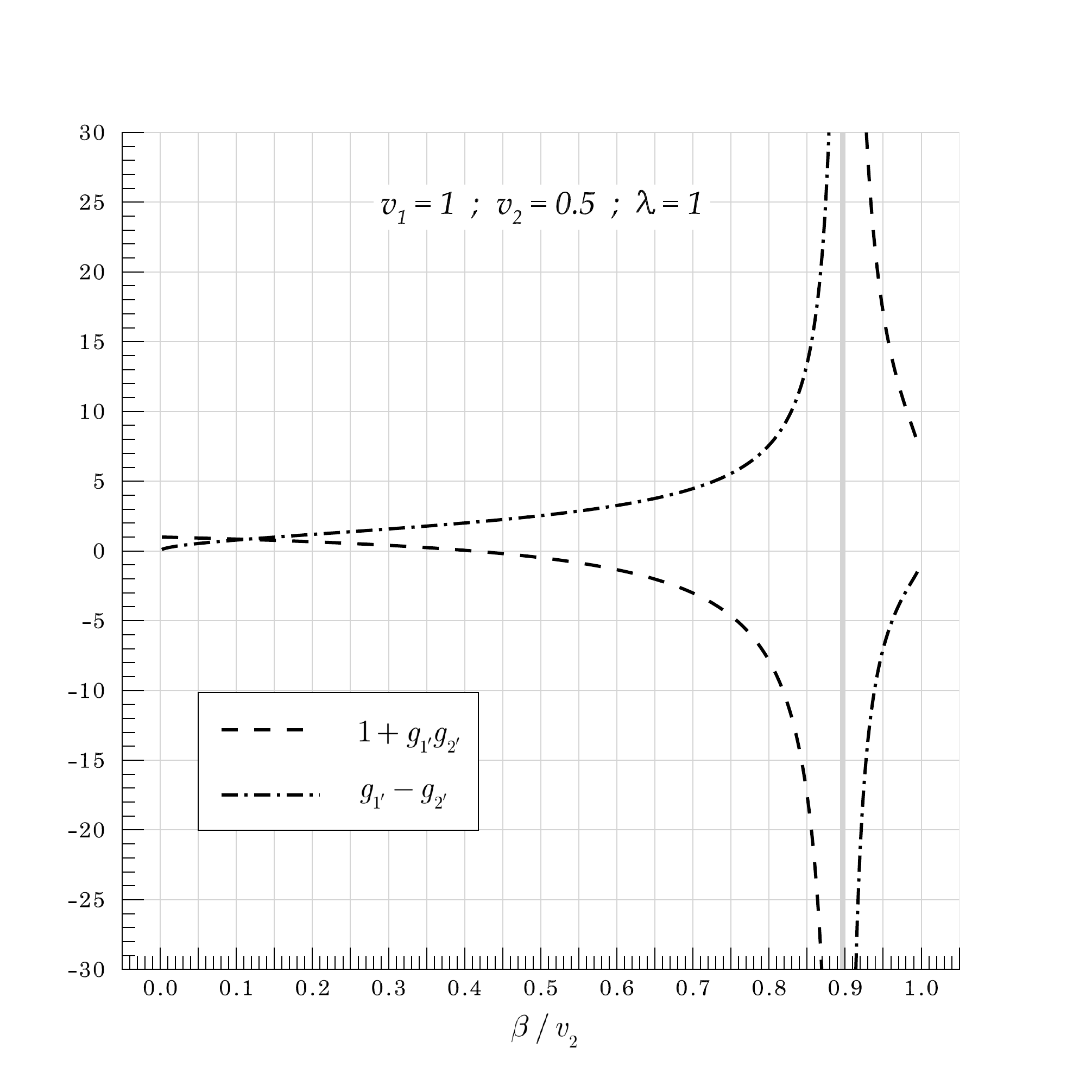}}}
   \subfloat[Determinant]  {\label{c1-D.fun} \resizebox{.47\textwidth}{!}{\includegraphics*[trim=33 20 60 60]{\gdir/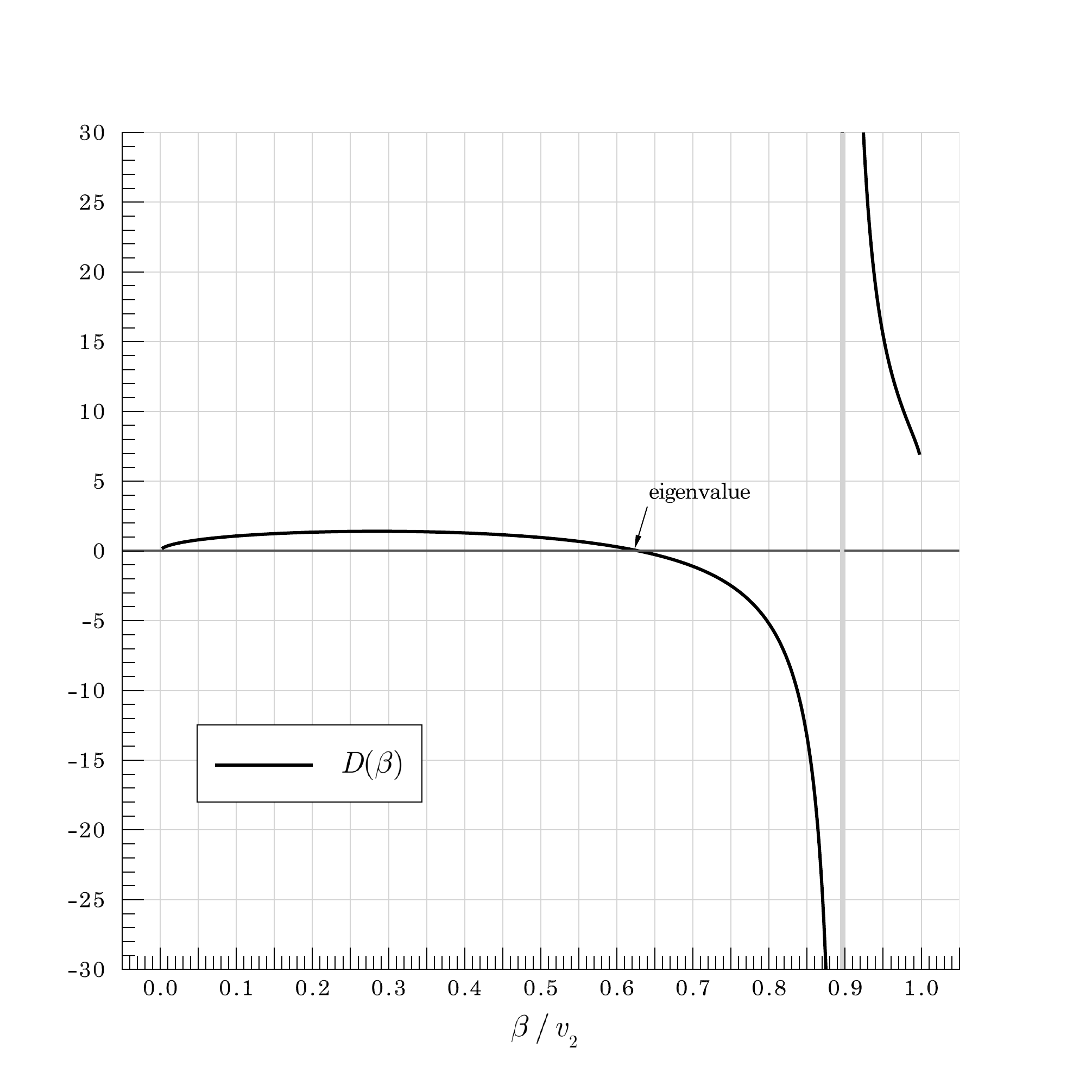}}}
   \caption{The coefficients of the trigonometric functions in \REq{eigenv} and the function $D(\beta)$ versus $\beta/v_{2}$ in the interval [0,1] featuring one eigenvalue in $[0.62, 0.63]$ and a vertical asymptote at $\sim 0.9$.\llpush\label{c1-D}}
\end{figure}
%
We return now to the original transcendental equation [\REq{eigenv}] and concentrate on the determination of its roots.
The strategy consists in plotting the function $D(\beta)$ versus $\beta/v_{2}$, detecting visually the intersections with the horizontal axis in order to extract initial-guess values for $\beta/v_{2}$ and then launching the numerical algorithm based on the Newton-Raphson method.
The latter involves the derivative $dD/d\beta$ whose determination requires a bit of careful attention to mathematical details but, in general, it works very well with convergence residuals of the order $10^{-10}$ achieved in just a few iterations.
We have discovered that the coefficients of the trigonometric functions in \REq{eigenv} and, by reflection, the function $D(\beta)$ may present vertical asymptotes for some specific triplets of $v_{1}, v_{2}, \lambda$, as shown in the example of \Rfi{c1-D}, but, fortunately, such occurrences do not hamper the convergence of the method's iteration procedure if the initial value of $\beta/v_{2}$ is appropriately chosen.
Nevertheless, we have probed \REq{eigenv} from different angles in order to obtain alternative forms freed from unaesthetic infinities inside the interval [0,1].
An effective cure, we found out, consists in introducing the reciprocal factors
\begin{equation}\label{oog}
   \subeqn{\gamma_{s} = \frac{1}{g_{s}}}{s=1',2'}{0}
\end{equation}
whose substitution in \REq{eigenv} leads to another transcendental equation
\begin{equation}\label{eigenv.rf}
   D^{\circ}(\beta) = \left( \gal\gar + 1 \right) \sin(2\sqrt{\beta}) + \left( \gar - \gal \right) \cos(2\sqrt{\beta}) = 0
\end{equation}%
\begin{figure}[h]
\dgcap{.06}{.06}
   \subfloat[Coefficients] {\label{c1-Dc.coef}\resizebox{.47\textwidth}{!}{\includegraphics*[trim=33 20 60 60]{\gdir/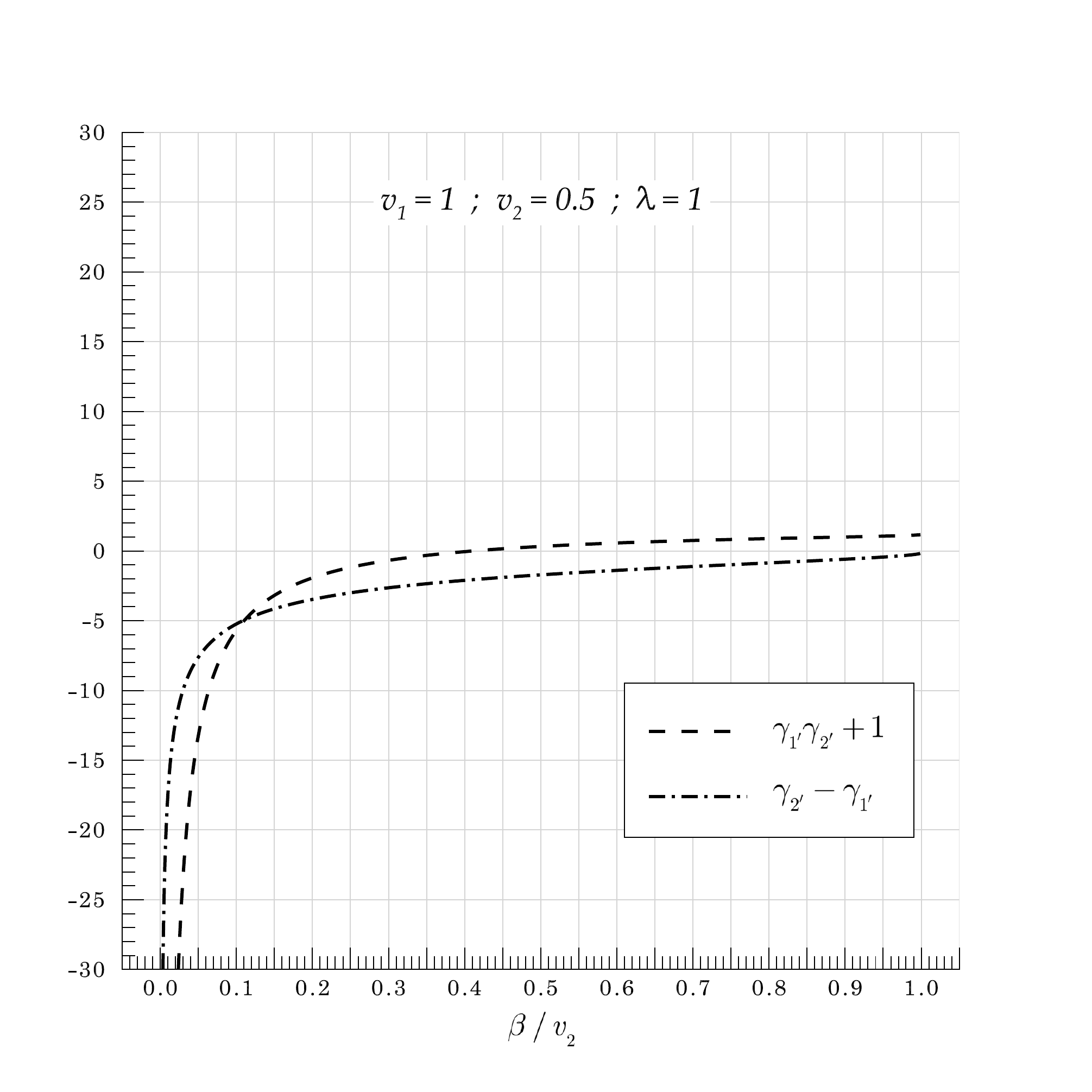}}}
   \subfloat[Determinant]  {\label{c1-Dc.fun} \resizebox{.47\textwidth}{!}{\includegraphics*[trim=33 20 60 60]{\gdir/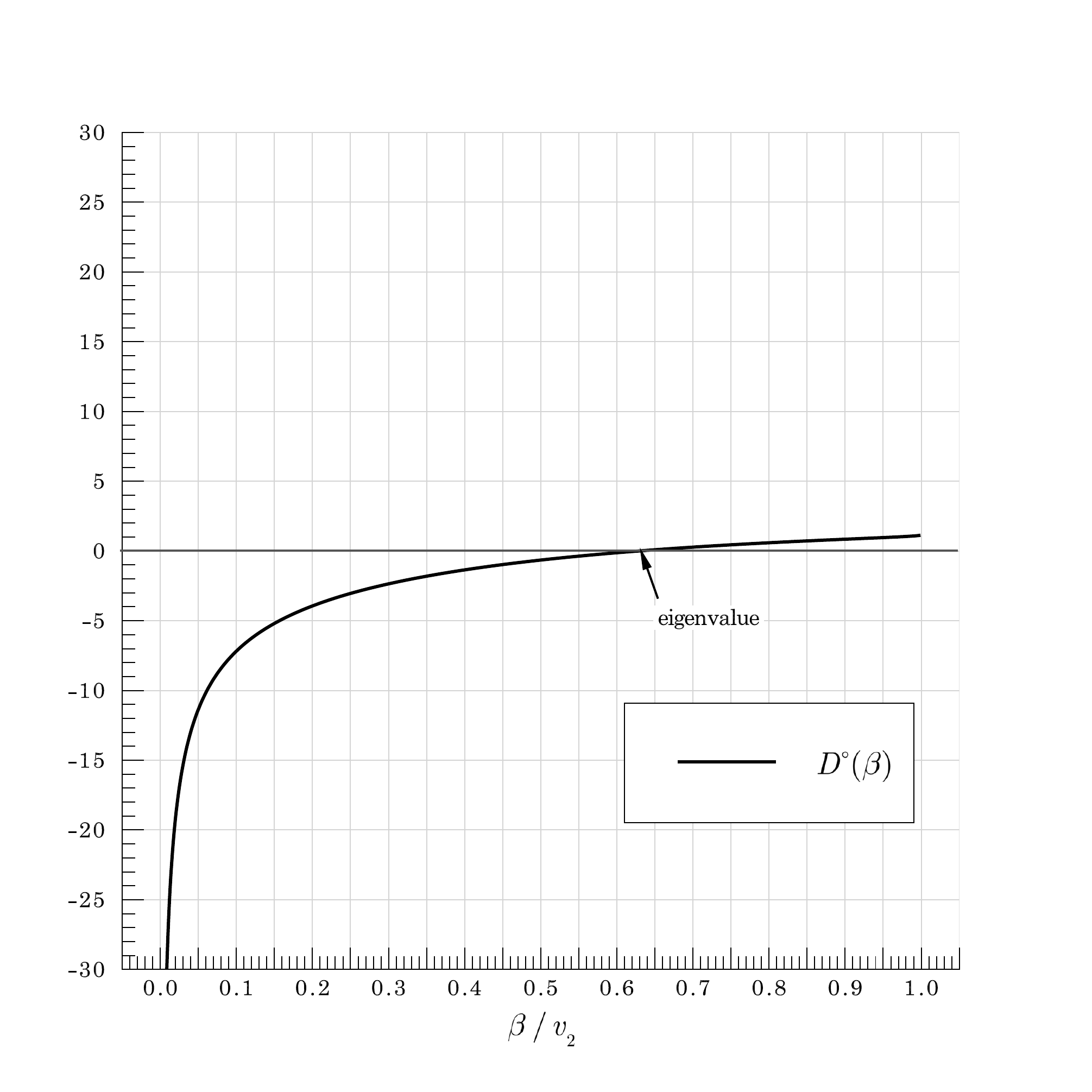}}}
   \caption{The coefficients of the trigonometric functions in \REq{eigenv.rf} and the function $D^{\circ}(\beta)$ versus $\beta/v_{2}$ in the interval [0,1].\llpush\label{c1-Dc}}
\end{figure}
%
\hspace*{-1ex}on the basis of which the graphs of \Rfi{c1-D} evolve into those of \Rfi{c1-Dc} that give evidence of how the curves acquire a beneficent monotonicity and are better behaved; true, there is still a vertical asymptote at $\beta=0$ but it is innocuous because its position is frozen with respect to the values of the triplet $v_{1}, v_{2}, \lambda$.
Further improvement is possible and can be achieved by following the guidelines of the smart idea proposed by Sprung and coauthors \cite{ds1992ejp} in 1992.
Let us write for brevity
\begin{subequations}\label{coeff} \seqn
  \begin{align}
     C & = \gal\gar + 1 \label{coeff.C} \\[.25\baselineskip]
     S & = \gar - \gal  \label{coeff.S}
  \end{align}
and define the normalization factor
  \begin{equation}     \label{coeff.R}
     R = \sqrt{C^{2} + S^{2}} = \sqrt{ \left( 1 + \gal^{2} \right) \left( 1 + \gar^{2} \right) }
  \end{equation}
\end{subequations}
Then the ratios $C/R$ and $S/R$ are bound within the interval [-1,\tsplus 1] and permit the introduction of the angle $\varphi$ defined by
\begin{subequations}\label{angle} \seqn
  \begin{align}
    \cos\varphi & = - \frac{C}{R} \label{cosphi} \\[.25\baselineskip]
    \sin\varphi & = - \frac{S}{R} \label{sinphi}
  \end{align}
\end{subequations}
The minus-sign choice in \REqq{angle} counteracts the negativity of the ratio $C/R$ which tends to -1 when $\beta$ approaches zero and compels the convenient initial condition \itm{\varphi(\beta\rightarrow0)=0}.\footnote{The positive-sign alternative leads to the initial condition \itm{\varphi(\beta\rightarrow0)=\pi}, an unnecessary complication that may introduce a risk of confusion when we need to invert \REqq{angle} to obtain the angle $\varphi$.}
In the sequel, we will imply the dependence of $\varphi$ on $\beta, v_{1}, v_{2}, \lambda$ and will explicit it only if and as required by the context.
The division of \REq{eigenv.rf} by the normalization factor $R$ produces an ulterior version of transcendental equation
\begin{equation}\label{eigenv.trig}
   D^{\ast}(\beta) = \frac{1}{R}D^{\circ}(\beta) = -\cos\varphi \sin(2\sqrt{\beta}) - \sin\varphi \cos(2\sqrt{\beta}) = - \sin\left( 2\sqrt{\beta} + \varphi(\beta) \right) = 0
\end{equation}
We can obviously go one step further and pull out from \REq{eigenv.trig} the solution in terms of angles
\begin{subequations}\label{eigenv.trig.angle} \seqn
	\begin{equation}\label{eigenv.trig.angle.kpi}
	   \subeqn{2\sqrt{\beta} + \varphi(\beta) = n \pi}{n=1,2,\ldots}{0}
	\end{equation}
	or
	\begin{equation}\label{eigenv.trig.angle.onpi}
	   \subeqn{\theta(\beta) = \frac{2\sqrt{\beta} + \varphi(\beta)}{\pi} = n}{n=1,2,\ldots}{0}
	\end{equation}
	after a slightly more convenient rearrangement.
\end{subequations}
Typical graphs of the functions $D^{\ast}(\beta)$ and $\theta(\beta)$ are shown in \Rfi{c1-Da};
\begin{figure}[h]
   \captionsetup{justification=centerlast,margin={.2\textwidth,.2\textwidth}}
   \resizebox{.47\textwidth}{!}{\includegraphics*[trim=33 20 60 60]{\gdir/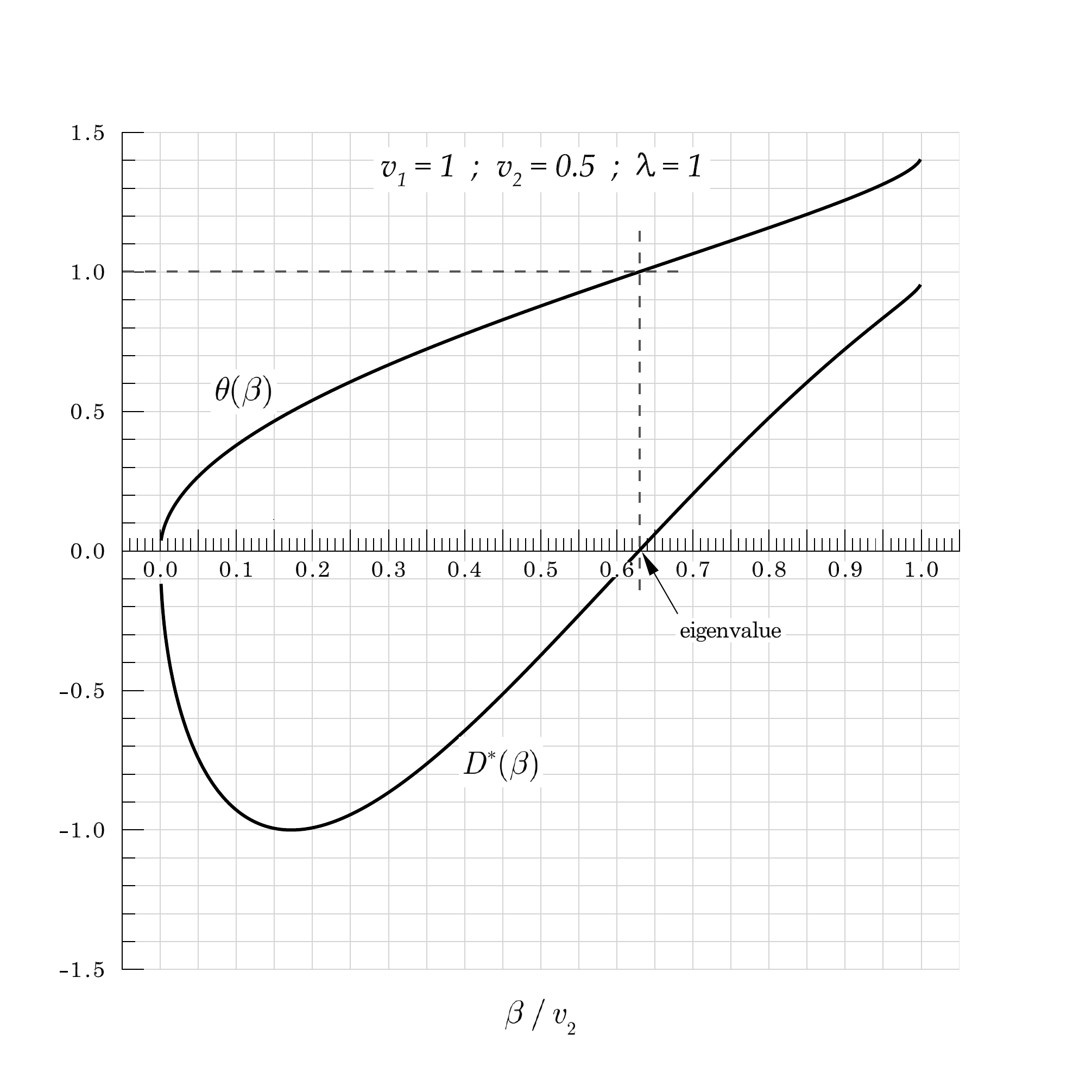}}
   \caption{The functions $D^{\ast}(\beta)$ and $\theta(\beta)$ versus $\beta/v_{2}$ in the interval [0,1]; in correspondence to the eigenvalue, the former function vanishes [\REq{eigenv.trig}] and the latter function attains the integer value $n=1$ [\REq{eigenv.trig.angle.onpi}].\llpush\label{c1-Da}}
\end{figure}
%
we believe that they are definitely more visually representative and elegant than those of the functions $D(\beta)$ and $D^{\circ}(\beta)$ illustrated in \Rfid{c1-D.fun}{c1-Dc.fun}, respectively; nevertheless, in our experience, the Newton-Raphson method works well with each one of the mentioned functions.
We believe appropriate to remark two important aspects.
First, we have to keep in mind that \REqd{eigenv.trig}{eigenv.trig.angle.onpi} are still transcendental equations, although they look apparently simpler with respect to \REqd{eigenv}{eigenv.rf}; the complexity is hidden behind the angle $\varphi$ and involves all the formulae, encountered previously, which we need to navigate through to obtain it.
Second, \REq{eigenv.trig.angle.kpi} tells us that the angle $\varphi$ can be interpreted as a quantitative measure of the conceptual difference between the eigenvalue spectrum of our \twp\ (\Rfi{ndtwp}) and that of the infinite \swp, for which we are led to anticipate that \mbox{$\varphi\rightarrow 0$} from the visual inspection of \REq{eigenv.trig.angle.kpi}, in terms of well's finiteness, asymmetry and trapezoidal shape.

We selected two test cases to validate both the Newton-Raphson algorithm to find the roots of the described transcendental equations and the finite-difference numerical method that solves \REqq{ndevp} and produces collaterally the eigenvalues;
Reed \cite{bcr1990ajp} considered an electron ($m = 9.1093837015\cdot10^{-31}$ kg) in a finite symmetrical \swp\ of depth \mbox{$V_{0}=V_{1}=V_{2} = 100$ eV} and semi-width \itm{L=1}\AAo\;;
\begin{figure}[h]
   \subfloat[Reed \cite{bcr1990ajp}]                       {\label{evdg.reed}\resizebox{.47\textwidth}{!}{\includegraphics*[trim=33 20 60 60]{\gdir/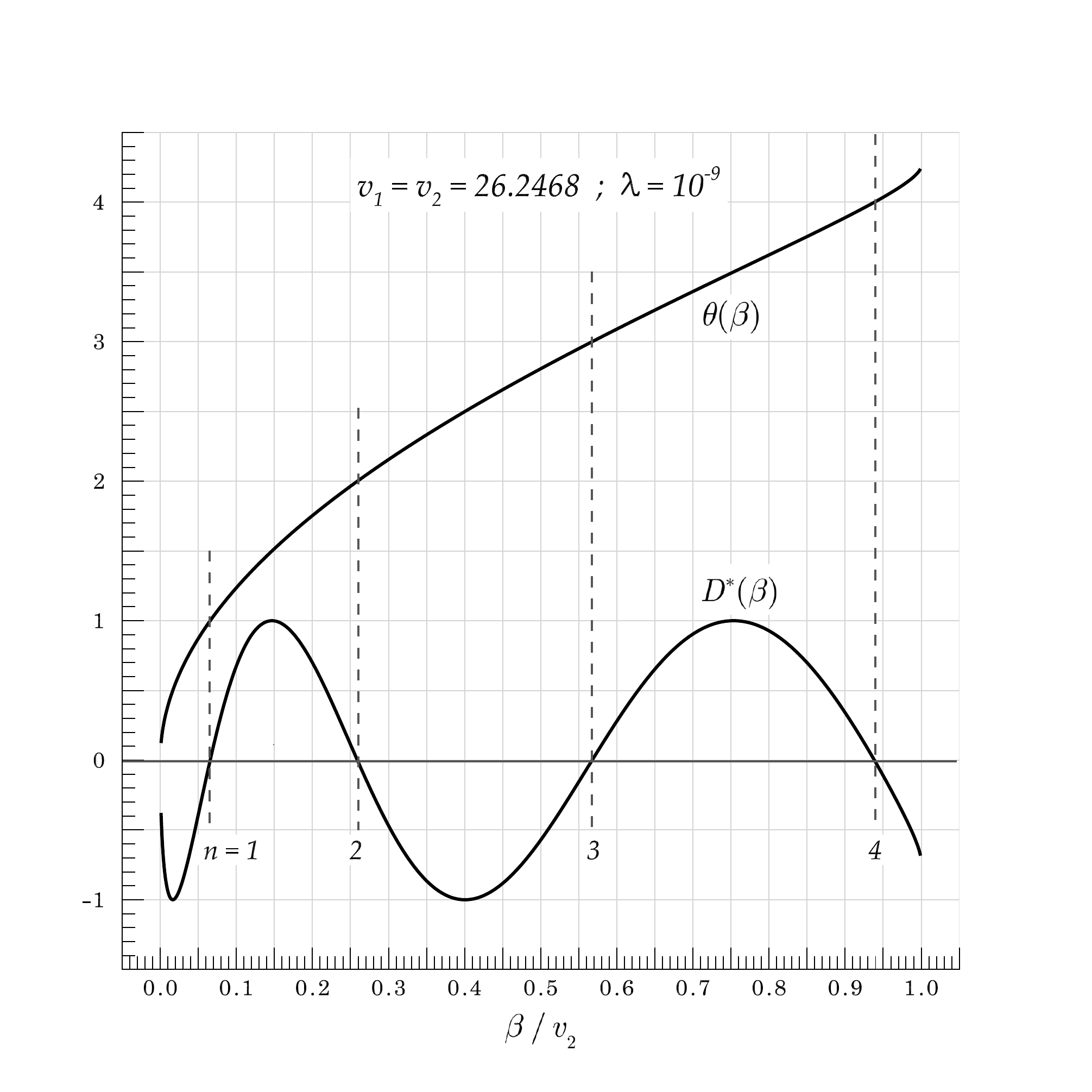}}}
   \subfloat[de Alcantara and Griffiths \cite{oda2006ajp}] {\label{evdg.dabg}\resizebox{.47\textwidth}{!}{\includegraphics*[trim=33 20 60 60]{\gdir/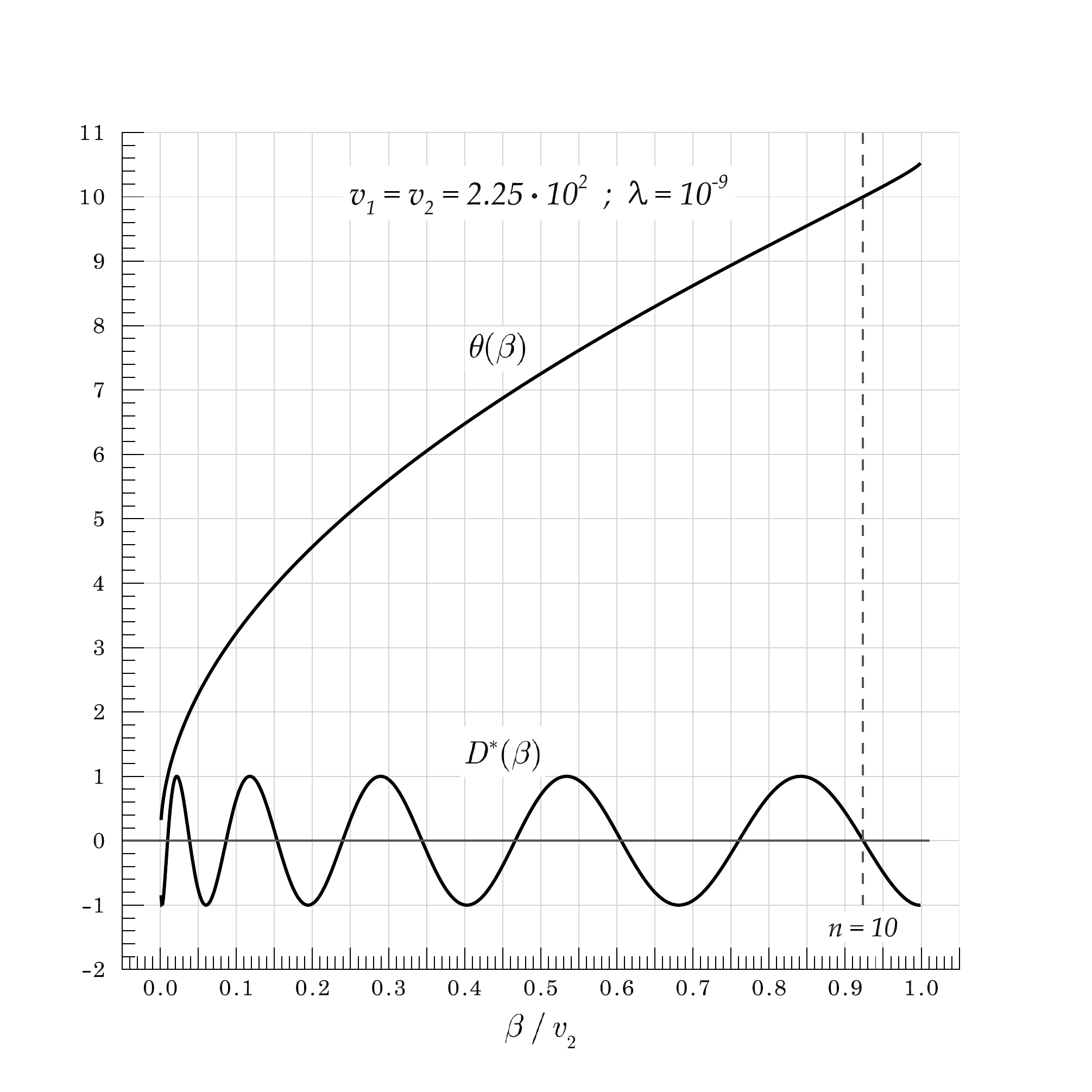}}}
   \caption{Eigenvalue-detection graphs for the selected test cases.\label{evdgtc}}
\end{figure}
the four eigenvalues he found by utilizing a bisection method are listed at page 504 (bottom of the left column) of his article.
From \REq{ndv1.2}, with the recommended\footnote{\url{https://www.nist.gov/pml/fundamental-physical-constants}} values \mbox{$\hbar = 1.054571817\cdot 10^{-34}$ J$\cdot$s}, \mbox{1 J$\,= 6.24150907446076\cdot 10^{18}$ eV} and \mbox{1 \AAo\ $ = 10^{-10}$ m}, we obtained \mbox{$v_{1}=v_{2}\simeq 26.2468$} and set \itm{\lambda=10^{-9}} to simulate the potential's squareness.
The eigenvalue-detection graph shown in \Rfi{evdg.reed} confirms the existence of four eigenvalues; the $D^{\ast}(\beta)$ curve is a stretched sinusoid similar to the one provided by Reed in his \mbox{Fig. 1} at page 504 of \cite{bcr1990ajp}.
The dashed lines emphasize graphically how the zeros of $D^{\ast}(\beta)$ correspond systematically to integer values of $\theta(\beta)$ in compliance with \REq{eigenv.trig.angle.onpi}.
Our results are tabulated in the upper section of \Rta{evtc}; columns 4 and 5 from left contain those obtained, respectively, with the Newton-Raphson algorithm via \REq{eigenv.trig} and with the finite-difference numerical method.
Columns 6 and 7 contain the data generated from those of columns 3 and 4 post-processed to match Reed's format.
De Alcantara and Griffiths \cite{oda2006ajp} also considered a generic particle in a finite symmetrical \swp\ and specified directly the characteristic numbers \itm{\sqrt{v_{1}}=\sqrt{v_{2}}=15} ($z_{0}$ in their notation); the ten eigenvalues they found are tabulated in Table I at page 44 (bottom of left column) of their article.
The eigenvalue-detection graph shown in \Rfi{evdg.dabg} confirms the existence of ten eigenvalues and the values we found are listed in the lower section of \Rta{evtc}, again in columns 4 and 5;
the data of columns 3 and 6 correspond to de Alcantara and Griffiths' format.
For both test cases, the Newton-Raphson algorithm and the finite-difference numerical method are in full agreement and our eigenvalues match all the significant digits of the eigenvalues reported in the original articles.\\ \hspace*{-10em}
\begin{table}[h]
\dgcap{.15}{.15}
\caption{\color{black}Test cases for the validation of the algorithm based on the Newton-Raphson method [\REq{eigenv.trig}] and of the finite-difference numerical method.
            The main header contains our notation; the sub-headers contain (in parentheses) the original notation adopted in the indicated references.
            All our calculations were carried out with $\lambda=10^{-9}$.\llpush\label{evtc}}%
\vspace*{-.8\baselineskip}\resizebox{.69\textwidth}{!}{\includegraphics*[trim=0 0 0 0]{\gdir/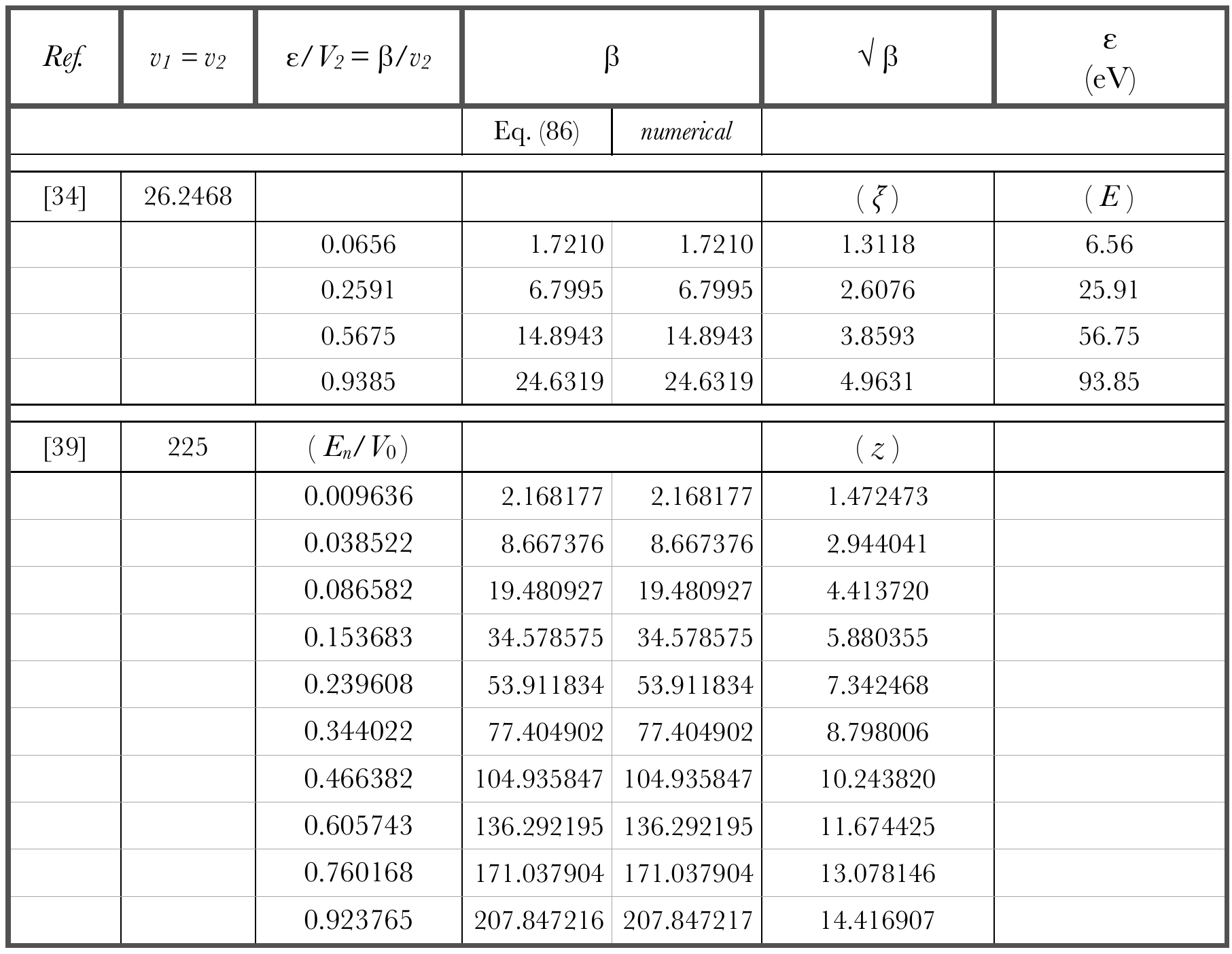}}
\end{table}
%

\subsubsection{Existence of eigenvalues \label{eev}}

It is well evidenced in the literature \cite{am11961,ll1977} that eigenvalues may not exist for unsymmetrical {\swp s} with sufficiently deep gap $(v_{1}-v_{2})$.
The transcendental equations involving the angle $\varphi$ [\REqd{eigenv.trig}{eigenv.trig.angle.onpi}] suggest that such an occurrence can happen also for {\twp s}.
Indeed, there can be triplets of the characteristic numbers $v_{1}, v_{2}, \lambda$ in correspondence to which
\begin{subequations}\label{noesev} \seqn
\begin{equation}\label{eigenv.trig.noev}
  - \sin\left( 2\sqrt{\beta} + \varphi(\beta) \right) < 0
\end{equation}
and
\begin{equation}\label{eigenv.trig.angle.onpi.noev}
   \subeqn{\theta(\beta) = \frac{2\sqrt{\beta} + \varphi(\beta)}{\pi} < 1}{}{0}
\end{equation}
\end{subequations}
when $\beta/v_{2}$ ranges in the interval [0,1]; if these conditions are fulfilled then, again, eigenvalues do not exist.
The inequalities indicated in \REqq{noesev} are exemplified in \Rfi{c7-Ds} for the triplet \itm{v_{1}=1, v_{2}=0.15, \lambda=1}; an increase of either the well's gap \itm{(v_{1}-v_{2})} by lowering \itm{v_{2}} from 0.2 to 0.15 (\Rfi{c7-Ds.v}) or the well's steepness by reducing \itm{\lambda} from 1.5 to 1.0 (\Rfi{c7-Ds.l}) expels the intersection points outside the interval [0,1] and makes the eigenvalue disappear.
Which physical interpretation should we attach to the absence of eigenvalues?
A simple and straightforward one: that, notwithstanding both the receptive mathematical structure of the \SEq\ [\REq{seq}] towards variable separation [\REq{vst}] and the benevolent imprimatur of the boundary conditions [\REq{evp.s.bc}], still separated-variable solutions are not allowed by the potential.
\begin{figure}[h]
\dgcap{.05}{.05}
   \subfloat[Effect due to well's gap]       {\label{c7-Ds.v}\resizebox{.47\textwidth}{!}{\includegraphics*[trim=33 20 60 60]{\gdir/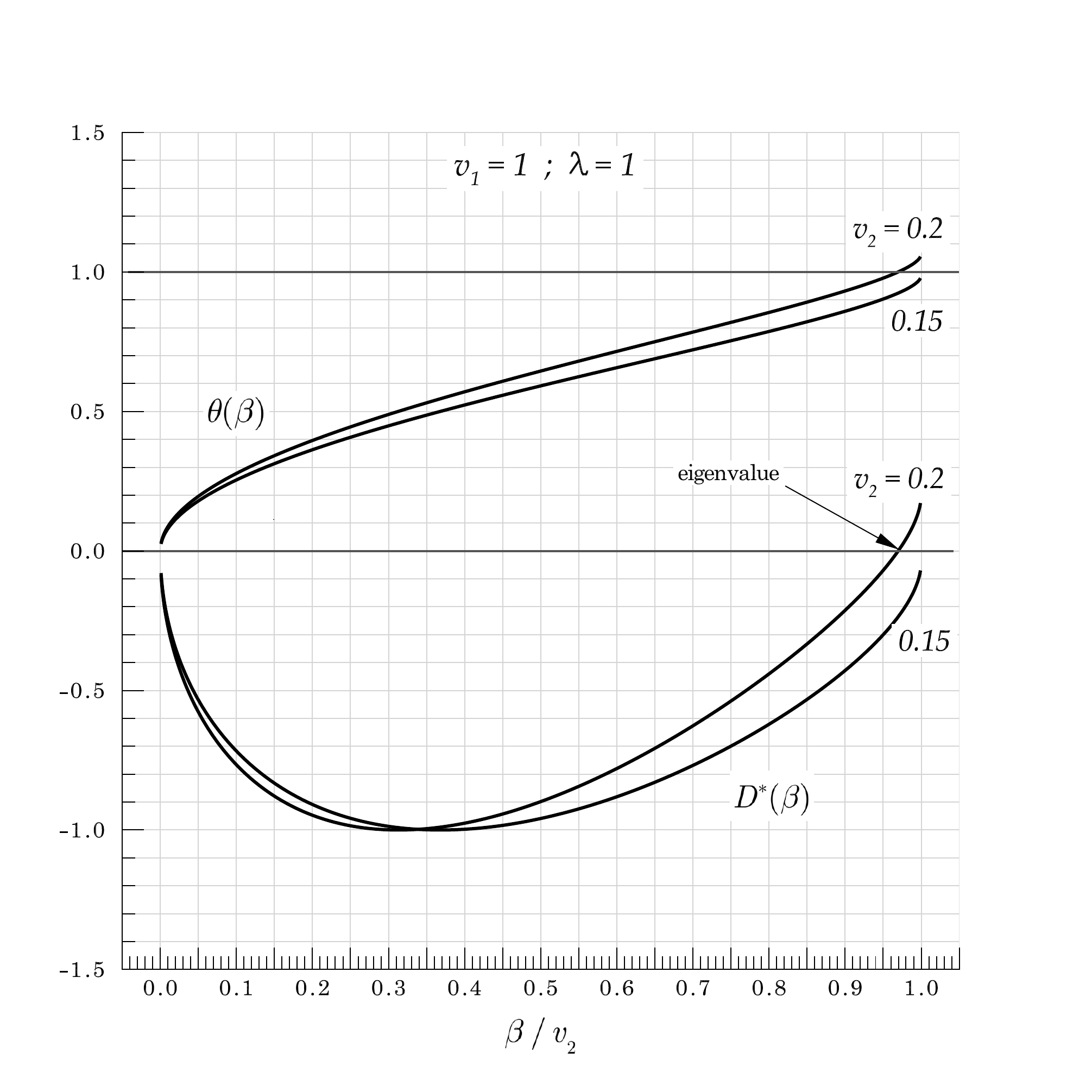}}}
   \subfloat[Effect due to well's steepness] {\label{c7-Ds.l}\resizebox{.47\textwidth}{!}{\includegraphics*[trim=33 20 60 60]{\gdir/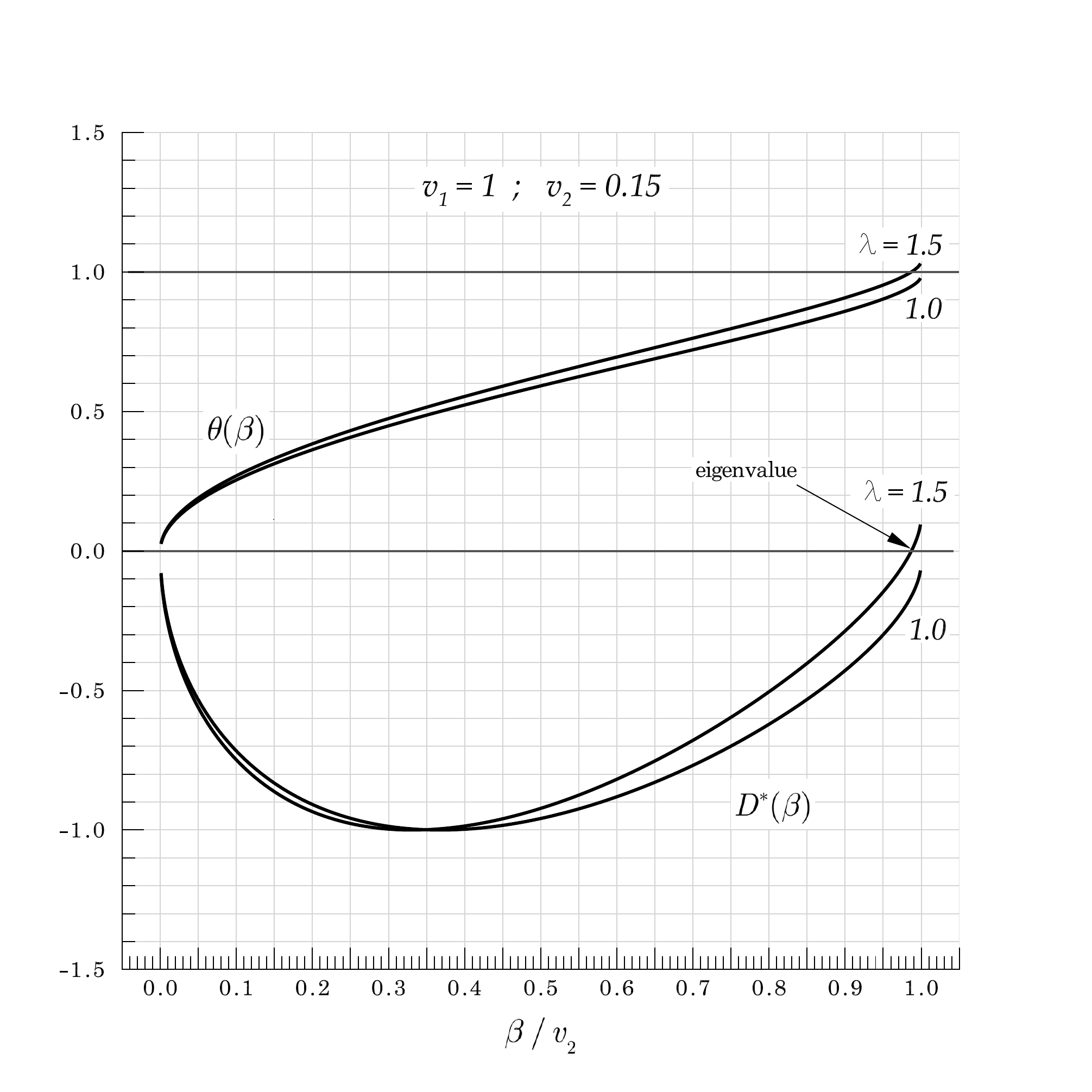}}}
   \caption{Graphical evidence of the possibility that eigenvalues do not exist for some triplets of the characteristic numbers $v_{1}, v_{2}, \lambda$.\llpush\label{c7-Ds}}
\end{figure}
The condition for the absence of eigenvalues can be formulated by noticing from \Rfid{c1-Da}{evdgtc} that the function $\theta(\beta)$ is monotonically increasing with respect to the ratio $\beta/v_{2}$ in [0,1]
\begin{equation}\label{theta.mi} 
   \theta(\beta) = \frac{2\sqrt{\beta} + \varphi(\beta)}{\pi} \leq \frac{2\sqrt{v_{2}} + \left[ \varphi(\beta,v_{1},v_{2},\lambda)\right]_{\beta=v_{2}}}{\pi}
\end{equation}
Therefore, if
\begin{equation}\label{cae} 
   \frac{2\sqrt{v_{2}} + \left[ \varphi(\beta,v_{1},v_{2},\lambda)\right]_{\beta=v_{2}}}{\pi} < 1
\end{equation}
then the inequality indicated in \REq{eigenv.trig.angle.onpi.noev} is verified \textit{a fortiori}.
\REqb{cae} is the condition whose fulfillment entails the absence of eigenvalues.
Unfortunately, its exploitation is not feasible in any analytical fashion in the general case of arbitrary triplets of the characteristic numbers $v_{1}, v_{2}, \lambda$ due to the mathematical cumbersomeness of the factors $f_{1'}, g_{1'}, f_{2'}, g_{2'}$ [\REq{fbar}, \REq{gleft}, \REq{f2p}, \REq{gright}] that constitute the input to the algorithm to extract the angle $\left[ \varphi(\beta,v_{1},v_{2},\lambda)\right]_{\beta=v_{2}}$ based on \REqq{coeff} and \REqq{angle}; a numerical treatment is always implied and necessary.
This inconvenience notwithstanding, it is possible to show that also for symmetrical {\twp s} at least one eigenvalue always exists, precisely as it happens for symmetrical {\swp s} \cite{dth1964,ll1977,dg2005}.
Indeed, if we set \itm{v_{1}=v_{2}=v, \lambda\neq0} and \itm{\beta=v} as required by \REq{cae} then the following cascade of simplifications takes place.
The overlined values [\REqd{mp.left.11p}{mp.right.2p2}] vanish \itm{(\etab=\zetab=0)} and the circumflexed values [\REqd{mp.left.1p0}{mp.right.02p}] coincide
\begin{equation}\label{cv.stwp}
   \etah = \zetah = - \left( \lambda \sqrt{v} \right)^{2/3}
\end{equation}
We deduce right away from \REq{cv.stwp} that the angle $\left[ \varphi(\beta,v,\lambda)\right]_{\beta=v}$ we are looking for is not going to depend on $\lambda$ and $v$ separately but on the product appearing on the right-hand side of \REq{cv.stwp}, product that defines formally the new characteristic number
\begin{equation}\label{u.dv} 
  u = \left( \lambda \sqrt{v} \right)^{2/3}
\end{equation}
Its usefulness will become apparent in a few lines from here.
Further coincidence takes place for the factors $f_{1'}, f_{2'}$ [\REqd{fbar}{f2p}]
\begin{equation}\label{ff.stwp}
   f_{1'}=f_{2'}=\frac{\Bi'(0)}{\Ai'(0)} = f(0) \simeq -1.73205
\end{equation}
for the factors $\gl, -\gr$ [\REqd{gleft}{gright}]
\begin{equation}\label{gf.stwp}
  \gl = - \gr = - \sqrt{u}\frac{ \Bi(-u) - f(0)\, \Ai(-u) }{ \Bi'(-u) - f(0)\, \Ai'(-u) }
\end{equation}
and, consequently, for their reciprocals [\REq{oog}]
\begin{equation}\label{gaf.stwp}
  \gal = - \gar
\end{equation}
With these simplifications, the algorithm based on \REqq{coeff} and \REqq{angle} becomes somewhat lighter computationally  and furnishes the angle
$\left[\varphi(\beta,v,\lambda)\right]_{\beta=v}$ required in the eigenvalue-absence condition [\REq{cae}] adapted to the present case
\begin{equation}\label{cae.stwp} 
   \frac{2\sqrt{v} + \left[ \varphi(\beta,v,\lambda)\right]_{\beta=v}}{\pi} < 1
\end{equation}
%
%
The nice feature of the angle $\left[\varphi(\beta,v,\lambda)\right]_{\beta=v}$ being dependent only on the lately defined characteristic number $u$ [\REq{u.dv}] suggests
the clever move to extract $\sqrt{v}$ from \REq{u.dv}
\begin{equation}\label{sqrtv}
  \sqrt{v} = \frac{u^{3/2}}{\lambda}
\end{equation}
to substitute it into \REq{cae.stwp} and rearrange the condition into the separated form
\begin{equation}\label{cae.stwp.s}
  \frac{2}{\lambda} + \frac{\left[ \varphi(\beta,v,\lambda)\right]_{\beta=v} - \pi}{u^{3/2}} < 0
\end{equation}
The first term on the left-hand side of \REq{cae.stwp.s} depends only on the potential's steepness and is unconditionally positive.
The responsibility for positivity or negativity falls on the second term; this term, however, turns out to be a universal function
\begin{equation}\label{Hu.f}
  H(u) = \frac{\left[ \varphi(\beta,v,\lambda)\right]_{\beta=v} - \pi}{u^{3/2}}
\end{equation}
of the characteristic number $u$ whose graph, illustrated in \Rfi{Hu}, reveals to be also positive and monotonic.
These valuable features of the function $H(u)$ ensure the falseness of the inequality in \REq{cae.stwp.s} and sanction the conclusion that eigenvalues certainly exist for symmetrical \twp s.
More graphical evidence supporting this conclusion is illustrated in \Rfi{c2-theta}.
The rightmost curve \itm{(\lambda=10^{-9})} corresponds essentially to a symmetrical \swp\ and, therefore, it has always at least one intersection with the level \mbox{$\theta(\beta)=1$}, no matter how shallow the well's depth is.
If the well's steepness decreases then the potential becomes a symmetrical \twp; the curve shifts leftward, so does the intersection, and, again, we can conclude \textit{a fortiori} that also  a symmetrical \twp\ possesses at least one eigenvalue.
\begin{figure}[h]
   \subfloat[The universal function $H(u)$ is always positive.]        {\label{Hu}      \resizebox{.47\textwidth}{!}{\includegraphics*[trim=10 20 60 60]{\gdir/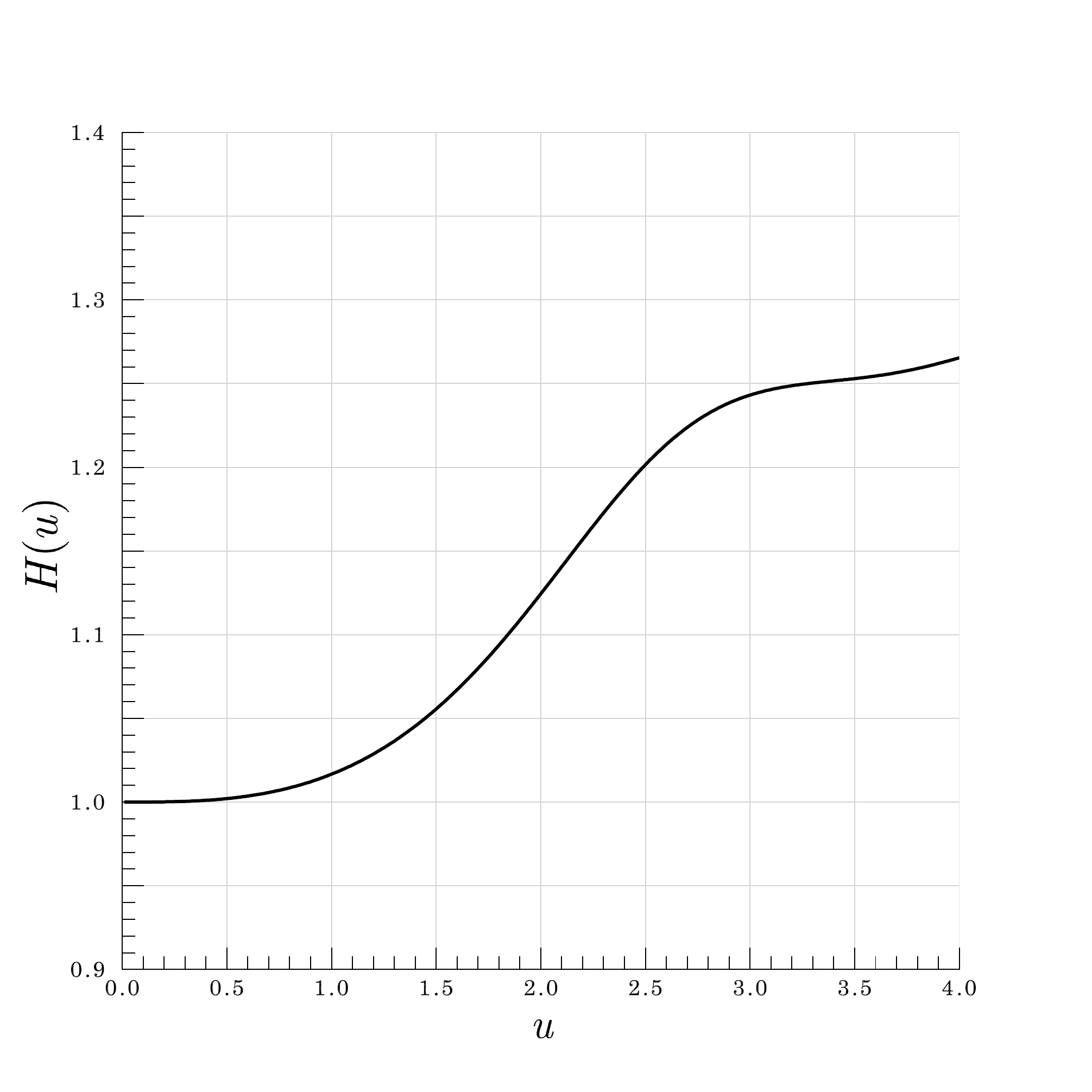}}}
   \subfloat[The intersection \mbox{$\theta(\beta)=1$} always exists.] {\label{c2-theta}\resizebox{.47\textwidth}{!}{\includegraphics*[trim=10 20 60 60]{\gdir/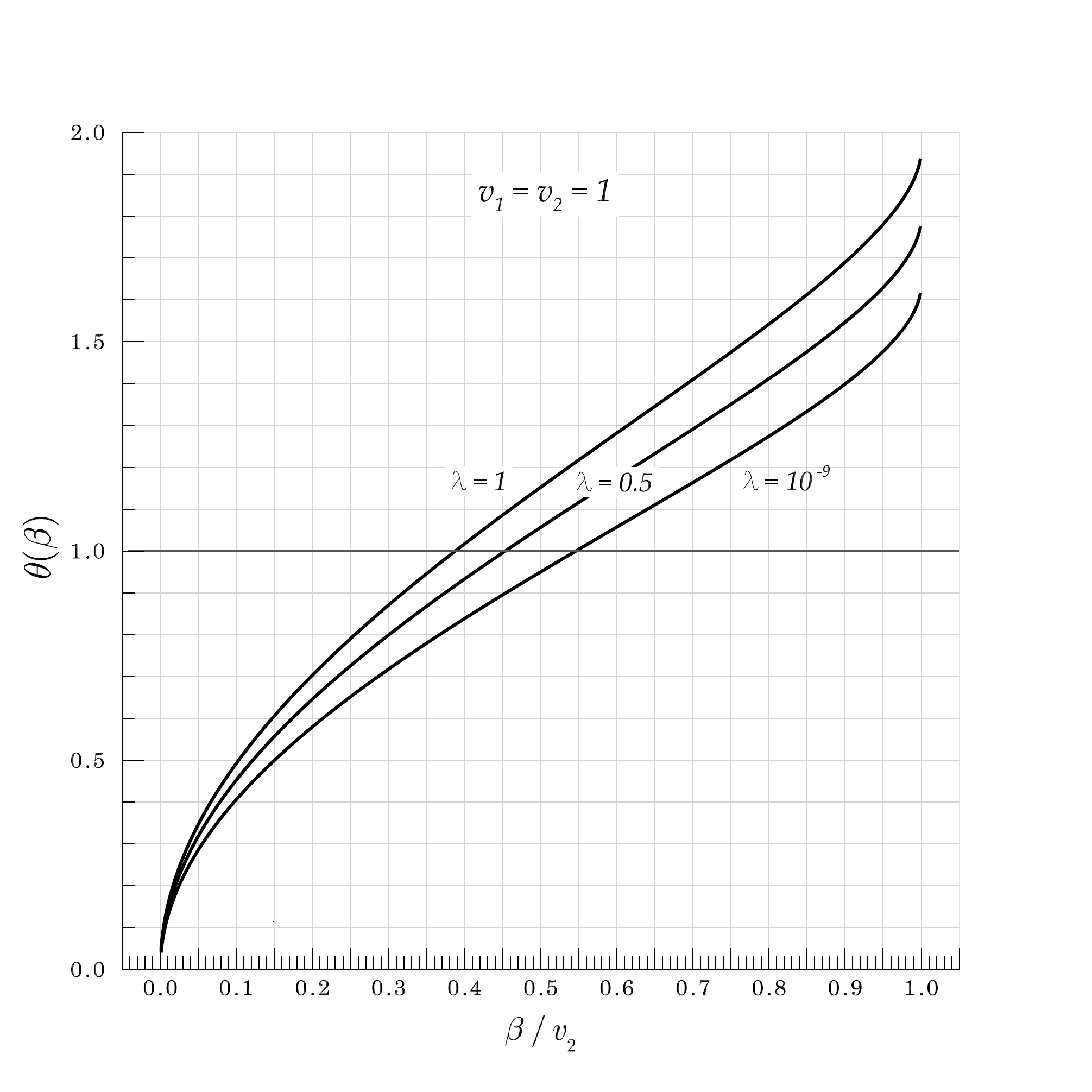}}}
   \caption{Graphical evidence that at least one eigenvalue always exists for symmetrical {\twp s}.\label{c2-theta.old}}
\end{figure}

%
%


\subsubsection{Eigenfunction's coefficients \label{efc}}

The successive step, after the calculation of the eigenvalues, consists in the determination of the eigenfunction's coefficients.
On account of the determinant's vanishing [\REq{eigenv}], the algebraic system composed by \REqdt{A0B0.1.1}{A0B0.2.1} coalesce into one single equation connecting the coefficients $A_{0}, B_{0}$.
Accordingly, an instinctive manner to proceed could comprise the following sequence of operations: (a) decide which of the two coefficients should be assumed independent and solve either \REqt{A0B0.1.1} or \REqt{A0B0.2.1} for the dependent one; (b) determine the other coefficients in terms of the independent one from the group of equations listed in the beginning of the paragraph following \REq{B2p}, after setting aside \REqd{A0B0.1}{A0B0.2}, of course; (c) obtain the independent coefficient from the exploitation of the eigenfunction's normalization condition [\REq{efn.nd}].
And, indeed, this sequence would work smoothly and swiftly for unsymmetrical wells; yet, we found out that failure is lurking behind operation (a) if the potential well is symmetrical.
This is the particular case in which eigenfunction's parity, even and odd, must be explicitly contemplated; it is thoroughly discussed in the literature for symmetrical {\swp s} but it turns up also for symmetrical {\twp s}.
Let us see the details.
The well symmetry implies the following simplifications.
The overlined and the circumflexed values [\REqq{mp.left} and \REqq{mp.right}] come, respectively, to coincide
\begin{subequations}\label{etazeta.symw} \seqn
  \begin{align}
     \etab & = \zetab   \label{etazeta.b.symw} \\[.25\baselineskip]
     \etah & = \zetah   \label{etazeta.h.symw}
  \end{align}
\end{subequations}
and so do the factors $f_{1'}$ and $f_{2'}$ [\REqd{fbar}{f2p}]
\begin{equation}\label{f1pf2p.symw}
   f_{1'} = f_{2'}
\end{equation}
The factors $\gl$ and $\gr$ [\REqd{gleft}{gright}] become mathematically opposite
\begin{equation}\label{glgr.symw}
   \gl = - \gr \rightarrow g
\end{equation}
The algebraic system in \REq{A0B0.m} simplifies to the form
\begin{subequations}\label{A0B0.as.symw} \seqn
    \begin{equation}\label{A0B0.m.sw}
      \begin{bmatrix}
         \,\sin(\sqrt{\beta}) + g \cos(\sqrt{\beta}) \quad & g \sin(\sqrt{\beta}) - \cos(\sqrt{\beta})\, \\[1.5ex]
           \sin(\sqrt{\beta}) - g \cos(\sqrt{\beta}) \quad & g \sin(\sqrt{\beta}) + \cos(\sqrt{\beta})
      \end{bmatrix} \cdot
      \begin{bmatrix} A_0 \\[1.5ex] B_{0} \end{bmatrix} = 0
    \end{equation}
and its determinant provides the factorized transcendental equation\footnote{\REqb{eigenv.symw} is obviously the simplified form which \REq{eigenv} reduces to with the help of the trigonometric formulae
\begin{displaymath}
   \sin(2\sqrt{\beta}) = \frac{2 \tan(\sqrt{\beta})}{1+ \tan^{2}(\sqrt{\beta})} \qquad\qquad  \cos(2\sqrt{\beta}) = \frac{1 - \tan^{2}(\sqrt{\beta})}{1 + \tan^{2}(\sqrt{\beta})}
\end{displaymath}
and if the identity indicated in \REq{glgr.symw} is enforced.}
\begin{equation}\label{eigenv.symw}
   [D(\beta)]_{\mathrm{sw}} = - 2 \left[ \sin(\sqrt{\beta}) + g \cos(\sqrt{\beta}) \right] \cdot \left[ g \sin(\sqrt{\beta}) - \cos(\sqrt{\beta}) \right] = 0
\end{equation}
\end{subequations}
\REqqb{A0B0.as.symw} generate the even-parity solution
\begin{subequations}\label{A0B0.as.symw.ep} \seqn
  \begin{align}
      & A_{0} = 0                   \label{A0B0.as.symw.ef.ep.c}  \\[.25\baselineskip]
      & g \tan(\sqrt{\beta}) = 1    \label{A0B0.as.symw.efd.ep.ev}
  \end{align}
\end{subequations}
and the odd-parity solution
\begin{subequations}\label{A0B0.as.symw.op} \seqn
  \begin{align}
      & B_{0} = 0                    \label{A0B0.as.symw.ef.op.c}  \\[.25\baselineskip]
      & g \cot(\sqrt{\beta}) = -1    \label{A0B0.as.symw.efd.op.ev}
  \end{align}
\end{subequations}
\REqdb{A0B0.as.symw.ef.ep.c}{A0B0.as.symw.ef.op.c} tell how unwise the operation (a) mentioned in the beginning of this section would be without knowing beforehand which parity situation we are dealing with.\footnote{This is precisely the trap which one of us (DG) walked head-on into. An unexpected and mystifying sign change of the eigenfunction at the junction point \jp{0}{2'} for the second excited eigenstate shown in \Rfi{c3-ef2} was the unequivocal omen that something had gone wrong with the calculation of the eigenfunction's coefficients and triggered the debugging investigation that lead to the understanding of the details explained in the text. A good lesson learned from a mistake.}
The coefficient-vanishing possibility is the reason, mentioned just below \REq{gleft}, behind our decision to keep \REq{A0B0.1} in that form instead of solving it for one of the two coefficients.
This turn of events is particularly critical when a numerical method, such as the Newton-Raphson method we used, is adopted to calculate the eigenvalues because the necessity of parity distinction is basically invisible to the numerical algorithm that operates on the transcendental equation, that being any of \REq{eigenv} or \REq{eigenv.trig} or \REq{eigenv.trig.angle.onpi}.
The three characteristic numbers $v_{1}, v_{2}, \lambda$ constitute all that the numerical algorithm needs to know to grind out the eigenvalue and the circumstance of symmetrical well is handled as mechanically as that of unsymmetrical well.
There is no automatic mechanism built in the algorithm that, in the former circumstance, raises a parity-distinction flag to be remembered and taken into account at the moment of calculating the eigenfunction coefficients.
The if-then-else situation created by the necessity of parity distinction for symmetrical wells must be programmed in the algorithm.
It is doable and is not a serious preoccupation, of course, but it is a perhaps rather tedious inconvenience.
Luckily, there is a simple stratagem to circumvent it.\footnote{This is a bright example of how knowledge in one department of science, tensor algebra in this case, can help to inspire ideas in another one. A second-order tensor $A_{ij}$ can always be separated in a symmetric part $A^{s}_{ij}=(A_{ij}+A_{ji})/2$ and an antisymmetric part $A^{a}_{ij}=(A_{ij}-A_{ji})/2$; then, addition of the parts returns the tensor $A_{ij}=A^{s}_{ij}+A^{a}_{ij}$ while subtraction returns the tensor's transpose $A_{ji}=A^{s}_{ij}-A^{a}_{ij}$. Alright, it is not exactly the same situation we are dealing with because the coefficients $A_{0}, B_{0}$ are independent but it is the spark that inspired \REq{C0D0}.}
Let us introduce two new coefficients defined as
\begin{equation}\label{C0D0}
   \begin{bmatrix} C_{0} \\ D_{0} \end{bmatrix} = \frac{1}{2} \begin{bmatrix} \,1 & 1 \\ 1 & -1\, \end{bmatrix} \cdot \begin{bmatrix} A_0 \\ B_{0} \end{bmatrix}
                                                = \begin{bmatrix} \dfrac{A_0 + B_{0}}{2} \\[2ex] \dfrac{A_0 - B_{0}}{2} \end{bmatrix}
\end{equation}
These coefficients never vanish, even if the well is symmetric; in that case, they are either opposite ($C_{0}=-D_{0}$) or equal ($C_{0}=D_{0}$) if the parity is even ($A_{0}=0$) or odd ($B_{0}=0$), respectively.
Therefore, one of them can always be expressed in term of the other one without fear of disrupting the coefficient-calculation procedure.
Now, we can invert \REq{C0D0}
\begin{equation}\label{iC0D0}
   \begin{bmatrix} A_{0} \\ B_{0} \end{bmatrix} = \begin{bmatrix} \,1 & 1 \\ 1 & -1\, \end{bmatrix} \cdot \begin{bmatrix} C_0 \\ D_{0} \end{bmatrix}
                                                = \begin{bmatrix} C_0 + D_{0} \\[1ex] C_0 - D_{0} \end{bmatrix}
\end{equation}
and substitute \REq{iC0D0} into the algebraic system in \REq{A0B0.m} to derive an analogous system but in terms of the new coefficients
\begin{equation}\label{C0D0.as}
  \begin{bmatrix}
     \,\sin(\sqrt{\beta}) + \gl \cos(\sqrt{\beta}) \quad & \gl \sin(\sqrt{\beta}) - \cos(\sqrt{\beta})\, \\[1.5ex]
       \sin(\sqrt{\beta}) - \gr \cos(\sqrt{\beta}) \quad & \gr \sin(\sqrt{\beta}) + \cos(\sqrt{\beta})
  \end{bmatrix} \cdot
  \begin{bmatrix} \,1 & 1 \\[1.5ex] 1 & -1\, \end{bmatrix} \cdot \begin{bmatrix} C_0 \\[1.5ex] D_{0} \end{bmatrix} = 0
\end{equation}
It is a straightforward consequence of matrix algebra, and an easy exercise to verify, that the determinant of this new algebraic system is proportional\footnote{The determinant of the product of matrices is the product of the determinants of the matrices and
\begin{displaymath}det\begin{bmatrix} \,1 & 1 \\ 1 & -1\, \end{bmatrix} = -2\end{displaymath}} to that of the old one [\REq{A0B0.m}] and, therefore they share the same transcendental equation [\REq{eigenv}].
If we select the coefficient $C_{0}$ as independent then \REq{C0D0.as} gives
  \begin{equation} \label{D00}
      D_{0} = - C_{0} \, \frac{\left( 1+\gl \right)\sin(\sqrt{\beta}) - \left( 1-\gl \right)\cos(\sqrt{\beta})}
                                {\left( 1-\gl \right)\sin(\sqrt{\beta}) + \left( 1+\gl \right)\cos(\sqrt{\beta})}
            = - C_{0} \, \frac{\left( 1+\gr \right)\sin(\sqrt{\beta}) + \left( 1-\gr \right)\cos(\sqrt{\beta})}
                                {\left( 1-\gr \right)\sin(\sqrt{\beta}) - \left( 1+\gr \right)\cos(\sqrt{\beta})}
  \end{equation}
In the case of symmetrical wells, the simplifications in \REq{glgr.symw}, \REq{A0B0.as.symw.efd.ep.ev}, \REq{A0B0.as.symw.efd.op.ev} apply and the fractions in \REq{D00} reduce to \itm{\,-1} for even parity or \itm{\,+1} for odd parity.
The coefficients $A_{0}, B_{0}$ follow from \REq{iC0D0}, harmlessly in case of symmetrical wells, and then the formulae discussed in \Rse{ecjp} become operative to determine the remaining required coefficients.
We summarize them here for convenience:
\begin{equation}\taglabel{A1p}{\eqsubone}\label{A1p.1}
    A_{1'} = - B_{1'}\cdot f_{1'}
\end{equation}
\begin{equation}\taglabel{B1}{\eqsubone}\label{B1.1}
    B_{1} =   B_{1'}                        \left[ \Bi(\etab)  - f_{1'}\,\Ai(\etab) \right] \exp[(1+\lambda)\sqrt{k_{1}}]
          = -   \dfrac{B_{1'}}{\sqrt{\etab}}\left[ \Bi'(\etab) - f_{1'}\,\Ai'(\etab) \right] \exp[(1+\lambda)\sqrt{k_{1}}] \\[.25\baselineskip]
\end{equation}
\begin{equation}\taglabel{B1p}{\eqsubone}\label{B1p.1}
   B_{1'} = \frac{-A_{0} \sin(\sqrt{\beta}) + B_{0} \cos(\sqrt{\beta})}{\Bi(\etah)  - f_{1'}\,\Ai(\etah)}
          = - \sqrt{-\etah} \frac{ A_{0} \cos(\sqrt{\beta}) + B_{0} \sin(\sqrt{\beta})}{\Bi'(\etah) - f_{1'}\,\Ai'(\etah)} \\[.45\baselineskip]
\end{equation}
\begin{equation}\taglabel{A2p}{\eqsubone}\label{A2p.1}
    A_{2'} = - B_{2'}\cdot f_{2'} \\[.45\baselineskip]
\end{equation}
\begin{equation}\taglabel{B2}{\eqsubone}\label{B2.1}
    B_{2} =   B_{2'}                        \left[ \Bi(\zetab)  - f_{2'}\,\Ai(\zetab) \right] \exp[(1+\lambda)\sqrt{k_{2}}]
          = -  \dfrac{B_{2'}}{\sqrt{\zetab}}\left[ \Bi'(\zetab) - f_{2'}\,\Ai'(\zetab) \right] \exp[(1+\lambda)\sqrt{k_{2}}] \\[.35\baselineskip]
\end{equation}
\begin{equation}\taglabel{B2p}{\eqsubone}\label{B2p.1}
   B_{2'} = \frac{A_{0} \sin(\sqrt{\beta}) + B_{0} \cos(\sqrt{\beta})}{\Bi(\zetah)  - f_{2'}\,\Ai(\zetah)}
          = \sqrt{-\zetah} \frac{ A_{0} \cos(\sqrt{\beta}) - B_{0} \sin(\sqrt{\beta})}{\Bi'(\zetah) - f_{2'}\,\Ai'(\zetah)} \\[.25\baselineskip]
\end{equation}
Two remarks are in order with a view to carry out calculations with these equations.
First, the exponentials in \REqdt{B1.1}{B2.1} call for attention; they are latent numerical troublemakers because they can definitely overflow calculations when the characteristic numbers $v_{1}, v_{2}$ are sufficiently great [\REq{negbetas}].
A good cure to make the exponentials harmless is to merge them with the exponentials of the corresponding eigenfunction's components [\REqd{ndevp.ef.1}{ndevp.ef.2}]; as preparatory work, we define the auxiliary coefficients
\begin{equation}\label{B1t}
   \tilde{B}_{1} = B_{1'} \left[ \Bi(\etab) - f_{1'}\,\Ai(\etab) \right] = - \dfrac{B_{1'}}{\sqrt{\etab}}\left[ \Bi'(\etab) - f_{1'}\,\Ai'(\etab) \right]
\end{equation}
\begin{equation}\label{B2t}
   \tilde{B}_{2} = B_{2'} \left[ \Bi(\zetab) - f_{2'}\,\Ai(\zetab) \right] = - \dfrac{B_{2'}}{\sqrt{\zetab}}\left[ \Bi'(\zetab) - f_{2'}\,\Ai'(\zetab) \right]
\end{equation}
and rewrite \REqdt{B1.1}{B2.1} as
\newcommand{\eqsubtwo}{$_{2}$}
\begin{equation}\taglabel{B1}{\eqsubtwo}\label{B1.2}
    B_{1} = \tilde{B}_{1} \exp[(1+\lambda)\sqrt{k_{1}}]
\end{equation}
\begin{equation}\taglabel{B2}{\eqsubtwo}\label{B2.2}
    B_{2} =   \tilde{B}_{2} \exp[(1+\lambda)\sqrt{k_{2}}]
\end{equation}
Second, the double expressions in most of the equations between \REq{D00} and \REq{B2t} are obviously analytically equivalent; yet, numerical operations are always burdened with round-off errors and expectedly the numerical outputs from corresponding double expressions differ slightly.
In order to contain somehow the impact of round-off errors and to make both expressions count, we calculated the corresponding coefficient as arithmetic average of the numerical outputs from the double expressions; for example, the coefficient $\tilde{B}_{1}$ defined in \REq{B1t} was actually calculated as
\begin{equation}\taglabel{B1t}{\eqsubone}\label{B1t.1}
   \tilde{B}_{1} = \frac{B_{1'}}{2}\left\{ \left[ \Bi(\etab) - f_{1'}\,\Ai(\etab) \right] - \dfrac{1}{\sqrt{\etab}}\left[ \Bi'(\etab) - f_{1'}\,\Ai'(\etab) \right]  \right\}
\end{equation}
Likewise for the other concerned coefficients.

The linear dependence on $C_{0}$ originated in \REq{D00} propagates to all the other coefficients; the completion of the task of this section, therefore, requires the determination of this last coefficient.
In order to achieve that, we must assemble the global eigenfunction from the zonal components [respectively: \REq{ndevp.ef.1} with \REqt{B1.2}; \REq{ndevp.s.1p.i} with \REqt{A1p.1}; \REq{ndevp.s.0.i}; \REq{ndevp.s.2p.i} with \REqt{A2p.1}; \REq{ndevp.ef.2} with \REqt{B2.2}]
\begin{equation}\label{ef.all}
   \phi(\xi) =
                 \begin{cases}
                       \tilde{B}_{1} \exp\left[ \left(  \xi + 1 + \lambda\right) \sqrt{k_{1}} \right]  \qquad & \text{zone 1} \\[1.5ex]
                        B_{1'}  \left[ \Bi(\eta) - f_{1'} \,\Ai(\eta) \right]
                        \qquad\quad    \left[\eta   = - \left(\dfrac{v_{1}}{\lambda}\right)^{1/3} \left( \xi + 1 + \dfrac{\lambda}{v_{1}}\beta \right) \right] \qquad & \text{zone 1'}    \\[1.5ex]
                        A_{0} \sin(\xi\sqrt{\beta}) + B_{0} \cos(\xi\sqrt{\beta})                            \qquad &  \text{zone 0}    \\[1.5ex]
                        B_{2'}  \left[ \Bi(\zeta) - f_{2'} \,\Ai(\zeta) \right]
                        \qquad\quad    \left[\zeta = + \left(\dfrac{v_{2}}{\lambda}\right)^{1/3} \left( \xi - 1 - \dfrac{\lambda}{v_{2}}\beta \right) \right] \qquad & \text{zone 2'}    \\[1.5ex]
                       \tilde{B}_{2} \exp\left[ \left( -\xi + 1 + \lambda\right) \sqrt{k_{2}} \right]  \qquad & \text{zone 2} \\
                 \end{cases}
\end{equation}
and pass its square through the integral of the eigenfunction's normalization condition [\REq{efn.nd}].
The integral splits in five contributions, one for each zone.
The contributions of the zones with constant potential can be easily obtained analytically; instead, the contributions of the zones with linear potential are refractory to analytical handling and require recourse to numerical integration, a minor formality with modern programming languages.
The integration-operation algebra calls for moderate skills and particular attention to the differentials' transformations [\REqq{ivt.d}] in the zones \itm{s=1', 2'} but it is rather straightforward; so, we skip the details and jump directly to the final result
\begin{equation}\label{C0}
   \frac{\tilde{B}_{1}^{2}}{2\sqrt{k_{1}}} +
   B_{1'}^{2} \sqrt{\frac{\etab}{k_{1}}} \cdot J_{1'}(\etah,\etab) +
   A_{0}^{2} \left[ 1 - \frac{\sin(2\sqrt{\beta})}{2\sqrt{\beta}}  \right] +
   B_{0}^{2} \left[ 1 + \frac{\sin(2\sqrt{\beta})}{2\sqrt{\beta}}  \right] +
   B_{2'}^{2} \sqrt{\frac{\zetab}{k_{2}}} \cdot J_{2'}(\zetah,\zetab) +
   \frac{\tilde{B}_{2}^{2}}{2\sqrt{k_{2}}} = 2
\end{equation}
in which
\begin{subequations}\label{Jint}\seqn
  \begin{align}
      J_{1'}(\etah,\etab)    & = \int_{\etah}^{\etab}   \left[ \Bi(\eta)  - f_{1'} \Ai(\eta)  \right]^{2}\,d\eta      \label{J1p}  \\
      J_{2'}(\zetah,\zetab)  & = \int_{\zetah}^{\zetab} \left[ \Bi(\zeta) - f_{2'} \Ai(\zeta) \right]^{2}\,d\zeta     \label{J2p}
  \end{align}
	are the integrals that require numerical evaluation.
\end{subequations}
\REqb{C0} balances the number of equations with the number of coefficients and, in so doing, fixes the coefficient $C_{0}$.

\subsubsection{Eigenfunctions\label{ef}}

With the coefficients in hand, the analytical eigenfunctions can be calculated straightforwardly from \REq{ef.all}; the numerical eigenfunctions are provided by the finite-difference method briefly described at the end of \Rse{ndf}.
We show two validation examples.
The first one, in \Rfi{c1-ef1}, illustrates the eigenfunction of the single eigenstate belonging to the unsymmetrical well \itm{v_{1}=1, v_{2}=0.5, \lambda=1} whose eigenvalue-detection graph is displayed in \Rfi{c1-Da}.
Numerical results (solid circles) superpose to analytical results (lines) very satisfactorily.
We adopted a thinner line for the analytical eigenfunction's curves in the zones with linear potential in order to emphasize the smooth transition among the zones with constant potential and to appreciate graphically eigenfunction's and its first derivative's continuity at the junction points; we have systematically used the data-representation style adopted in \Rfi{c1-ef1} in all forthcoming figures related to eigenfunctions.
Analytical and numerical approaches concur also about the eigenvalue: they both give $\beta=0.31447$.
The second example is relative to the symmetrical well \itm{v_{1}=v_{2}=10, \lambda=0.5} and is illustrated in \Rfi{c3-all}.
The eigenvalue-detection graph (\Rfi{c3-Da}) reveals the existence of three eigenstates whose eigenfunctions are shown in \Rfii{c3-all}b--d.
\begin{figure}[h]
\dgcap{.2}{.2}
   \resizebox{.47\textwidth}{!}{\includegraphics*[trim=33 20 60 60]{\gdir/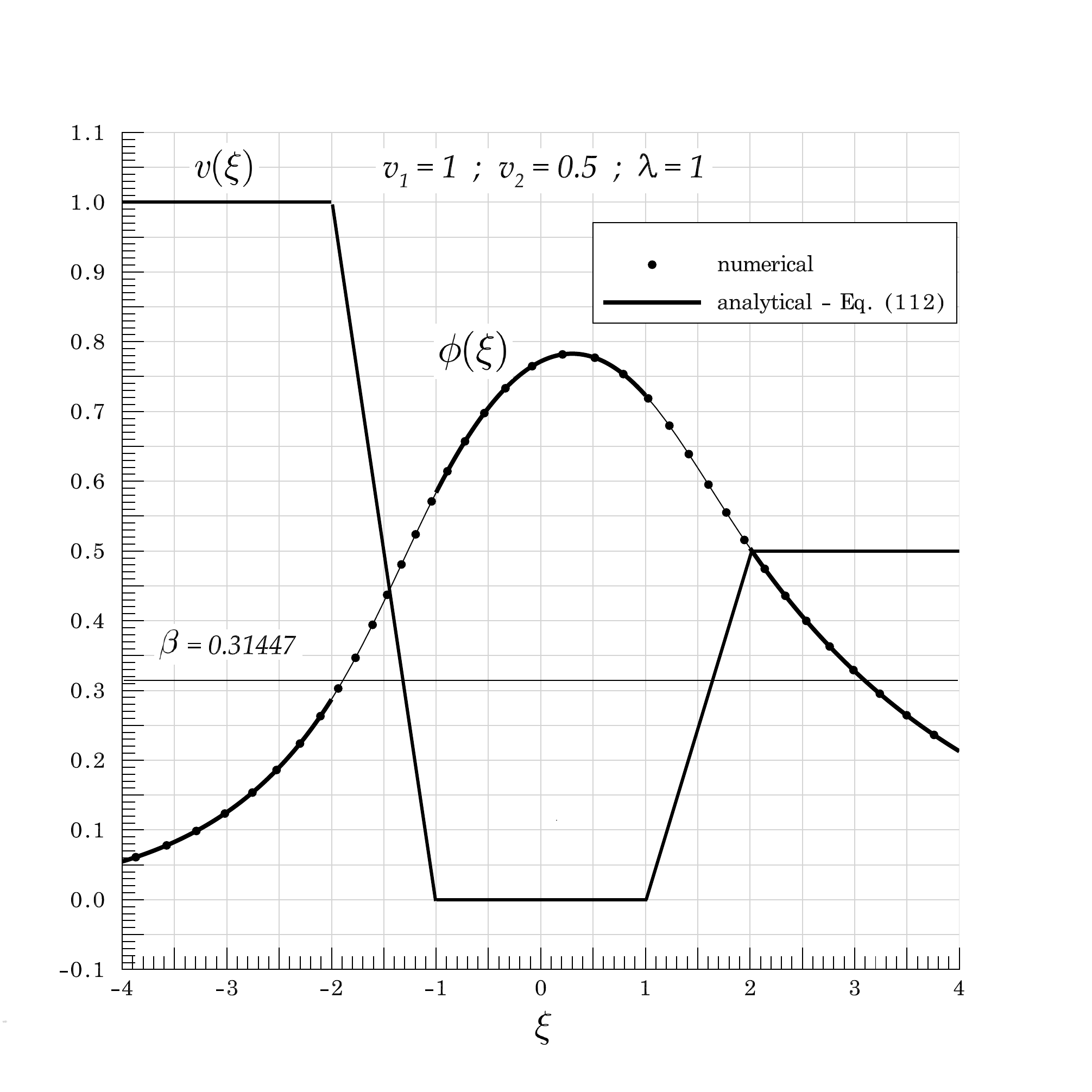}}
   \caption{\color{black} Eigenfunction of the single eigenstate belonging to the unsymmetrical well $v_{1}=1, v_{2}=0.5, \lambda=1$; the eigenvalue-detection graph is displayed in \Rfi{c1-Da}.\llpush\label{c1-ef1}}
\end{figure}
\begin{figure}[h]
\dgcap{.05}{.05}
   \subfloat[Eigenvalue-detection graph] {\label{c3-Da} \resizebox{.47\textwidth}{!}{\includegraphics*[trim=33 20 60 60]{\gdir/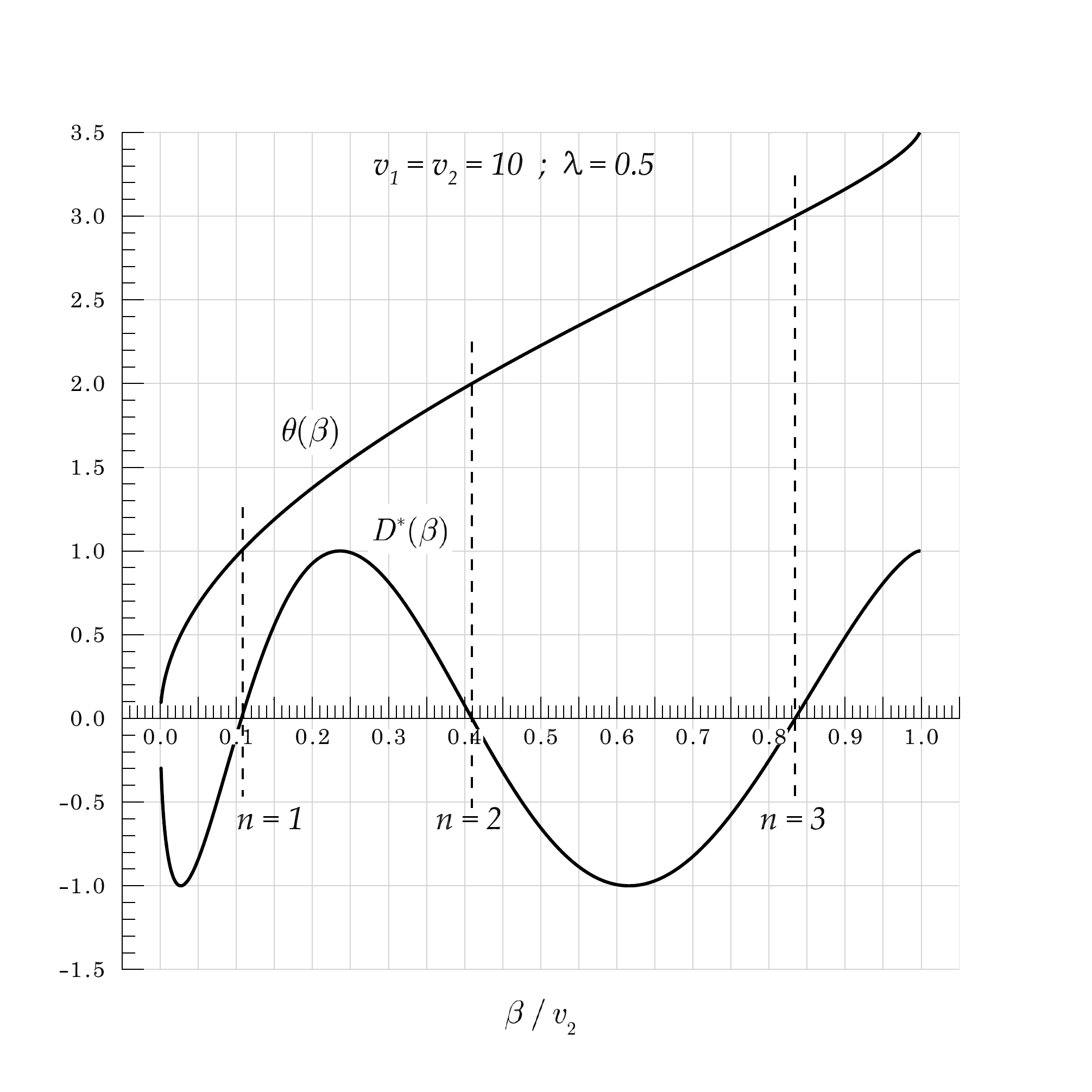}}}
   \\
   \subfloat[Ground state]               {\label{c3-ef1}\resizebox{.31\textwidth}{!}{\includegraphics*[trim=33 20 60 60]{\gdir/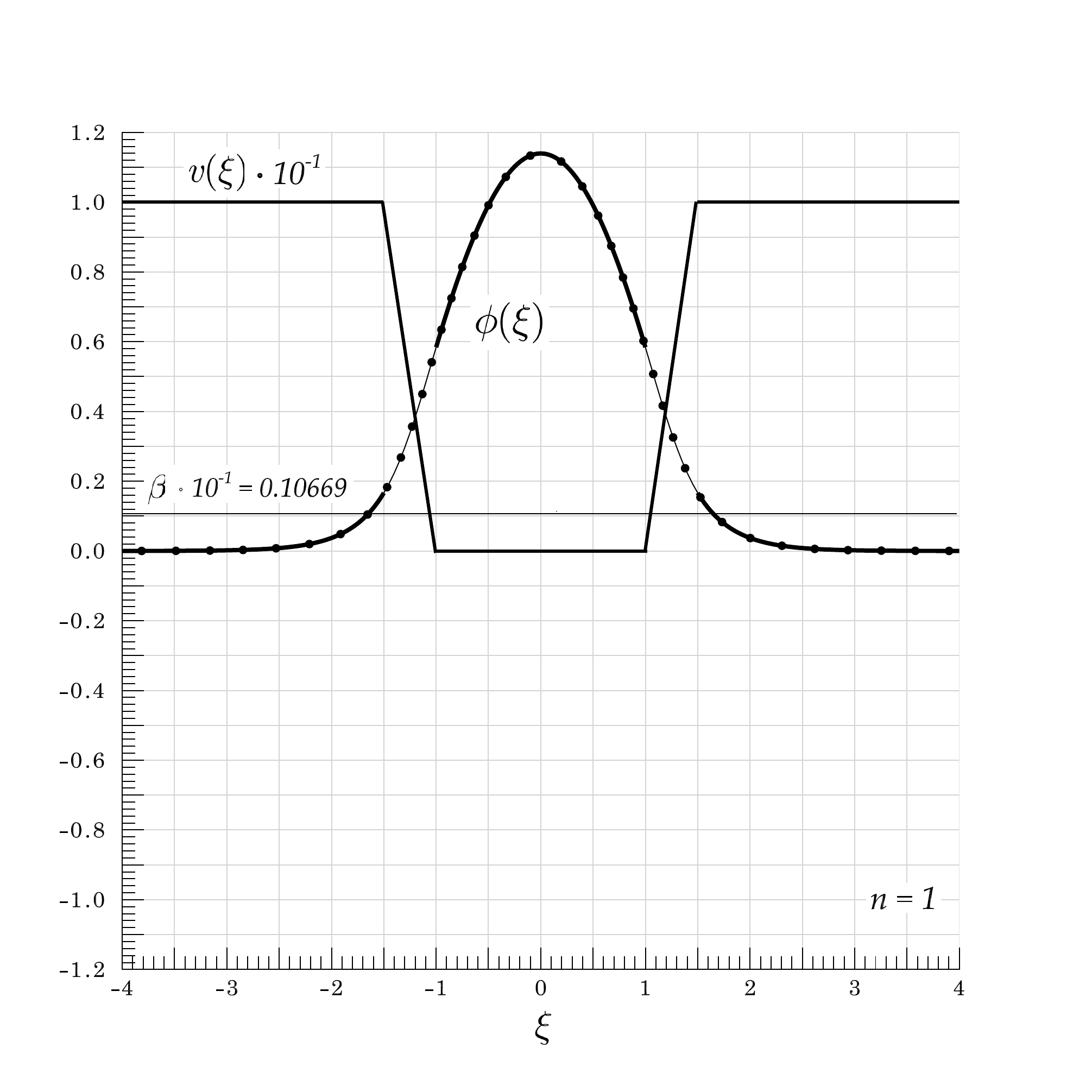}}}
   \subfloat[First excited state]        {\label{c3-ef2}\resizebox{.31\textwidth}{!}{\includegraphics*[trim=33 20 60 60]{\gdir/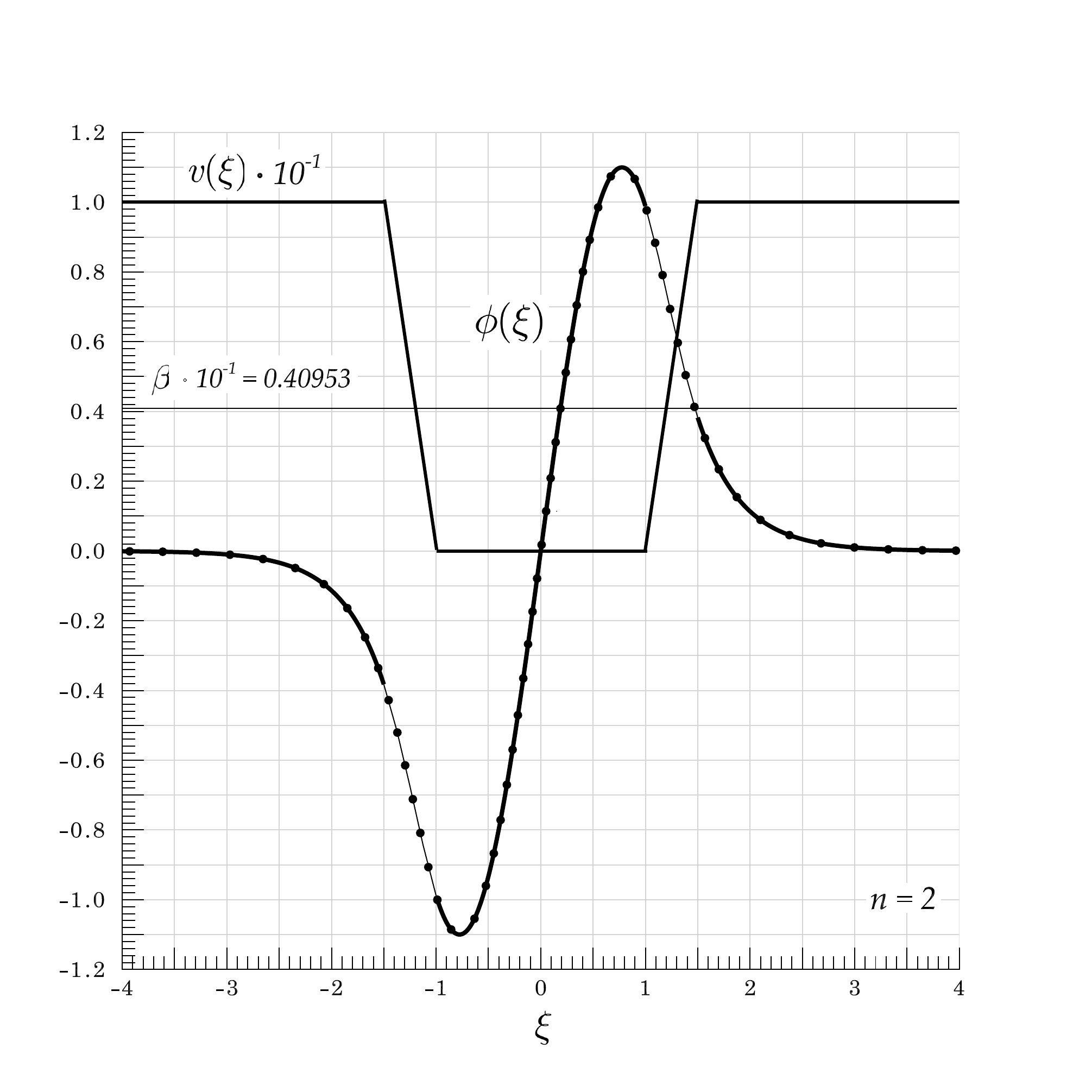}}}
   \subfloat[Second excited state]       {\label{c3-ef3}\resizebox{.31\textwidth}{!}{\includegraphics*[trim=33 20 60 60]{\gdir/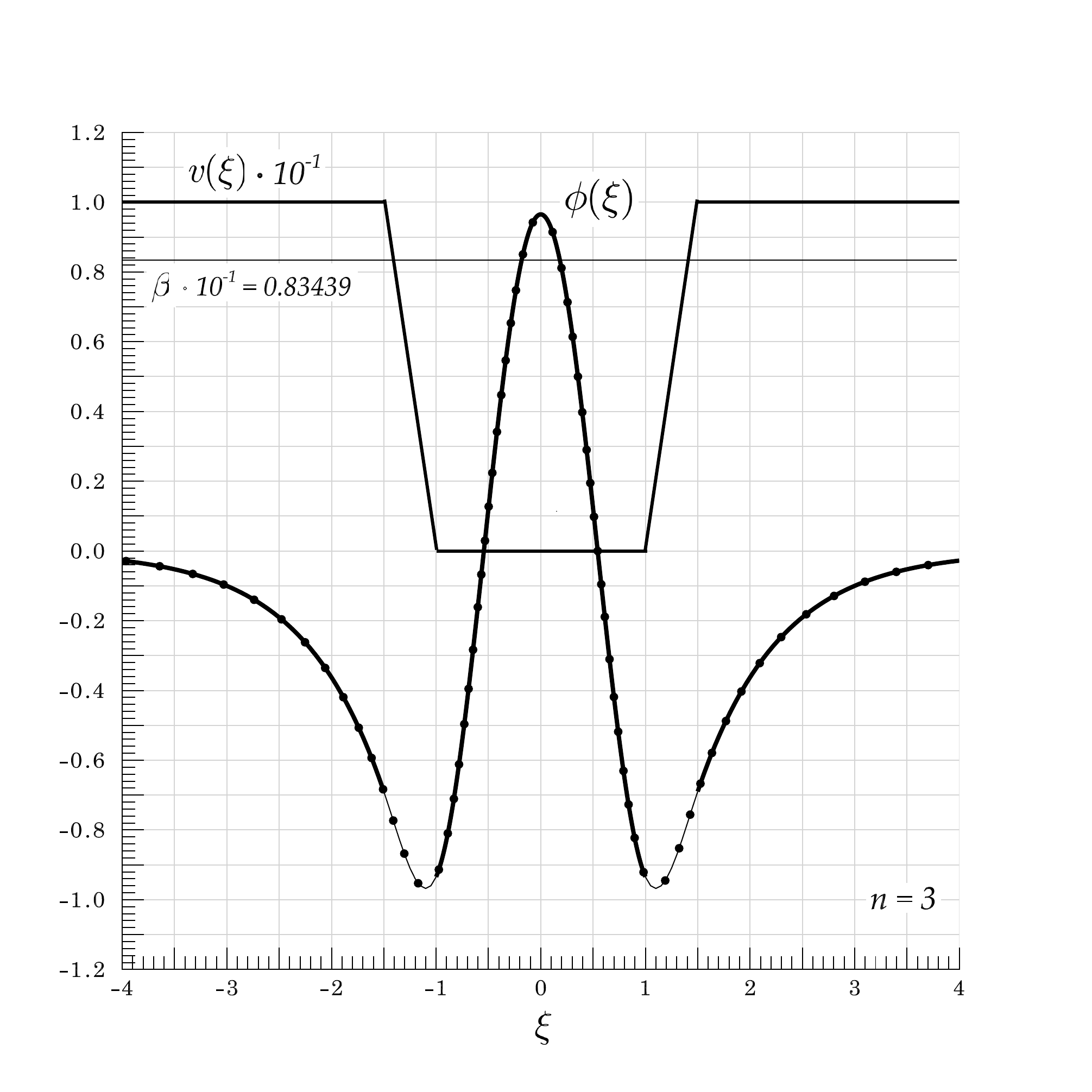}}}
   \caption{Eigenvalue-detection graph and eigenfunctions of the three eigenstates belonging to the symmetrical well $v_{1}=v_{2}=10, \lambda=0.5$.\llpush\label{c3-all}}
\end{figure}
The expected continuity of the eigenfunction's second derivative at the junction points, an aspect which will become of full relevance in the upcoming \Rse{ef.s}, is hardly verifiable visually from the graphs but this graphical limitation is of little concern because it can be surmounted analytically by repeatedly differentiating \REq{ef.all} to obtain first derivative
\begin{equation}\label{def.all}
   \pd{}{\phi}{\xi} =
                 \begin{cases}
                       + \sqrt{k_{1}} \tilde{B}_{1} \exp\left[ \left(  \xi + 1 + \lambda\right) \sqrt{k_{1}} \right]        \qquad & \text{zone 1}  \\[1.5ex]
                       - \sqrt{k_{1}} B_{1'} \dfrac{\Bi'(\eta)  - f_{1'} \,\Ai'(\eta)} {\sqrt{\etab}}   \qquad & \text{zone 1'} \\[1.5ex]
                       \sqrt{\beta} \left[  A_{0} \cos(\xi\sqrt{\beta}) - B_{0} \sin(\xi\sqrt{\beta})  \right]              \qquad & \text{zone 0}  \\[1.5ex]
                       + \sqrt{k_{2}} B_{2'} \dfrac{\Bi'(\zeta) - f_{2'} \,\Ai'(\zeta)}{\sqrt{\zetab}}  \qquad & \text{zone 2'} \\[1.5ex]
                       - \sqrt{k_{2}} \tilde{B}_{2} \exp\left[ \left( -\xi + 1 + \lambda\right) \sqrt{k_{2}} \right]        \qquad & \text{zone 2}  \\
                 \end{cases}
\end{equation}
and second derivative
\begin{equation}\label{d2ef.all}
   \pd{2}{\phi}{\xi} =
                 \begin{cases}
                       + k_{1} \tilde{B}_{1} \exp\left[ \left(  \xi + 1 + \lambda\right) \sqrt{k_{1}} \right]                   \qquad & \text{zone 1}  \\[1.5ex]
                       + k_{1} \dfrac{\eta}{\etab}  \, B_{1'} \left[ \Bi(\eta)  - f_{1'} \,\Ai(\eta)  \right] \qquad & \text{zone 1'} \\[1.5ex]
                       - \beta \left[  A_{0} \sin(\xi\sqrt{\beta}) + B_{0} \cos(\xi\sqrt{\beta})  \right]                \qquad & \text{zone 0}  \\[1.5ex]
                       + k_{2} \dfrac{\zeta}{\zetab}\, B_{2'} \left[ \Bi(\zeta) - f_{2'} \,\Ai(\zeta) \right] \qquad & \text{zone 2'} \\[1.5ex]
                       + k_{2} \tilde{B}_{2} \exp\left[ \left( -\xi + 1 + \lambda\right) \sqrt{k_{2}} \right]                   \qquad & \text{zone 2}  \\
                 \end{cases}
\end{equation}
and by evaluating \REq{d2ef.all} at the junction points.
In \REqd{def.all}{d2ef.all} and for the zones 1' and 2', we have used the derivatives transformations indicated in \REqq{ivt.der} and replaced the terms $\,{v_{s}}/{\lambda}\;(s=1,2)$ by inverting \REqd{mp.left.11p}{mp.right.2p2}; additionally in \REq{d2ef.all}, we have expressed the second derivatives of the Airy functions according to the corresponding differential equations [\REqd{airyde.1p}{airyde.2p}].
Let us see now what happens, for example, at the junction point \jp{1}{1'} whereat \itm{\xi=-(1+\lambda)} and \itm{\eta=\etab\,}: \REq{d2ef.all} (zones 1 and 1') gives
\begin{equation}\label{d2ef.11p}
   \pdatb{2}{\phi}{\xi}{\xi=-(1+\lambda)} =
                 \begin{cases}
                       + k_{1} \tilde{B}_{1}                    \qquad & \text{zone 1}  \\[1.5ex]
                       + k_{1} B_{1'} \left[ \Bi(\etab)  - f_{1'} \,\Ai(\etab)  \right] \qquad & \text{zone 1'} \\
                 \end{cases}
\end{equation}
which, with due account of \REq{B1t} (left equality), 
confirms the continuity of the eigenfunction's second derivative at the junction point under consideration.
Similar processes apply to and same confirmations are reached for the other junction points.
\FloatBarrier

\subsection{Wavefunction's general solution\label{gs}}

The determination of the eigenfunctions completes the study of the eigenvalue problem and we can concentrate again on the time-dependent problem [\REqd{seq}{seq.ic}].
The standard paradigm requires to assemble a specific solution
\begin{equation}\label{vst.n}
   \wf_{n}(x,t) = \Phi_{n}(0)\cdot \exp\left(-i \frac{\epsilon_{n} t}{\hbar}\right)\cdot\psi_{n}(x)
\end{equation}
for each eigenstate according to wavefunction's variable separation [\REq{vst}], based on the integral [\REq{evp.t.i}] of the temporal problem and the eigenfunction $\psi_{n}(x)$, and then to build up the wavefunction's general solution as a linear combination of the eigenstates' contributions
\begin{equation}\label{wgs}
   \wf(x,t) = \sum_{n=1}^{N }\wf_{n}(x,t) = \sum_{n=1}^{N } c_{n}\cdot \exp\left(-i \frac{\epsilon_{n} t}{\hbar}\right)\cdot\psi_{n}(x)
\end{equation}
In \REq{wgs}, $N$ represents the total number of eigenstates permitted by the potential.
We consider appropriate to recall here the discussion centered around \REqq{noesev} and the conclusion drawn from it: the absence ($N=0$) of eigenstates is a possibility (\Rfi{c7-Ds}) and, correspondingly, the quantum-mechanical problem does not entail separated-variable solutions; thus, the significance of \REqd{vst.n}{wgs} fades away.
The existence of eigenstates ($N>0$) grants the applicability of \REqd{vst.n}{wgs} and the determination of the coefficients $c_{n}$, which have absorbed the constants $\Phi_{n}(0)$, constitutes our next task.
For that purpose, we have at our disposal the initial-wavefunction condition [\REq{seq.ic}] and the moment has come to exploit it.

In principle, the initial wavefunction $F(x)$ should be looked at as arbitrary to some extent although, in spite of its presumed arbitrariness, it cannot escape two important constraints attached to the initial time ($t=0$): it has to be consistent with both the normalization condition [\REq{wfn}]
\begin{equation}\label{wfn.i}
   \intmpi{\wfc(x,0)\cdot\wf(x,0)}{x} = \intmpi{\cco{F}(x)\cdot F(x)}{x} = 1
\end{equation}
and the boundary conditions [\REqq{seq.bc}]
\begin{subequations}\label{seq.bc.0} \seqn
  \begin{align}
     G_{1}[F(-\infty),F_{x}(-\infty),F(+\infty),F_{x}(+\infty,)] = 0 \label{seq.bc.0.1}\\[.25\baselineskip]
     G_{2}[F(-\infty),F_{x}(-\infty),F(+\infty),F_{x}(+\infty,)] = 0 \label{seq.bc.0.2}
  \end{align}
\end{subequations}
which, more specifically for our problem [\REq{seq.bc.i.v}], reduce to
\begin{equation}\label{seq.bc.i.v.0}
   F(-\infty) = F(+\infty) = 0
\end{equation}
The substitution of the general solution [\REq{wgs}] into the initial condition [\REq{seq.ic}] gives
\begin{equation}\label{wgs.0}
   \sum_{n=1}^{N } c_{n} \psi_{n}(x) = F(x)
\end{equation}
It is then seemingly rather straightforward from a mathematical point of view to take advantage of eigenfunctions' orthonormality [\REqd{ef.ortho}{efn}] to invert \REq{wgs.0} and to obtain the coefficients
\begin{equation}\label{coeff.wgs}
  \subeqn{c_{m} = \intmpi{\cco{\psi}_{m}(x) \cdot F(x)}{x}}{m=1,\ldots,N}{}
\end{equation}
And that is fine, of course.
However, we wish to look at \REq{wgs.0} from a slightly different angle with respect to the standard one of the literature and point out an aspect that, we believe, is hardly emphasized in quantum-mechanics textbooks, at least in those we have %
consulted.\footnote{For example, Griffiths \cite{dg2005} dealt with the method, that he colorfully called ``Fourier's trick'', to obtain the coefficients $c_{n}$ in Sec. 2.2, at page 30 of his textbook, dedicated to the infinite \swp, a potential with an infinite number of eigenstates; but there is no mention to the ``Fourier's trick'' in Sec. 2.6 at page 78 where the finite \swp\ is considered, a potential that gives rise to a finite number of eigenstates (\Rfi{evdgtc}).
A similar situation can be found also in Bransden and Joachain's textbook \cite{bb2000}.}
If the number $N$ of eigenstates is finite then \REq{wgs.0} must be read from right to left: the initial wavefunction cannot be arbitrary but must conform to the mathematical structure of a linear combination of eigenfunctions, say
\begin{equation}\label{iwf.lcef}
   \subeqn{F(x) = \sum_{r=1}^{N} \alpha_{r} \psi_{r}(x)}{r=1,\ldots,N}{}
\end{equation}
with
\begin{equation}\label{salpha}
   \sum_{r=1}^{N} \cco{\alpha_{r}} \alpha_{r} = 1
\end{equation}
in compliance with \REq{wfn.i}, as necessary condition for the existence of separated-variable solutions [\REq{wgs}].
Then
\begin{equation}\label{coeff.wgs.a}
  \subeqn{c_{m} = \alpha_{m}}{m=1,\ldots,N}{}
\end{equation}
and
\begin{equation}\label{wgs.a}
   \wf(x,t) =  \sum_{n=1}^{N } \alpha_{n}\cdot \exp\left(-i \frac{\epsilon_{n} t}{\hbar}\right)\cdot\psi_{n}(x)
\end{equation}
These considerations are brought forth with dramatic evidence by the potential of \Rfi{c1-ef1} which produces only \itm{N=1} eigenstate (\Rfi{c1-Da}).
In that case, \itm{c_{1}=\alpha_{1}=1};
if the particle occupies initially the unique eigenstate shown in \Rfi{c1-ef1} [$F(x)=\psi_{1}(x)$] then its wavefunction is simply
\begin{equation}\label{wgs.oes}
   \wf(x,t) = \wf_{1}(x,t) =  \exp\left(-i \frac{\epsilon_{1} t}{\hbar}\right)\cdot\psi_{1}(x)
\end{equation}
and the particle will continue to occupy that unique eigenstate forever.
Otherwise [$F(x)\neq\psi_{1}(x)$] there are no other separated-variable solutions and the differential-equation problem [\REq{seq}, \REq{seq.ic}, \REq{seq.bc.i.v}] requires numerical integration.
In general, separated-variable solutions to the \SEq\ with finite-well potentials do not exist for arbitrary initial wavefunctions; they do exist only for properly structured initial wavefunctions [\REq{iwf.lcef}].
We believe it is even more instructive didactically to press the argument into graphical evidence by considering the triangular-shaped function
\begin{equation}\label{iwf.ts}
   F(x) = \sqrt{\frac{3}{2L}} \cdot
    \begin{cases}
             0             & \quad        \xi \le -1  \\[1ex]
             \xi + 1   & \quad -1 \le \xi \le  0  \\[1ex]
             1 - \xi   & \quad  0 \le \xi \le +1  \\[1ex]
             0             & \quad +1 \le \xi         
   \end{cases}
\end{equation}
This is a perfectly legitimate initial wavefunction because it complies with both normalization [\REq{wfn.i}] and boundary conditions [\REq{seq.bc.i.v.0}].
It generates the coefficients
\begin{equation}\label{coeff.wgs.ts}
  \subeqn{c_{n} = \sqrt{3} \, B_{0n}\frac{1-\cos(\sqrt{\beta_{n}})}{\beta_{n}} }{n=1,\ldots,N}{}
\end{equation}
from \REq{coeff.wgs} with due account of the adopted variable scaling [\REqq{nd}] and eigenfunction's analytical expression [\REq{ef.all}].
We have carried out calculations of the inital condition [\REq{wgs.0}] for the \twp\ of \Rfi{c3-all} which includes \itm{N=3} eigenstates and for the \swp\ considered by de Alcantara and Griffiths \cite{oda2006ajp} which includes \itm{N=10} eigenstates (\Rfi{evdg.dabg}).
\Rfib{iwf.c3} refers to the former potential and illustrates how poorly the left-hand side of \REq{wgs.0} approximates the triangular-shaped initial wavefunction; in particular, the sum of the quantum-state probabilities $P_{n} = c_{n}^2$, tabulated in the figure, differs appreciably from unity.
The situation corresponding to the latter potential is shown in \Rfi{iwf.dabg} and reveals a noticeable improvement in accuracy due to the existence of more eigenstates but the match is not rigorously exact.
\begin{figure}[h]
\dgcap{.05}{.05}
   \subfloat[\twp\ illustrated in \Rfi{c3-all}.] {\label{iwf.c3}                      \resizebox{.47\textwidth}{!}{\includegraphics*[trim=33 20 60 60]{\gdir/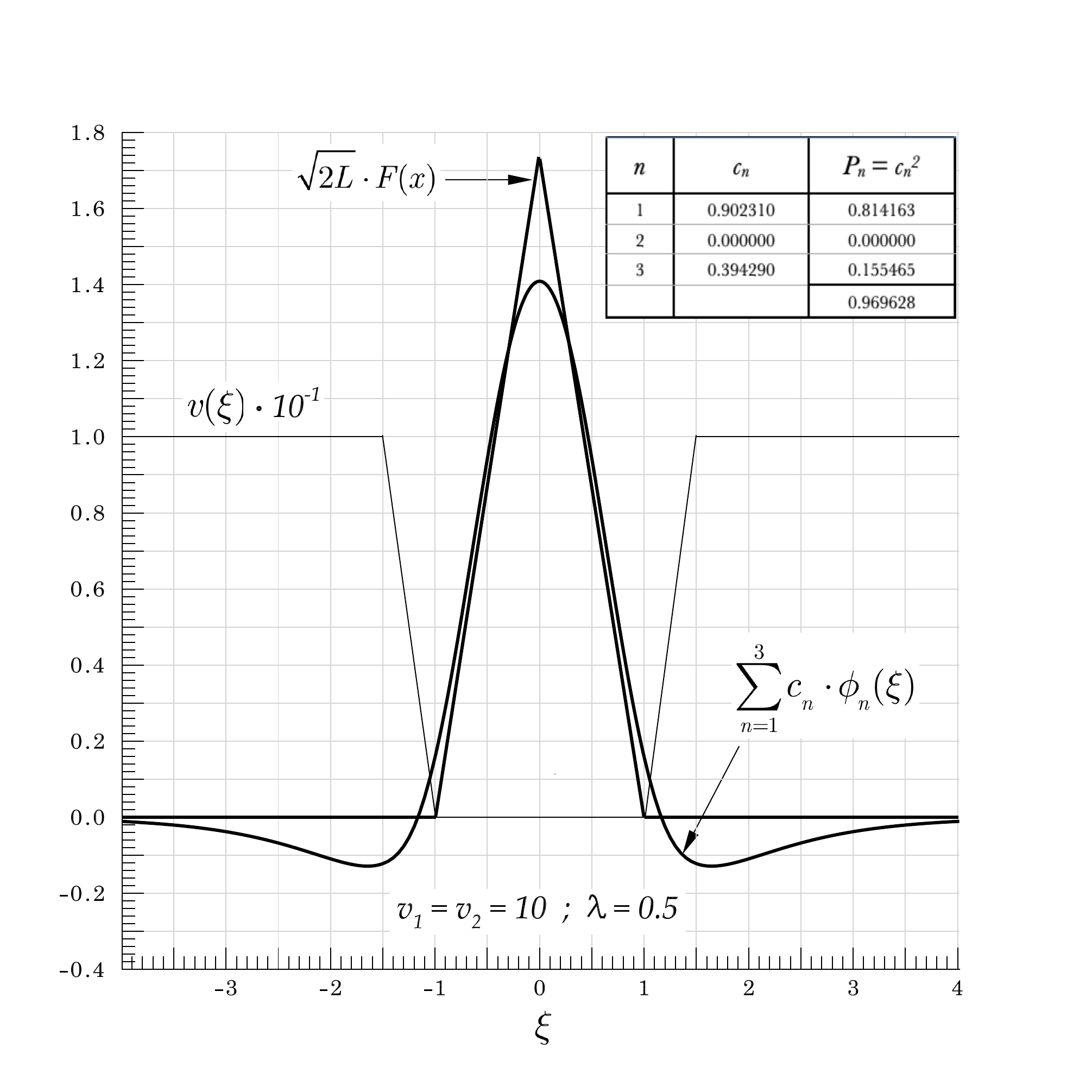}}}
   \subfloat[\swp\ of de Alcantara and Griffiths \cite{oda2006ajp}] {\label{iwf.dabg} \resizebox{.47\textwidth}{!}{\includegraphics*[trim=33 20 60 60]{\gdir/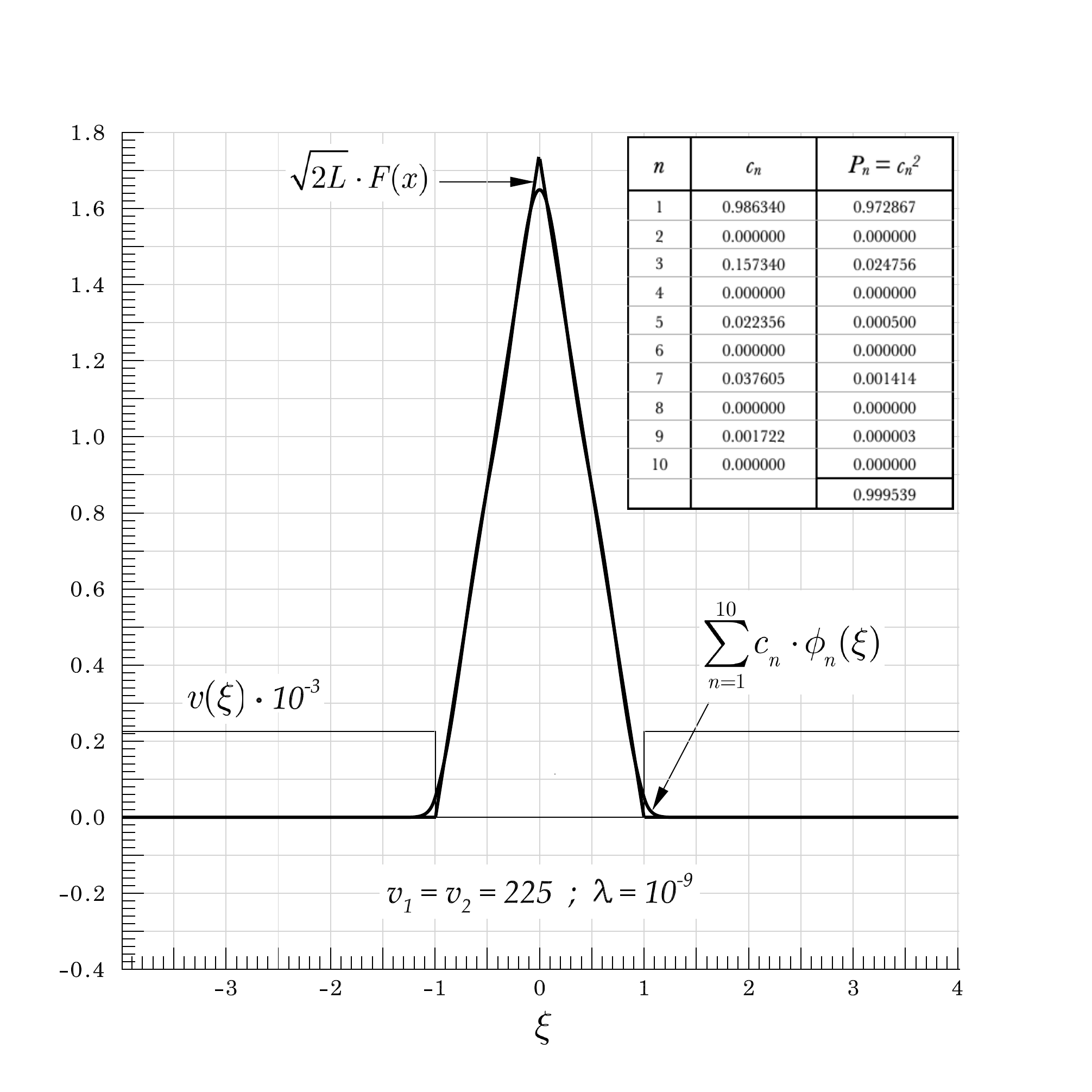}}}
   \caption{Numerical test of accuracy of \REq{wgs.0} with a triangular-shaped initial wavefunction for two potentials with finite number of eigenstates.\llpush\label{iwf}}
\end{figure}
The inversion operation from \REq{wgs.0} to \REq{coeff.wgs} to obtain the coefficients $c_{n}$ if the initial wavefunction $F(x)$ is arbitrary acquires physical significance and works exactly only if the number of eigenstates is infinite $(N\rightarrow\infty)$,\footnote{Persico \cite{ep1936} is the only author we found who touched upon this matter. In Sec. 9 at page 105 of his textbook, he clearly explained the necessity of $N\rightarrow\infty$ to confer physical significance to \REq{wgs.0} if the initial wavefunction $F(x)$ is arbitrary and, in the footnote ($^{1}$) of the mentioned page, he referred the reader to Courant and Hilbert \cite{ch1924,ch2004} who treated the subject of series expansions of arbitrary functions in all its mathematical splendor in chapter II of their referenced textbooks.
In the English translation \cite{ep1950} of Persico's textbook, Sec. 9 is at page 98 and the footnote is number 5 at page 99.}
and that happens only to infinite-well potentials.
Then the separated-variable wavefunction
\begin{equation}\label{wgs.Ninf}
   \wf(x,t) = \sum_{n=1}^{\infty}\wf_{n}(x,t) = \sum_{n=1}^{\infty} c_{n}\cdot \exp\left(-i \frac{\epsilon_{n} t}{\hbar}\right)\cdot\psi_{n}(x)
\end{equation}
is truly a general solution built as a series expansion based on infinite eigenfunctions that constitute a complete set in the sense explained by Griffith \cite{dg2005}.

\section{The square-well potential as limit when \itm{\,\boldsymbol{\lgz}}\label{ec.swph}}

\subsection{Introductory remarks}
With the completion of the study of the \twp, we have acquired all the elements necessary to deal with the implications of a vanishing $\lambda$ and we are ready to explore the circumstances under which the \twp\ (\Rfi{ndtwp}) turns into a \swp\ (\Rfi{ndswp}).
If \lgzt, geometrically the potential's ramps become vertical and the zones with linear potential shrink to points;
analytically, overlined and circumflexed values [\REqd{mp.left}{mp.right}] coincide and vanish  \itm{(\etah = \etab = 0 ;\, \zetah = \zetab = 0)},
the variables $\eta$ and $\zeta$ freeze at \itm{\eta=\zeta=0} [\REq{ivt.range}], the original variable $\xi$ gets nailed down at the fixed values $\xi=\mp 1$ [\REqd{ivt.1p}{ivt.2p}], and
the potential's functional definitions [\REq{ndevp.twp}, second and fourth line from top] go into mathematical indeterminate forms of the kind $0/0$, which is another way of saying that the potential turns into a multi-valued function spanning all values comprised in $[0,v_{1}]$ at \itm{\xi=-1} and in $[0,v_{2}]$ at \itm{\xi=+1}.
The task ahead of us consists mainly in finding out how the collapse of the zones $s=1',2'$ affects the eigenvalue spectrum  and the eigenfunctions obtained for the \twp\ (\Rse{tevp}), particularly its repercussions on the solutions in the collapsed zones (\Rse{z1p2p}).
The main questions whose answers are of particular interest to us regard whether or not the collapsed eigenvalue spectrum checks with the one ensuing from the \sta\ (\Rse{ev.lgtz}), and what happens to the continuity property of eigenfunction and its derivatives if the potential's junction points become jump points (\Rse{ef.s}).
%
\begin{figure}[h]
   \resizebox{.47\textwidth}{!}{\includegraphics*[trim=33 20 60 60]{\gdir/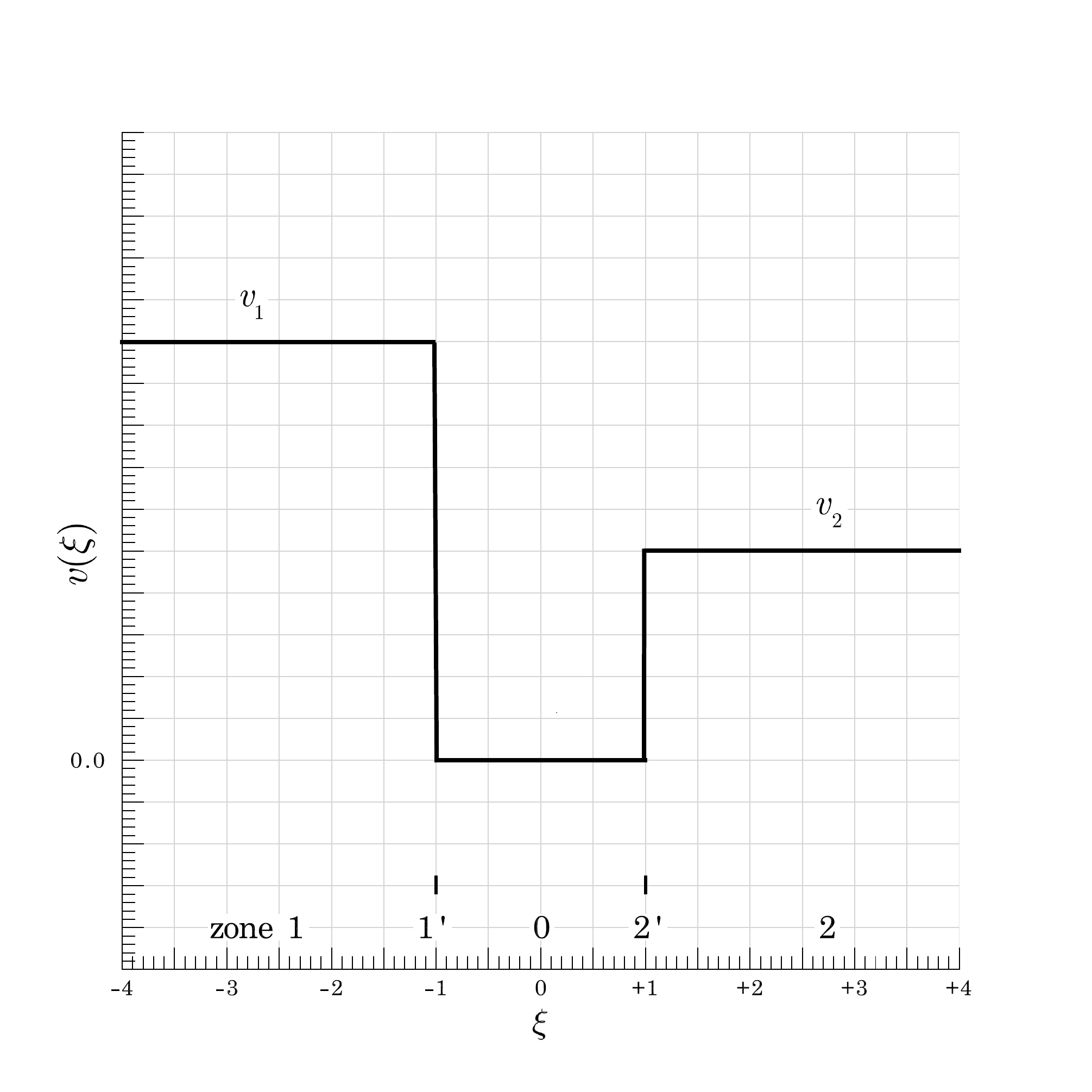}}
   \caption{Nondimensional \swp.\label{ndswp}}
\end{figure}
%

Before embarking in the accomplishment of the described task, it is convenient to forge briefly a few preparatory tools meant to facilitate the forthcoming mathematical operations.

\subsection{Mathematical tools\label{mt}}

The factors $f_{1'}$ and $f_{2'}$ [\REqd{fbar}{f2p}] can be both collected into the generic function
\begin{equation}\label{fz}
    f(z) = \frac{\sqrt{z}\,\,\Bi(z) + \Bi'(z)}{\sqrt{z}\,\,\Ai(z) + \Ai'(z) }
\end{equation}
whose dummy variable $z$ represents the overlined values \itm{\etab\,,\zetab\,} [\REqd{mp.left.11p}{mp.right.2p2}].
Similarly to what we did already with \REq{fbar.i}, we rearrange \REq{fz} into the convenient identity
\begin{equation}\label{fz.1}
    \Bi(z) - f(z)\,\Ai(z) = -\frac{ \Bi'(z) - f(z)\,\Ai'(z) }{\sqrt{z}}
\end{equation}
The factors $\gl$ and $-\gr$ [\REqd{gleft}{gright}] can also be absorbed 
into the generic function
\begin{equation}\label{gw}
   g(w,z) = - \sqrt{-w} \frac{\Bi(w) - f(z)\,\Ai(w)}{\Bi'(w) - f(z)\,\Ai'(w)}
\end{equation}
whose dummy variable $w$ represents the circumflexed values \itm{\etah\,,\zetah\,} [\REqd{mp.left.1p0}{mp.right.02p}].

The ratio of the dummy variables is not affected by $\lambda$, as it is easily verified by member-to-member division of \REqd{mp.left}{mp.right} respectively
\begin{equation}\label{wonz}
   \subeqn{\frac{w}{z} \equiv \frac{\etah}{\etab}\; \mbox{or}\; \frac{\zetah}{\zetab} \rightarrow - \frac{\beta}{k_{s}} }{s=1,2}{}
\end{equation}
so, with the shrinking \lgzt, the dummy variables are forced to vanish \itm{(z,w \rightarrow 0)}, because of what they represent, but their ratio stays finite.
The function $f(z)$ goes into the numerical constant
that we have already met in \REq{ff.stwp}.
The left-hand side of \REq{fz.1} attains the numerical constant
\begin{equation}\label{fz.1.lhs.s}
    \Lambda = \llgz \left[ \Bi(z) - f(z)\,\Ai(z)  \right] = \Bi(0) - f(0)\,\Ai(0) \simeq 1.22985
\end{equation}
and so must do the apparently indeterminate form on the right-hand side
\begin{equation}\label{fz.1.rhs.s}
    \llgz \left[-\frac{ \Bi'(z) - f(z)\,\Ai'(z) }{\sqrt{z}}\right] = \Lambda
\end{equation}
A corroborating check, perhaps more convincing and certainly more elegant from a mathematical point of view, of the trueness of \REq{fz.1.rhs.s} consists in processing the limit according to de L'H\^{o}pital's theorem, an exercise that we did for the sake of completeness\footnote{We admit that mathematical curiosity pushed as well to some extent.} and were pleased to see its outcome to fall inline with \REq{fz.1.rhs.s}.
Other recurrent limits are similar to \REqd{fz.1.lhs.s}{fz.1.rhs.s} but the dummy variables are mixed, as in the numerator and denominator of the function \itm{g(w,z)} for example.
The limit of the numerator of \REq{gw} is easy
\begin{equation}\label{fz.1.lhs.s.m}
    \llgz \left[ \Bi(w) - f(z)\,\Ai(w)  \right] = \Bi(0) - f(0)\,\Ai(0) = \Lambda
\end{equation}
The limit of the other one
\begin{subequations}\label{fz.1.rhs.s.mixed}\seqn
    \begin{equation}\label{fz.1.rhs.s.m}
        \llgz \left[-\frac{ \Bi'(w) - f(z)\,\Ai'(w) }{\sqrt{-w}}\right]
    \end{equation}
    requires a bit of attention.
    We must first adapt the square root by taking advantage of \REq{wonz}
    \begin{equation}\label{sqrw}
       \sqrt{-w} = \sqrt{z}\,\sqrt{\frac{\beta}{k_{s}}}
    \end{equation}
    so that we can preliminarily transform \REq{fz.1.rhs.s.m}
    \begin{equation}\label{fz.1.rhs.s.m.0}
        \llgz \left[-\frac{ \Bi'(w) - f(z)\,\Ai'(w) }{\sqrt{-w}}\right] =  \sqrt{\frac{k_{s}}{\beta}} \,\llgz \left[-\frac{ \Bi'(w) - f(z)\,\Ai'(w) }{\sqrt{z}}\right]
    \end{equation}
    and then proceed with the limit on the right-hand side of \REq{fz.1.rhs.s.m.0}
    \begin{equation}\label{fz.1.rhs.s.m.1}
        \llgz \left[-\frac{ \Bi'(w) - f(z)\,\Ai'(w) }{\sqrt{z}}\right] =
        \llgz \left[-\frac{ \Bi'(z) - f(z)\,\Ai'(z) }{\sqrt{z}} \cdot \frac{ \Bi'(w) - f(z)\,\Ai'(w) }{ \Bi'(z) - f(z)\,\Ai'(z) }\right]
        = \Lambda\cdot 1 = \Lambda
    \end{equation}
    which leads to the final result
    \begin{equation}\label{fz.1.rhs.s.m.f}
        \llgz \left[-\frac{ \Bi'(w) - f(z)\,\Ai'(w) }{\sqrt{-w}}\right] = \sqrt{\frac{k_{s}}{\beta}}\,\Lambda
    \end{equation}
\end{subequations}
With the mixed-variable limits [\REqd{fz.1.lhs.s.m}{fz.1.rhs.s.m.f}] in hand, the important limit of the function $g(w,z)$ follows easily
    \begin{equation}\label{gw.s}
       \subeqn{\quad\llgz g(w,z) = \llgz \frac{\Bi(w)-f(z)\Ai(w)}{-\dfrac{\Bi'(w)-f(z)\Ai'(w)}{\sqrt{-w}}}=\sqrt{\frac{\beta}{k_{s}}}}{s=1,2}{1.5}
    \end{equation}

\subsection{Eigenvalues\label{ev.lgtz}}
Our first check consists in the verification of the retrieval of the same spectrum produced by the \sta.
If \lgzt, according to the limit indicated in \REq{gw.s}, the factors $\gl$ and $\gr$ become
\begin{subequations}\label{gf.sp} \seqn
  \begin{equation}\label{gleft.sp}
     \gl = + \sqrt{\frac{\beta}{k_{1}}}
  \end{equation}
  \begin{equation}\label{gright.sp}
     \gr = - \sqrt{\frac{\beta}{k_{2}}}
  \end{equation}
\end{subequations}
and the transcendental equation [\REq{eigenv}] that produces the eigenvalues goes into the slightly simpler form
\begin{equation}\label{eigenv.sp}
   [D(\beta)]_{\mathrm{swp}} = \llgz D(\beta) = \left( 1 - \frac{\beta}{ \sqrt{k_{1} k_{2}} } \right) \sin(2\sqrt{\beta}) + \left( \sqrt{\frac{\beta}{k_{1}}} + \sqrt{\frac{\beta}{k_{2}}} \right) \cos(2\sqrt{\beta}) = 0
\end{equation}
As example to verify that \REq{eigenv.sp} is indeed in line with the transcendental equations proposed in the literature, we take the \swp\ considered by Reed \cite{bcr1990ajp}; in his case, \itm{v_{1}=v_{2}=v\,}; \itm{k_{1}=k_{2}=k=v-\beta} and the simplified transcendental equation [\REq{eigenv.sp}] reduces even further to
\begin{equation}\label{eigenv.sp.reed}
   [D(\beta)]_{\mathrm{Reed}} = \left( 1 - \frac{\beta}{k} \right) \sin(2\sqrt{\beta}) + 2 \sqrt{\frac{\beta}{k}} \cos(2\sqrt{\beta}) = 0
\end{equation}
By taking into account the notation conversions based on Reed's definitions and collected in \Rta{nctr}, it is straightforward to prove that \REq{eigenv.sp.reed} coincides exactly with Reed's Eq.~(15) that we reproduce here
\begin{equation*}
  \frac{1}{k}\,[D(\beta)]_{\mathrm{Reed}} \equiv f(\xi,K) = (K^{2} - 2 \xi^{2}) \sin(2\xi) + 2\xi \sqrt{K^{2} - \xi^{2}} \cos(2\xi) = 0
\end{equation*}
for the reader's convenience.
\begin{table}[h]
\dgcap{.0}{.0}
   \caption{\color{black}Notation conversions relative to Reed's transcendental equation, Eq.~(15) in \cite{bcr1990ajp}. \label{nctr}}
   \vspace*{-.8\baselineskip}
   \resizebox{.69\textwidth}{!}{\includegraphics*[trim=0 0 0 0]{\gdir/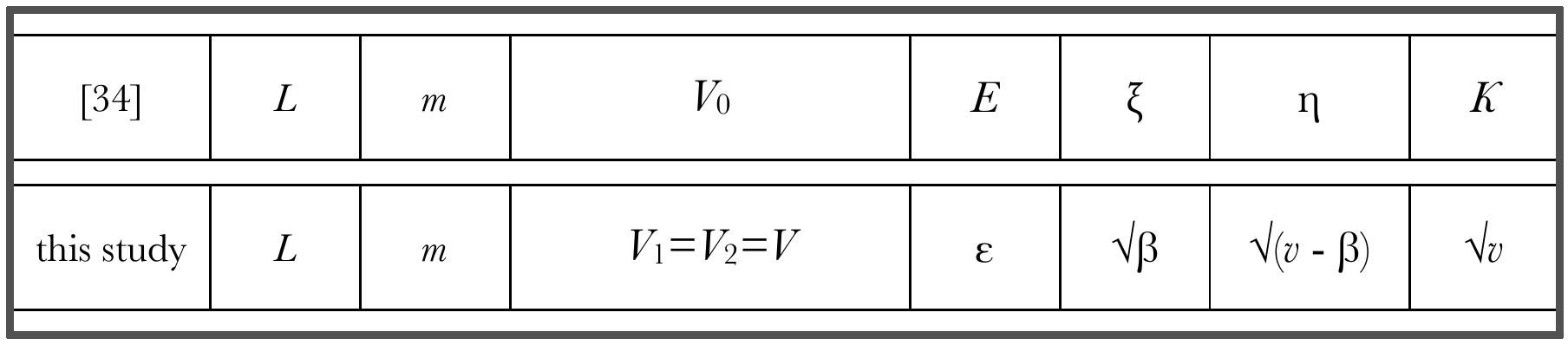}}
\end{table}
%

Further verification can be achieved with regard to the determination of the angle $\varphi$ needed in \REqq{eigenv.trig.angle}.
We start again from \REqq{gf.sp}; then, in cascade, we evaluate the reciprocal factors [\REq{negbetas}, \REq{oog}]
\begin{subequations}\label{oogf.sp} \seqn
  \begin{equation}\label{oogleft.sp}
     \gal = + \sqrt{\frac{k_{1}}{\beta}} = + \sqrt{\frac{v_{1}}{\beta}-1}
  \end{equation}
  \begin{equation}\label{oogright.sp}
     \gar = - \sqrt{\frac{k_{2}}{\beta}} = - \sqrt{\frac{v_{2}}{\beta}-1}
  \end{equation}
\end{subequations}
and determine coefficients $C,S$ and normalization factor $R$ [\REqq{coeff}]
\begin{subequations}\label{coeff.sp} \seqn
  \begin{align}
     C & = 1 - \sqrt{\left(\frac{v_{1}}{\beta} - 1 \right)\left(\frac{v_{2}}{\beta} - 1 \right)} \label{coeff.C.sp} \\[.25\baselineskip]
     S & =   - \sqrt{\frac{v_{1}}{\beta} - 1 } - \sqrt{\frac{v_{2}}{\beta} - 1 }                 \label{coeff.S.sp} \\[.25\baselineskip]
     R & = \sqrt{\frac{v_{1}}{\beta}\frac{v_{2}}{\beta}}                                         \label{coeff.R.sp}
  \end{align}
\end{subequations}
With these simplifications, the equations that fix the angle $\varphi$ [\REqq{angle}] acquire the interesting structure
\begin{subequations}\label{angle.sp} \seqn
  \begin{align}
    \cos\varphi & = \sqrt{1-\frac{\beta}{v_{1}}}\sqrt{1-\frac{\beta}{v_{2}}} - \sqrt{\frac{\beta}{v_{1}}} \sqrt{\frac{\beta}{v_{2}}}  \label{cosphi.sp} \\[.25\baselineskip]
    \sin\varphi & = \sqrt{\frac{\beta}{v_{2}}}\sqrt{1-\frac{\beta}{v_{1}}}   + \sqrt{\frac{\beta}{v_{1}}}\sqrt{1-\frac{\beta}{v_{2}}} \label{sinphi.sp}
  \end{align}
\end{subequations}
The square-root terms \itm{\sqrt{\beta/v_{s}}} and \itm{\sqrt{1-\beta/v_{s}}} \;\;$(s=1,2)$
are both contained in $[0,1]$ and the sum of their squares adds up to unity; therefore, they uniquely identify an angle $\omega_{s}$ in $[0,\pi/2]$ which can be extracted by setting
\begin{subequations}\label{omega.sp} \seqn
  \begin{align}
    \cos\omega_{s} & = \sqrt{1-\frac{\beta}{v_{s}}}  \label{cosome.sp} \\[.25\baselineskip]
    \sin\omega_{s} & = \sqrt{\frac{\beta}{v_{s}}}    \label{sinome.sp}
  \end{align}
\end{subequations}
We can now take advantage of \REqq{omega.sp} to make the angles $\omega_{1}, \omega_{2}$ appear in \REqq{angle.sp} and carry on with their manipulation
\begin{subequations}\label{angleome.sp} \seqn
  \begin{align}
    \cos\varphi & = \cos\omega_{1} \cos\omega_{2} - \sin\omega_{1} \sin\omega_{2} = \cos\left(\omega_{1}+\omega_{2}\right)  \label{cosphi.sp.o} \\[.25\baselineskip]
    \sin\varphi & = \sin\omega_{2} \cos\omega_{1} + \sin\omega_{1} \cos\omega_{2} = \sin\left(\omega_{1}+\omega_{2}\right) \label{sinphi.sp.o}
  \end{align}
\end{subequations}
to reach the simple result
\begin{equation}\label{angle.sp.sum}
  \varphi = \omega_{1} + \omega_{2}
\end{equation}
If we now invert
\REq{sinome.sp}
  \begin{equation}
    \omega_{s} = \arcsin\sqrt{\frac{\beta}{v_{s}}}    \label{invsinome.sp}
  \end{equation}
%
\begin{table}[h]
\dgcap{.0}{.0}
   \caption{Notation conversions relative to Messiah's, ter Haar's, and Landau and Lifchitz's transcendental equations \cite{am11961,dth1964,ll1977}; see footnote \ref{mthl}.\llpush\label{nctmtl}}
   \vspace*{-.8\baselineskip}
   \resizebox{.99\textwidth}{!}{\includegraphics*[trim=0 0 0 0]{\gdir/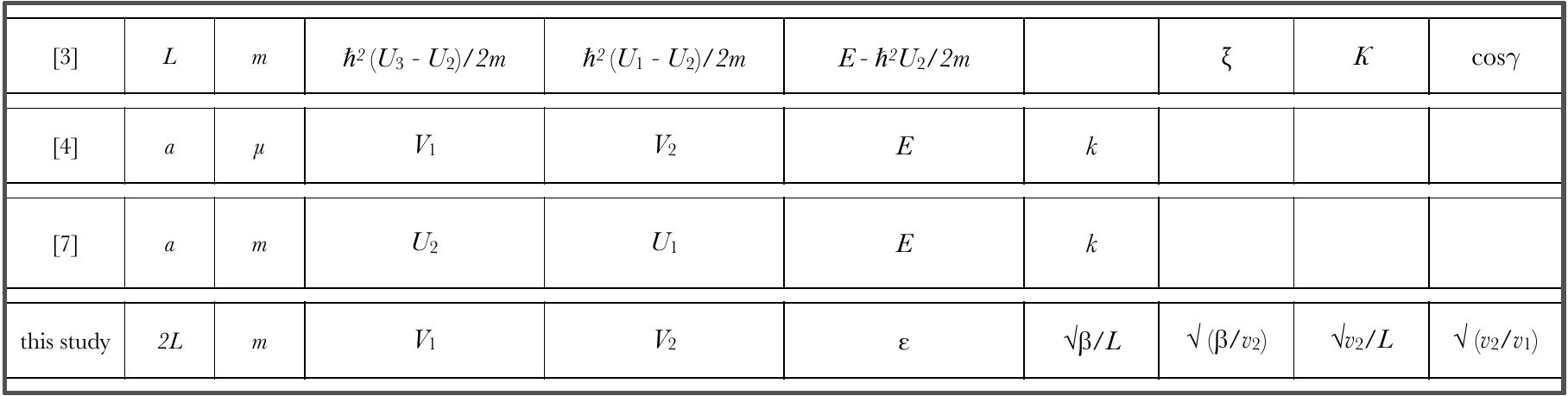}}
\end{table}
%
\FloatBarrier\noindent
then the angle $\varphi$ becomes
\begin{equation}\label{angle.sp.sum.as}
  \varphi = \arcsin\sqrt{\frac{\beta}{v_{1}}} + \arcsin\sqrt{\frac{\beta}{v_{2}}}
\end{equation}
and, by substituting \REq{angle.sp.sum.as} into \REq{eigenv.trig.angle.kpi}, we reach the transcendental equation
	\begin{equation}\label{eigenv.trig.angle.kpi.mtl}
	   \subeqn{2\sqrt{\beta} + \arcsin\sqrt{\frac{\beta}{v_{1}}} + \arcsin\sqrt{\frac{\beta}{v_{2}}} = n \pi}{n=1,2,\ldots}{0}
	\end{equation}
that matches exactly those proposed by Messiah \cite{am11961}, ter Haar \cite{dth1964} and Landau and Lifchitz \cite{ll1977};\footnote{The transcendental equations appear respectively: Messiah's in Sec. 6 of chapter III, just above Fig. III.4 at page 90 of \cite{am11961}, ter Haar's at the bottom of page 66 of \cite{dth1964}, and  Landau and Lifchitz's in the discussion relative to problem 2, Eq.~(1) at page 66 of \cite{ll1977}.\label{mthl}}
verification is straightforward via the notation conversions collected in \Rta{nctmtl} for the reader's convenience.

Another verification, that deserves mentioning, concerns the eigenvalue-absence condition [\REq{cae}] which becomes formally
\begin{equation}\label{cae.swp} 
   \frac{2\sqrt{v_{2}} + \left[ \varphi(\beta,v_{1},v_{2},\lambda\rightarrow0)\right]_{\beta=v_{2}}}{\pi} < 1
\end{equation}
The angle $\varphi$ appearing in \REq{cae.swp} descends from \REq{angle.sp.sum.as} with $\beta=v_{2}$
\begin{equation}\label{angle.sp.sum.as.bv2}
  \left[ \varphi(\beta,v_{1},v_{2},\lambda\rightarrow0)\right]_{\beta=v_{2}} = \arcsin\sqrt{\frac{v_{2}}{v_{1}}} + \frac{\pi}{2}
\end{equation}
The substitution of \REq{angle.sp.sum.as.bv2} into the eigenvalue-absence condition [\REq{cae.swp}] leads to the final form
\begin{equation}\label{cae.swp.f} 
   2\sqrt{v_{2}} < \frac{\pi}{2} - \arcsin\sqrt{\frac{v_{2}}{v_{1}}}
\end{equation}
in full agreement with the Landau and Lifchitz's condition\footnote{Landau and Lifchitz give the condition for eigenvalues' existence rather than absence; therefore, their Eq.~(2) contains the operator $\geq$ instead of the operator $<$, as in our \REq{cae.swp.f}.} given in their Eq.~(2) at page 66 of \cite{ll1977}.
Moreover, by taking into account the angular equivalence
\begin{equation}\label{av}
   \arccos\sqrt{\frac{v_{2}}{v_{1}}} = \frac{\pi}{2} - \arcsin\sqrt{\frac{v_{2}}{v_{1}}}
\end{equation}
we can reformulate \REq{cae.swp.f} in the form
\begin{equation}\label{cae.swp.f.m} 
   2\sqrt{v_{2}} < \arccos\sqrt{\frac{v_{2}}{v_{1}}}
\end{equation}
which coincides with Messiah's condition indicated at page 91 of \cite{am11961}.
In the case of a symmetrical \swp\ (\itm{v_{1}=v_{2}=v}), \REqd{cae.swp.f}{cae.swp.f.m} reduce to the simple inequality
\begin{equation}\label{cae.swp.f.sym} 
   2\sqrt{v} < 0
\end{equation}
which reconfirms the unconditional existence of eigenvalues because it is never verified.

\subsection{Eigenfunctions and derivatives\label{ef.s}}
The successful verifications we have carried out in \Rse{ev.lgtz} on transcendental equations imply reassurance regarding the eigenvalue spectrum: we retrieve exactly the same spectrum of the \sta.
With a comfortable sensation of being on the right track, we turn to next investigation which involves the eigenfunctions and their derivatives.


The simplification of the factors $\gl$ and $\gr$ [\REqq{gf.sp}] has a modest impact on the formulae for the calculation of the coefficients $A_{0}, B_{0}, D_{0}$ [\REqq{iC0D0}, \REq{D00}] but affects more markedly the other coefficients.
The most important are $B_{1'}$ and $B_{2'}$; with due account of the limits indicated in \REqd{fz.1.lhs.s.m}{fz.1.rhs.s.m.f}, they follow from \REqdt{B1p.1}{B2p.1}
\begin{equation}\label{B1p.1.s}
   B_{1'} = \frac{-A_{0} \sin(\sqrt{\beta}) + B_{0} \cos(\sqrt{\beta})}{\Lambda}
          = + \sqrt{\frac{\beta}{k_{1}}} \frac{ A_{0} \cos(\sqrt{\beta}) + B_{0} \sin(\sqrt{\beta})}{\Lambda}
\end{equation}
\begin{equation}\label{B2p.1.s}
   B_{2'} = \frac{A_{0} \sin(\sqrt{\beta}) + B_{0} \cos(\sqrt{\beta})}{\Lambda}
          = - \sqrt{\frac{\beta}{k_{2}}} \frac{ A_{0} \cos(\sqrt{\beta}) - B_{0} \sin(\sqrt{\beta})}{\Lambda}
\end{equation}
After them, the coefficients $\tilde{B}_{1}$ and $\tilde{B}_{2}$ follow from \REqd{B1t}{B2t}
\begin{equation}\label{B1t.s}
   \tilde{B}_{1} = B_{1'} \Lambda = -A_{0} \sin(\sqrt{\beta}) + B_{0} \cos(\sqrt{\beta}) = + \sqrt{\frac{\beta}{k_{1}}}\left[ A_{0} \cos(\sqrt{\beta}) + B_{0} \sin(\sqrt{\beta}) \right]
\end{equation}
\begin{equation}\label{B2t.s}
   \tilde{B}_{2} = B_{2'} \Lambda = A_{0} \sin(\sqrt{\beta}) + B_{0} \cos(\sqrt{\beta}) = - \sqrt{\frac{\beta}{k_{2}}} \left[ A_{0} \cos(\sqrt{\beta}) - B_{0} \sin(\sqrt{\beta}) \right]
\end{equation}
Finally, the equation meant to fix the coefficient $C_{0}$ [\REq{C0}] generated by the eigenfunction's normalization condition [\REq{efn.nd}] simplifies to
\begin{equation}\label{C0.s}
   \frac{\tilde{B}_{1}^{2}}{2\sqrt{k_{1}}} +
   A_{0}^{2} \left[ 1 - \frac{\sin(2\sqrt{\beta})}{2\sqrt{\beta}}  \right] +
   B_{0}^{2} \left[ 1 + \frac{\sin(2\sqrt{\beta})}{2\sqrt{\beta}}  \right] +
   \frac{\tilde{B}_{2}^{2}}{2\sqrt{k_{2}}} = 2
\end{equation}
because the terms involving the integrals [\REqq{Jint}] corresponding to the zones with linear potential vanish and do not contribute.



The mathematically coherent step to deduce the eigenfunction and its first and second derivatives for the \swp\ consists in passing to the limit for \itm{\lgz} those of the \twp.
The passage to the limit is smooth and unambiguous for eigenfunction [\REq{ef.all}]
\begin{equation}\label{ef.all.s}
   \phi(\xi) =
                 \begin{cases}
                       \tilde{B}_{1} \cdot \exp\left[ \left(  \xi + 1 \right) \sqrt{k_{1}} \right]  \qquad & \text{zone 1}                              \\[1.5ex]
                        B_{1'} \Lambda                                                              \qquad & \text{\jp{1}{1'}$\;\equiv\;$\jp{1'}{0}}    \\[1.5ex]
                        A_{0} \sin(\xi\sqrt{\beta}) + B_{0} \cos(\xi\sqrt{\beta})                   \qquad & \text{zone 0}                              \\[1.5ex]
                        B_{2'} \Lambda                                                              \qquad & \text{\jp{0}{2'}$\;\equiv\;$\jp{2'}{2}}    \\[1.5ex]
                       \tilde{B}_{2} \cdot \exp\left[ \left( -\xi + 1 \right) \sqrt{k_{2}} \right]  \qquad & \text{zone 2}                              \\
                 \end{cases}
\end{equation}
and first derivative [\REq{def.all}]
\begin{equation}\label{def.all.s}
   \pd{}{\phi}{\xi} =
                 \begin{cases}
                       + \sqrt{k_{1}} \tilde{B}_{1} \exp\left[ \left(  \xi + 1 \right) \sqrt{k_{1}} \right]        \qquad & \text{zone 1}                           \\[1.5ex]
                       + \sqrt{k_{1}} B_{1'} \Lambda                                                               \qquad & \text{\jp{1}{1'}$\;\equiv\;$\jp{1'}{0}} \\[1.5ex]
                       \sqrt{\beta} \left[  A_{0} \cos(\xi\sqrt{\beta}) - B_{0} \sin(\xi\sqrt{\beta})  \right]     \qquad & \text{zone 0}                           \\[1.5ex]
                       - \sqrt{k_{2}} B_{2'} \Lambda                                                               \qquad & \text{\jp{0}{2'}$\;\equiv\;$\jp{2'}{2}} \\[1.5ex]
                       - \sqrt{k_{2}} \tilde{B}_{2} \exp\left[ \left( -\xi + 1 \right) \sqrt{k_{2}} \right]        \qquad & \text{zone 2}                           \\
                 \end{cases}
\end{equation}
Their continuity is preserved through the shrunk zones at \itm{\xi\mp1} with the endorsement of \REqd{B1t.s}{B2t.s}.
The passage to the limit for the second derivative [\REq{d2ef.all}] is still smooth in the zones 1,0,2 but becomes indeterminate in the zones 1' and 2' due to the presence of the ratios \itm{\eta/\etab} and \itm{\zeta/\zetab}; different limits may be reached according to whether the variables $\eta$ and $\zeta$ approach either the overlined or the circumflexed values in the limit.
There is a simple way to circumvent this ambiguity.
Let us begin with the left zone. 
From \REq{d2ef.all}, we evaluate the second derivative first at the junction point \jp{1}{1'} where \itm{\xi=-(1+\lambda)\mbox{ and }\,\eta=\etab\,}
\begin{subequations}\label{s1p.11p} \seqn
	  \begin{equation}
	     \pdatb{2}{\phi}{\xi}{\xi=-(1+\lambda)} = + k_{1} B_{1'} \left[ \Bi(\etab)  - f_{1'} \,\Ai(\etab)  \right]   \label{d2ef.1p.11p}
	  \end{equation}
and then at the junction point \jp{1'}{0} where \itm{\xi=-1\mbox{ and }\,\eta=\etah\,}
	  \begin{equation}
	     \pdatb{2}{\phi}{\xi}{\xi=-1} = + k_{1} \dfrac{\etah}{\etab}  \, B_{1'} \left[ \Bi(\etah)  - f_{1'} \,\Ai(\etah)  \right]
	                                  = - \beta \, B_{1'} \left[ \Bi(\etah)  - f_{1'} \,\Ai(\etah)  \right]   \label{d2ef.1p.1p0}
	  \end{equation}
\end{subequations}
If \itm{\,\lgz} then the junction point \jp{1}{1'} shifts rightward and goes to superpose on the junction point \jp{1'}{0} at \itm{\xi=-1}; both \itm{\etab, \etah} vanish so that \REq{d2ef.1p.11p} gives
\begin{subequations}\label{s1p.11p.lgz} \seqn
	  \begin{equation}
	     \pdatb{2}{\phi}{\xi}{\xi=-1} = k_{1} B_{1'} \Lambda       \label{d2ef.1p.11p.lgz}
	  \end{equation}
but \REq{d2ef.1p.1p0} yields instead
	  \begin{equation}
	     \pdatb{2}{\phi}{\xi}{\xi=-1} = - \beta B_{1'} \Lambda       \label{d2ef.1p.1p0.lgz}
	  \end{equation}
\end{subequations}
The comparison between \REqq{s1p.11p.lgz} tells that the zone shrinking introduces a discontinuity in the second derivative
\begin{equation}\label{d2ef.11p.dis}
   \Delta \pdatb{2}{\phi}{\xi}{\xi=-1} = - \beta B_{1'} \Lambda - k_{1} B_{1'} \Lambda = - v_{1} B_{1'} \Lambda = - v_{1} \tilde{B}_{1}
\end{equation}
obtained by subtracting \REq{d2ef.1p.11p.lgz} from \REq{d2ef.1p.1p0.lgz} and taking into account the definition of \REq{negbetas} with \itm{s=1}.
The procedure for the right zone resembles in all aspects the one we followed for the left zone and leads to a similar result: the shrinking introduces the second-derivative discontinuity
\begin{equation}\label{d2ef.2p2.dis}
   \Delta \pdatb{2}{\phi_{2'}}{\xi}{\xi=+1} = + v_{2} B_{2'} \Lambda = + v_{2} \tilde{B}_{2}
\end{equation}
Inspection of \REqd{d2ef.11p.dis}{d2ef.2p2.dis} reveals that the second-derivative discontinuities are proportional only to the characteristic numbers \itm{v_{1}, v_{2}}; the other values of the potential belonging to the vertical segments do not play any role and, therefore, presence or omission of those segments are irrelevant.
For all purposes and intents, the \dswp\ obtained from the \swp\ of \Rfi{ndswp} by removing the vertical segments is equivalent; in other words, nothing new with respect to the \sta, exactly as predicted by the teacher's reassurance mentioned in \Rse{intro}. 
So, where is the difference? The difference is that both eigenfunction's and first derivative's continuity and irrelevance of the potential's vertical segments are unproven \textit{working assumptions} in the \sta\ which the standard teacher's reassurance is based upon whereas they are \textit{proven results} obtained in a physically as well as mathematically consistent manner in the study path we have followed.
A perspicacious reader may wonder whether or not our, admittedly long, detour via the \twp's study is maybe unjustified, if not even pedantic, mathematical sophistry.
We concede that there could be some legitimacy in such a reflection were it not for the existence of claims \cite{db1979ajp} that \sta's explanations of eigenfunction's and its first derivative's continuity at \swp's jump points are unsatisfactory from a mathematical point of view.
The potential trueness of those claims would confer worthiness to our detour with the \twp\ because, within its context, the debated continuity is a proven result.
Thus, if the textbook explanations are really unsatisfactory then there must exist other justifying reasons within the \sta's context awaiting for discovery.
We deal with these issues in \Rse{ccjp}.

\section{On the continuity conditions at the square-well potential's jump points \label{ccjp}}
Let us consider again the \swp\ of \Rfi{ndswp} and suppose we have solved the eigenvalue problem just from a mathematical point of view. 
The eigenfunction we would obtain is formally \REq{ef.all.s} without the second and fourth rows
\begin{equation}\label{ef.all.s.stp} 
   \phi(\xi) =
                 \begin{cases}
                       \tilde{B}_{1} \cdot \exp\left[ \left(  \xi + 1 \right) \sqrt{k_{1}} \right]  \qquad & \text{zone 1}                              \\[1.5ex]
                        A_{0} \sin(\xi\sqrt{\beta}) + B_{0} \cos(\xi\sqrt{\beta})                   \qquad & \text{zone 0}                              \\[1.5ex]
                       \tilde{B}_{2} \cdot \exp\left[ \left( -\xi + 1 \right) \sqrt{k_{2}} \right]  \qquad & \text{zone 2}                              \\
                 \end{cases}
\end{equation}
Of course, all the body of knowledge existing behind the coefficients $\tilde{B}_{1}, \tilde{B}_{2}$ that we acquired by studying the \twp\ (\Rse{ev.twp}) and what happens when \lgzt\ (\Rse{ec.swph}) would be absolutely invisible to us, particularly the existence of \REqd{B1t.s}{B2t.s}.
We would look at \REq{ef.all.s.stp} with the awareness that it contains four coefficients that must be fixed by assigning four conditions at the potential's jump points.
In this regard, the literature offers contrasting opinions. 
We take Bohm's words \cite[page 232]{db1989} to formulate the opinion overwhelmingly accepted in the \sta:
\begin{quote}
  Because the differential equation is of second order in $x$, it is necessary that both $\psi$ and its first derivative be continuous at the boundaries.
  This follows from the fact that $\psi, E$, and $V$ are all assumed to be finite.
  $\psi$ must be finite if its physical interpretation in terms of probability is to have meaning, whereas $E$ and $V$ must be finite, because infinite energies do not occur in nature.
  From the differential eq.~(2), we then conclude that $\tdt{2}{\psi}{x}$ is everywhere finite (but not necessarily continuous).
  $\tdt{2}{\psi}{x}$ can be finite, however, only if $\tdt{}{\psi}{x}$ is continuous.
  Thus, we obtain the first boundary condition.
  In order that $\tdt{}{\psi}{x}$ exist everywhere, however, as is implied by the mere use of a differential equation, it is also necessary that $\psi$ be continuous.
  This gives us the second boundary condition.
\end{quote}
Bohm's Eq.~(2) appears at page 230 and coincides with our \REq{evp.s} save for the energy notation \itm{(E\rightarrow\epsilon)}.
We take Branson's words \cite{db1979ajp} to represent the objection to the above opinion:
\begin{quote}
  The boundary conditions imposed on the Schr\"{o}dinger wave function at the edges of the well are:
  (a) the wave function vanishes, if the potential jump is infinite,
  (b) the wave function and its first derivative are continuous, if the potential jump is finite.
  \newline \ldots \newline
  In Sec. II we describe why most textbook explanations of conditions (b) are, in our view, unsatisfactory, and in the remaining sections we present arguments which are, we hope, more acceptable.
  \end{quote}
We definitely recommend the reader to familiarize with the mathematical arguments expounded by Branson in Sec. II of his paper; one of the ``more acceptable arguments'', proposed in his Sec.~V, is indeed the idea to consider the limit of a continuous potential such as our \twp.
Confronted with such an unsettled situation, we take a pragmatic stance: we listen to Branson's warning and assume eigenfunction's and its first derivative's differences formally prescribed at the jump points
\begin{subequations}\label{def.jp} \seqn
    \begin{align}
       \phi_{0}(-1) - \phi_{1}(-1)   & = \Delta\phi(-1)  = - A_{0} \sin(\sqrt{\beta}) + B_{0} \cos(\sqrt{\beta}) - \tilde{B}_{1} \label{def.jp.1}\\[.25\baselineskip]
       \phi'_{0}(-1) - \phi'_{1}(-1) & = \Delta\phi'(-1) =   \sqrt{\beta} \left[ A_{0} \cos(\sqrt{\beta}) + B_{0} \sin(\sqrt{\beta})\right] - \sqrt{k_{1}}\tilde{B}_{1} \label{defd.jp.1} \\[.25\baselineskip]
       \phi_{2}(+1) - \phi_{0}(+1)   & = \Delta\phi(+1)  =   \tilde{B}_{2} - A_{0} \sin(\sqrt{\beta}) - B_{0} \cos(\sqrt{\beta}) \label{def.jp.2} \\[.25\baselineskip]
       \phi'_{2}(+1) - \phi'_{0}(+1) & = \Delta\phi'(+1) = - \sqrt{k_{2}}\tilde{B}_{2} - \sqrt{\beta}\left[ A_{0} \cos(\sqrt{\beta}) - B_{0} \sin(\sqrt{\beta}) \right] \label{defd.jp.2}
    \end{align}
\end{subequations}
but without necessarily committing to Bohm's opinion of \textit{a priori} continuity
\begin{equation}\label{ef.def.c}
  \Delta\phi(-1) = \Delta\phi'(-1) = \Delta\phi(+1) = \Delta\phi'(+1) = 0
\end{equation}
Then, we proceed to the determination of the four coefficients by exploiting \REqq{def.jp} with the hope to encounter down the road a compelling physical reason to enforce mathematical continuity in order to save physical consistency.
Let us see what happens.

The first logical step consists in solving the system composed by \REqq{def.jp} for the four coefficients \itm{\tilde{B}_{1}, A_{0}, B_{0}, \tilde{B}_{2}}.
For the sake of notation simplification, first we conveniently predefine the auxiliary coefficients
\begin{subequations}\label{precoeff}\seqn
  \begin{align}
     \tilde{B}_{1e} & =  - A_{0} \sin(\sqrt{\beta}) + B_{0} \cos(\sqrt{\beta})                                           \label{Bt1e}  \\[.25\baselineskip]
     \tilde{B}_{1d} & =    \sqrt{\dfrac{\beta}{k_{1}}} \left[ A_{0} \cos(\sqrt{\beta}) + B_{0} \sin(\sqrt{\beta})\right] \label{Bt1d}  \\[.25\baselineskip]
     \tilde{B}_{2e} & =    A_{0} \sin(\sqrt{\beta}) + B_{0} \cos(\sqrt{\beta})                                           \label{Bt2e}  \\[.25\baselineskip]
     \tilde{B}_{2d} & =  - \sqrt{\dfrac{\beta}{k_{2}}} \left[ A_{0} \cos(\sqrt{\beta}) - B_{0} \sin(\sqrt{\beta})\right] \label{Bt2d}  
  \end{align}
\end{subequations}
and subsequently proceed to solve the system.
The coefficients $\tilde{B}_{1}, \tilde{B}_{2}$ are easily extracted
\begin{subequations}\label{coeff.def} \seqn
    \begin{equation}\label{Btilde1.def}
       \tilde{B}_{1} = \tilde{B}_{1e} - \Delta\phi(-1) = \tilde{B}_{1d} - \dfrac{\Delta\phi'(-1)}{\sqrt{k_{1}}}
    \end{equation}
    \begin{equation}\label{Btilde2.def}
       \tilde{B}_{2} = \tilde{B}_{2e} + \Delta\phi(+1) = \tilde{B}_{2d} - \dfrac{\Delta\phi'(+1)}{\sqrt{k_{2}}}
    \end{equation}
in terms of the coefficients $A_{0}, B_{0}$ hidden inside the auxiliary coefficients; in turn, the coefficients $A_{0}, B_{0}$ have to be determined from the algebraic system
\begin{equation}\label{A0B0.def}
  \begin{bmatrix}
     \,\sin(\sqrt{\beta}) + \sqrt{\dfrac{\beta}{k_{1}}} \cos(\sqrt{\beta}) \quad &   \sqrt{\dfrac{\beta}{k_{1}}} \sin(\sqrt{\beta}) - \cos(\sqrt{\beta})\, \\[3ex]
       \sin(\sqrt{\beta}) + \sqrt{\dfrac{\beta}{k_{2}}} \cos(\sqrt{\beta}) \quad & - \sqrt{\dfrac{\beta}{k_{2}}} \sin(\sqrt{\beta}) + \cos(\sqrt{\beta})\,
  \end{bmatrix} \cdot
  \begin{bmatrix} A_0 \\[3ex] B_{0} \end{bmatrix} =
  \begin{bmatrix} \,- \Delta\phi(-1) + \dfrac{\Delta\phi'(-1)}{\sqrt{k_{1}}}\, \\[3ex]  - \Delta\phi(+1) - \dfrac{\Delta\phi'(+1)}{\sqrt{k_{2}}} \end{bmatrix}
\end{equation}
\end{subequations}
and here we already encounter the first surprise: \REq{A0B0.def} indicates that the eigenvalue spectrum is continuous [compare with \REq{A0B0.m} with due account of \REqq{gf.sp}] because the algebraic system is not homogeneous due to the presence of the discontinuities on the right-hand side.
We concede that the expectation of a discrete eigenvalue spectrum qualifies as sufficiently physical motivation pushing in the direction of \REqq{ef.def.c}.
However, the rejoicing in the continuity camp is short lived because the push is not strong enough: the physical necessity for a discrete eigenvalue spectrum only requires the vanishing of the global terms 
\begin{subequations}\label{vrhs} \seqn
    \begin{align}
        - \Delta\phi(-1) + \dfrac{\Delta\phi'(-1)}{\sqrt{k_{1}}}  & = 0    \label{vrhs.jpm1}  \\[.25\baselineskip]
        - \Delta\phi(+1) - \dfrac{\Delta\phi'(+1)}{\sqrt{k_{2}}}  & = 0    \label{vrhs.jpp1}
      \end{align}
\end{subequations}
with respect to which the continuity conditions [\REqq{ef.def.c}] are just a particular case.
With the admission of \REqq{vrhs}, the algebraic system [\REq{A0B0.def}] becomes homogeneous,
its determinant coincides with the one we found for the \twp\ with $\lgz$ and, obviously, its vanishing [\REq{eigenv.sp}] generates the same discrete eigenvalue spectrum.
As a side note, we wish to point out that this occurrence clearly implies that the eigenvalues are real but we are forbidden to use this information within the perspective of this section to respect self-inclusiveness; however, we can certainly keep this expectation in mind for later in order to eventually check whether or not we are on the right track.
Thus, to continue, the requirement of a discrete eigenvalue spectrum does not rule out discontinuous eigenfunctions.
Nevertheless, it leads at least to a first improvement by reducing the number of independent differences [\REqq{def.jp}] from four to two via the imposition of \REqq{vrhs}; if, in this regard, we privilege the eigenfunction's discontinuities then we can write
\begin{subequations}\label{vrhs.ef} \seqn
    \begin{align}
        \Delta\phi'(-1)  & = + \sqrt{k_{1}} \Delta\phi(-1)     \label{vrhs.jpm1.ef}  \\[.25\baselineskip]
        \Delta\phi'(+1)  & = - \sqrt{k_{2}} \Delta\phi(+1)     \label{vrhs.jpp1.ef}
      \end{align}
\end{subequations}
The expressions of the coefficients [\REqd{Btilde1.def}{Btilde2.def}] also simplify with the aid of \REqq{vrhs.ef}; first of all, the auxiliary coefficients become
\begin{subequations}\label{Btilde.eda.def} \seqn
    \begin{equation}\label{Btilde1.eda.def}
       \tilde{B}_{1e} = \tilde{B}_{1d} = \dfrac{\tilde{B}_{1e} + \tilde{B}_{1d}}{2} \rightarrow \tilde{B}_{1a}
    \end{equation}
    \begin{equation}\label{Btilde2.eda.def}
       \tilde{B}_{2e} = \tilde{B}_{2d} = \dfrac{\tilde{B}_{2e} + \tilde{B}_{2d}}{2} \rightarrow \tilde{B}_{2a}
    \end{equation}
    and then
    \begin{equation}\label{Btilde1.f.def}
       \tilde{B}_{1} = \tilde{B}_{1a} - \Delta\phi(-1)
    \end{equation}
    \begin{equation}\label{Btilde2.f.def}
       \tilde{B}_{2} = \tilde{B}_{2a} + \Delta\phi(+1)
    \end{equation}
\end{subequations}
The calculation of the coefficients $A_{0}, B_{0}$ from the homogeneous version of \REq{A0B0.def} follows unaltered the description already given in \Rse{efc}, including the trick involving the coefficients $C_{0}, D_{0}$.
Again, the coefficient $C_{0}$ is fixed by the eigenfunction's normalization condition [\REq{efn.nd}] whose application on \REq{ef.all.s.stp} leads to the algebraic quadratic equation [compare with \REq{C0.s}, with account of \REqd{Btilde1.f.def}{Btilde2.f.def}, for vanishing discontinuities]
\begin{multline}\label{C0.s.def}
   \left\{ \frac{ \tilde{B}_{1a}^{2} }{ 2\sqrt{k_{1}} } + A_{0}^{2} \left[ 1 - \frac{ \sin(2\sqrt{\beta}) }{ 2\sqrt{\beta} } \right] +
           B_{0}^{2} \left[ 1 + \frac{ \sin(2\sqrt{\beta}) }{ 2\sqrt{\beta} } \right] + \frac{ \tilde{B}_{2a}^{2} }{ 2\sqrt{k_{2}} } \right\} \\[.25\baselineskip]
   - \left[ \frac{ \tilde{B}_{1a} }{ \sqrt{k_{1}} } \Delta\phi(-1) - \frac{ \tilde{B}_{2a}  }{ \sqrt{k_{2}} } \Delta\phi(+1) \right]
   + \left[ \frac{ \Delta\phi(-1)^{2} }{ 2\sqrt{k_{1}} } + \frac{ \Delta\phi(+1)^{2} }{ 2\sqrt{k_{2}} } \right] = 2
\end{multline}
As numerical example, we have chosen the ground state of the symmetrical \swp\ considered by Reed \cite{bcr1990ajp} ($n=1$ in \Rfi{evdg.reed}; \Rta{evtc}).
A comparison between the continuous eigenfunction (hollow squares) and the discontinuous eigenfunction (solid line) corresponding to $\Delta\phi(+1)=-\Delta\phi(-1)=0.5$ is shown in \Rfi{reed1.phi}; the squared eigenfunctions are shown in \Rfi{reed1.phi2} to illustrate the conservation of the geometrical area in compliance with the eigenfunction-normalization condition [\REq{efn.nd}].
\begin{figure}[h]
\dgcap{.05}{.05}
   \subfloat[Eigenfunctions]         {\label{reed1.phi}  \resizebox{.47\textwidth}{!}{\includegraphics*[trim=33 20 60 60]{\gdir/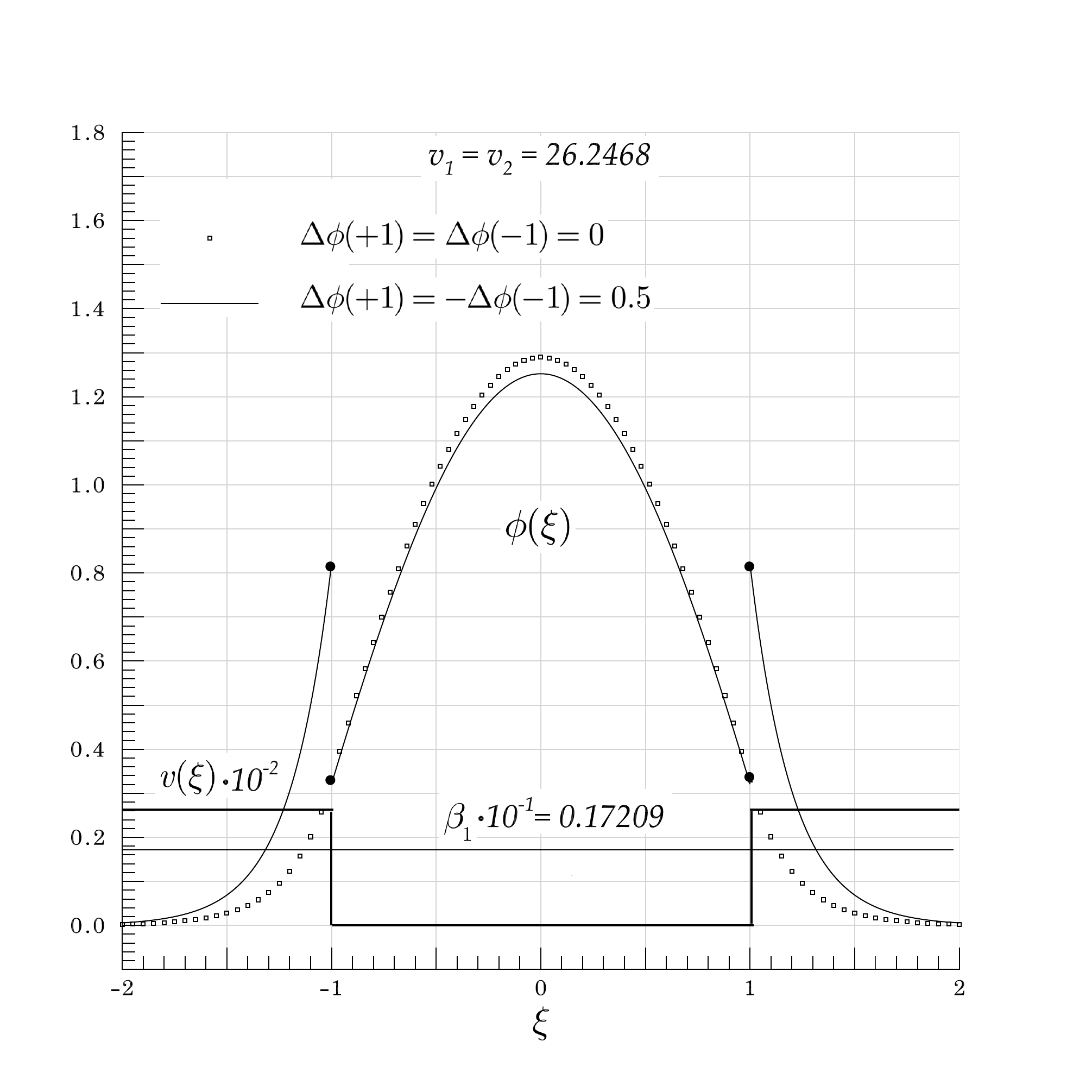}}}
   \subfloat[Squared eigenfunctions] {\label{reed1.phi2} \resizebox{.47\textwidth}{!}{\includegraphics*[trim=33 20 60 60]{\gdir/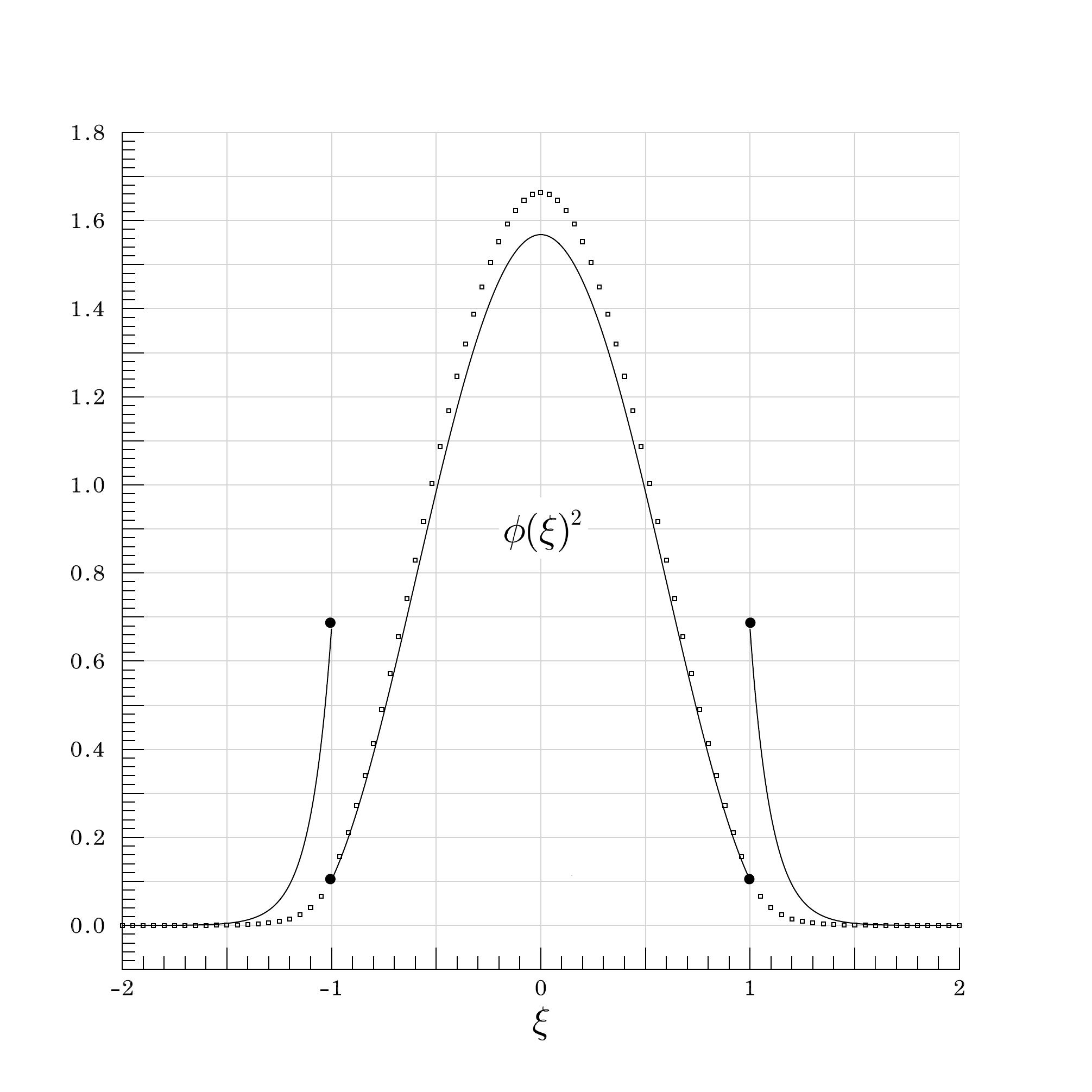}}}
   \caption{Comparison between continuous and discontinuous eigenfunctions for the ground state of the symmetrical \swp\ considered by Reed ($n=1$ in \Rfi{evdg.reed}; \Rta{evtc}).\llpush\label{reed1}}
\end{figure}
%
\Rfib{reed1.phi} seemingly leaves no doubt that, at least within a mathematical perspective, the discontinuous eigenfunction is as acceptable as the continuous one because they both satisfy same differential equation and boundary conditions.
In the same figure, we also see portrayed the flagrant groundlessness of Bohm's statement ``$\tdt{2}{\psi}{x}$ can be finite, however, only if $\tdt{}{\psi}{x}$ is continuous'' and the veracity of Branson's concern ``most textbook explanations of conditions (b) are, in our view, unsatisfactory'': the discontinuous eigenfunction (solid line) has everywhere a finite second derivative but the first derivative is discontinuous [\REqq{vrhs.ef}] at the jump points.
As a matter of fact, we can easily calculate the second-derivative discontinuities
\begin{subequations}\label{d.efd2} \seqn
    \begin{align}
        \Delta\phi''(-1)  & = \phi''_{0}(-1) - \phi''_{1}(-1) = - v_{1} \tilde{B}_{1} - \beta \Delta\phi(-1)     \label{d.efd2.jpm1}  \\[.25\baselineskip]
        \Delta\phi''(+1)  & = \phi''_{2}(+1) - \phi''_{0}(+1) = + v_{2} \tilde{B}_{2} - \beta \Delta\phi(+1)     \label{d.efd2.jpp1}
      \end{align}
\end{subequations}
[compare with \REqd{d2ef.11p.dis}{d2ef.2p2.dis}].
Well, there is not much to argue: the continuous eigenfunction comes accompanied by a ballast of infinite discontinuous eigenfunctions each one of which possesses the status of mathematical solution as legitimate as that of the continuous eigenfunction and we should be prepared to consider the wavefunction's general solution
\begin{equation}\label{wgs.swp}
   \wf(x,t) = \sum_{n=1}^{N}\sum_{r=1}^{\infty} c_{nr}\cdot \exp\left(-i \frac{\epsilon_{n} t}{\hbar}\right)\cdot\psi_{nr}(x)
\end{equation}
The index $r$ enumerates the infinite eigenfunctions that belong to the eigenvalue $\epsilon_{n}$, or its nondimensional counterpart $\beta_{n}$; we reserve the first place ($r=1$) for the continuous one.
In our opinion, the latter's selection and the others' disregard on the basis of unsatisfactory mathematical arguments, whether it may be seen either as an \textit{educated} guess by an optimist who sticks to the \sta\ or a sheer hit of luck by a pessimist who decides to go through the detour of the limit with $\lgzt$ of a \twp, for the purpose of shortcutting the teaching effort is not a didactically honest pass. 
Yet, the probable desperation generated by \REq{wgs.swp} in the continuity camp is once again short lived because a more attentive look at \Rfi{reed1.phi} reveals the second surprise: the blatant infringement of the conclusion, ``So, the eigenstates are not degenerate: for a specified eigenvalue there is one and only one eigenfunction'', that we drew when elaborating the proof of eigenfunction's uniqueness involving the Wronskian in the middle of \Rse{vsbc} from \REq{w} until just before \REq{ef.ortho}.
Indeed, in \Rfi{reed1.phi} we see two independent eigenfunctions corresponding to the same eigenvalue; as a matter fact, we can produce infinite independent eigenfunctions for the same eigenvalue by arbitrarily varying the discontinuities $\Delta\phi(-1), \Delta\phi(+1)$.
Can this infinite degeneracy be reconciled with the eigenfunction-uniqueness proof? No, it cannot!
A quick reexamination of the proof shows unequivocally that it breaks down with discontinuous eigenfunctions.
We must remember the flag planted near \REq{evp.u.4}, rewind the discourse to that equation, switch to nondimensional mode and adapt the notation $\psi_{1}, \psi_{2} \rightarrow \phi,\vartheta$ to the case of the \swp\ in \Rfi{ndswp}; then we have
\begin{equation}\label{evp.u.4.1}
  \pd{}{}{\xi}\left( \phi\,\pd{}{\vartheta}{\xi} -  \pd{}{\phi}{\xi}\, \vartheta  \right) = \pd{}{W}{\xi} = 0 \taglabel{evp.u.4}{\eqsubone}
\end{equation}
The Wronskian's discontinuities at the jump points implies that the integration of \REqt{evp.u.4.1} must now take place separately in the three zones and, consequently, the Wronskian turns out to be only piecewise constant.
In zone 1, the integration yields a vanishing Wronskian
\begin{subequations}\label{w1.swp} \seqn
    \begin{equation}\label{w.def.z1}
      W_{1} = W_{1}(-\infty) = W_{1}(-1) = 0
    \end{equation}
    Indeed, the boundary condition [\REq{ndevp.bc}, left] does away with the asymptotic value
    \begin{equation}\label{w1.swp.minf}
      W_{1}(-\infty) = \phi_{1}(-\infty)\,\pds{}{\vartheta_{1}}{\xi}{\xi=-\infty} - \pds{}{\phi_{1}}{\xi}{\xi=-\infty}\,\vartheta_{1}(-\infty) = 0
    \end{equation}
    while the eigenfunctions' exponential [\REq{ef.all.s.stp}, zone 1] duly marshals the value at the left jump point to be consistent with the asymptotic value
    \begin{equation}\label{w1.swp.m1}
      W_{1}(-1) = \phi_{1}(-1)\,\pds{}{\vartheta_{1}}{\xi}{\xi=-1} - \pds{}{\phi_{1}}{\xi}{\xi=-1}\,\vartheta_{1}(-1)
                = \phi_{1}(-1)\,\sqrt{k_{1}}\vartheta_{1}(-1) - \sqrt{k_{1}}\phi_{1}(-1)\,\vartheta_{1}(-1) = 0
    \end{equation}
\end{subequations}
So, $\phi_{1}$ and $\vartheta_{1}$ are not independent and one can be expressed in terms of the other
\begin{equation}\label{efld.z1}
  \vartheta_{1}(\xi) = a_{1}\,\phi_{1}(\xi)
\end{equation}
via a constant $a_{1}$ that we are free to choose either real or complex.
Expectedly by symmetry, the same situation occurs in zone 2
\begin{equation}\label{w.def.z2}
  W_{2} = W_{2}(+1) = W_{2}(+\infty) = 0
\end{equation}
\begin{equation}\label{efld.z2}
  \vartheta_{2}(\xi) = a_{2}\,\phi_{2}(\xi)
\end{equation}
Perhaps a bit unexpectedly, that happens in zone 0 too
\begin{equation}\label{w.def.z0}
  W_{0} = W_{0}(-1) = W_{0}(+1) = 0
\end{equation}
\begin{equation}\label{efld.z0}
  \vartheta_{0}(\xi) = a_{0}\,\phi_{0}(\xi)
\end{equation}
discontinuities notwithstanding.
The verification of \REq{w.def.z0} requires involvement, and wise manipulation, of the discontinuity definitions in \REqq{def.jp} and the utilization of \REqq{vrhs.ef} which play an absolutely crucial role to enforce the validity of \REq{w.def.z0}.
Now, eigenfunction's uniqueness requires
\begin{equation}\label{efu.swp}
   a_{1} = a_{0} = a_{2}
\end{equation}
but here, unfortunately, we hit an insurmountable mathematical barrier.
Let us write \REq{efld.z0} at the left jump point
\begin{equation}\label{efld.ljp.z0}
  \vartheta_{0}(-1) = a_{0}\,\phi_{0}(-1)
\end{equation}
and then introduce the corresponding discontinuities [\REq{def.jp.1}]
\begin{equation}\label{efld.ljp.z0.a}
  \vartheta_{1}(-1) + \Delta\vartheta(-1) = a_{0}\,\phi_{1}(-1) + a_{0}\,\Delta\phi(-1)
\end{equation}
The substitution of \REq{efld.z1} evaluated at the left jump point into \REq{efld.ljp.z0.a} and the requirement $a_{1} = a_{0}$ allow to extract the value
\begin{equation}\label{a0.z1}
   a_{0} = \dfrac{\Delta\vartheta(-1)}{\Delta\phi(-1)}
\end{equation}
for the constant $a_{0}$.
If we repeat specularly the same procedure for the right jump point then we reach another value
\begin{equation}\label{a0.z2}
   a_{0} = \dfrac{\Delta\vartheta(+1)}{\Delta\phi(+1)}
\end{equation}
irreconcilable with the former one because the eigenfunction's discontinuities at the jump points can be chosen arbitrarily.
Thus, the requirement for eigenfunction's uniqueness [\REq{efu.swp}] cannot be met, in full agreement with the graphical situation portrayed in \Rfi{reed1.phi}.

The loss of eigenfunction's uniqueness may seem not having helped us much to advance our investigation regarding acceptability or inacceptability of discontinuous eigenfunctions; nevertheless, maybe it has put us on the right track if we listen to the good lesson it teaches: we move on shaky territory when dealing with discontinuous potentials and proofs of properties we are accustomed to with continuous potentials deserve careful reconsideration.
For example, what about eigenvalues' realness and eigenfunctions' orthogonality?
We already encountered these properties in \Rse{vsbc}, near \REq{ef.ortho}, and referred the reader to the proofs given in the textbooks cited in the beginning of \Rse{intro}.
The well known standard proofing strategy leads to the basic step
\begin{equation}\label{ef.ortho.swp}
  \left( \epsilon_{n} - \cco{\epsilon_{m}} \right) \intmpi{\cco{\psi}_{mj}\,\psi_{nr}}{x} =
   \hpib \intmpi{\;\;\pd{}{}{x} \left( \psi_{nr} \pd{}{\cco{\psi}_{mj}}{x} - \cco{\psi}_{mj} \pd{}{\psi_{nr}}{x}\right)}{x}
\end{equation}
which we conveniently put in nondimensional form [\REqq{nd} and \REq{ndev}]
\begin{equation}\label{ef.ortho.swp.nd}
   \left( \beta_{n} - \cco{\beta}_{m} \right)\intmpi{\cco{\phi}_{mj}\,\phi_{nr}}{\xi} =
   \intmpi{\;\;\pd{}{}{\xi} \left( \phi_{nr} \pd{}{\cco{\phi}_{mj}}{\xi} - \cco{\phi}_{mj} \pd{}{\phi_{nr}}{\xi}\right)}{\xi}
\end{equation}
The integral on the right-hand side of \REq{ef.ortho.swp.nd} is the sum of three zonal contributions
\begin{equation}\label{ef.ortho.swp.nd.I}
   \intmpi{(\cdots)}{\xi} = \int_{-\infty}^{-1}\hspace*{-0.8em}(\cdots)_{1}\,d\xi + \int_{-1}^{+1} \hspace*{-0.8em}(\cdots)_{0}\,d\xi + \int_{+1}^{+\infty} \hspace*{-0.8em}(\cdots)_{2}\,d\xi
\end{equation}
in the case of the \swp\ in \Rfi{ndswp}.
Their integration requires patience and a bit of mathematical dexterity.
The necessary ingredients' list that makes the integration possible includes the boundary conditions [\REq{ndevp.bc}], the discontinuity definitions [\REqq{def.jp}], the constraints [\REqq{vrhs.ef}] levied on the first derivatives' discontinuities to have a discrete eigenvalue spectrum and the proper manipulation of the definitions indicated in \REq{negbetas}.
We skip the details and present directly the final result
\begin{equation}\label{ef.ortho.swp.nd.I.fin}
   \intmpi{\;\,\pd{}{}{\xi} \left( \phi_{nr} \pd{}{\cco{\phi}_{mj}}{\xi} - \cco{\phi}_{mj} \pd{}{\phi_{nr}}{\xi}\right)}{\xi} =
   \left( \beta_{n} - \cco{\beta}_{m} \right) \left[ T_{1}(n,r,m,j;-1) + T_{2}(n,r,m,j;+1) \right]
\end{equation}
in which for brevity
\begin{subequations}\label{Tterms} \seqn
    \begin{align}
      T_{1}(n,r,m,j;-1) & =
      \frac{ \left[ \phi_{1nr} \cco{\phi}_{1mj} - \left( \phi_{1nr} + \Delta\phi_{nr} \right) \left( \cco{\phi}_{1mj} + \Delta\cco{\phi}_{mj} \right) \right]_{\xi=-1} }
      { \sqrt{v_{1}-\cco{\beta}_{m}} + \sqrt{v_{1}- \beta_{n}} } \label{T1} \\[.25\baselineskip]
      T_{2}(n,r,m,j;+1) & =
      \frac{ \left[ \phi_{2nr} \cco{\phi}_{2mj} - \left( \phi_{2nr} - \Delta\phi_{nr} \right) \left( \cco{\phi}_{2mj} - \Delta\cco{\phi}_{mj} \right) \right]_{\xi=+1} }
      { \sqrt{v_{2}-\cco{\beta}_{m}} + \sqrt{v_{2}- \beta_{n}} } \label{T2}
    \end{align}
\end{subequations}
It is important to notice that the terms $T_{1}, T_{2}$ vanish identically for continuous eigenfunctions \itm{(r=j=1)} because \itm{(\Delta\phi_{n1})_{\xi=\pm 1}=(\Delta\cco{\phi}_{m1})_{\xi=\pm 1}=0} by definition.
The substitution of \REq{ef.ortho.swp.nd.I.fin} into \REq{ef.ortho.swp.nd} and a subsequent slight rearrangement lead to the generalization
\begin{equation}\label{ef.ortho.swp.nd.gen}
   \left( \beta_{n} - \cco{\beta}_{m} \right) \left[\intmpi{\cco{\phi}_{mj}\,\phi_{nr}}{\xi} - T_{1}(n,r,m,j;-1) - T_{2}(n,r,m,j;+1)\right] = 0
\end{equation}
of the condition commonly found in textbooks and from which we can draw new interesting conclusions.
If $n \neq m$ then we are looking at different eigenstates, the eigenvalues are different
\begin{equation}\label{diffev}
  \beta_{n} - \cco{\beta}_{m} \neq 0
\end{equation}
and it is the quantity in squared brackets that must obligatorily vanish; from that, we obtain
\begin{equation}\label{ndiffm}
  \intmpi{\cco{\phi}_{mj}\,\phi_{nr}}{\xi} =  T_{1}(n,r,m,j;-1) + T_{2}(n,r,m,j;+1)
\end{equation}
If $r=j=1$ then, as we already noted just after \REqq{Tterms}, the right-hand side of \REq{ndiffm} vanishes identically and we retrieve the familiar orthogonality \begin{equation}\label{ndiffm.r=s=1}
  \intmpi{\cco{\phi}_{m1}\,\phi_{n1}}{\xi} =  0
\end{equation}
of the continuous eigenfunctions; otherwise, that is  if $r \neq j$, the right-hand side of \REq{ndiffm} does not vanish and, in so doing, it brings in the lack of orthogonality among discontinuous eigenfunctions belonging to different eigenvalues.
If $n=m$ then \REq{ef.ortho.swp.nd.gen} simplifies to
\begin{equation}\label{ef.ortho.swp.nd.gen.n=m}
   \left( \beta_{n} - \cco{\beta}_{n} \right) \left[\intmpi{\cco{\phi}_{nj}\,\phi_{nr}}{\xi} - T_{1}(n,r,n,j;-1) - T_{2}(n,r,n,j;+1)\right] = 0
\end{equation}
Now, \REq{ef.ortho.swp.nd.gen.n=m} is applicable regardless of the values assumed by the subscripts $r,j$.
In the case of continuous eigenfunctions ($r=j=1$, $T_{1}=T_{2}=0$), \REq{ef.ortho.swp.nd.gen.n=m} reduces even further to the form
\begin{equation}\label{ef.ortho.swp.nd.gen.n=m.r=s=1}
   \left( \beta_{n} - \cco{\beta}_{n} \right) \intmpi{\cco{\phi}_{n1}\,\phi_{n1}}{\xi} = 0
\end{equation}
which, by taking into account that the integral never vanishes, reveals the realness of the eigenvalues
\begin{equation}\label{ef.ortho.swp.nd.gen.n=m.r=s=1.evr}
   \beta_{n} - \cco{\beta}_{n} = 0
\end{equation}
\REqb{ef.ortho.swp.nd.gen.n=m.r=s=1.evr} represents a very comforting result because it arises self-consistently and checks with the expectation we spoke of in between \REqq{vrhs} and \REqq{vrhs.ef}.
With the eigenvalues' realness assured by \REq{ef.ortho.swp.nd.gen.n=m.r=s=1.evr}, \REq{ef.ortho.swp.nd.gen.n=m} becomes inconclusive with respect to its term in square brackets in the case of discontinuous eigenfunctions ($r \neq 1, j\neq 1$), even when $r=j$, but that is an occurrence of no further relevance.

The sequence of conclusions drawn from the discussion hinged on \REq{ef.ortho.swp.nd.gen} should sharpen our discernment of the matter we are investigating because they raise two severe warnings: the first is related to the integral appearing in \REq{ef.ortho.swp}, and evaluated in \REq{ef.ortho.swp.nd.I.fin}, because it intervenes also in, and obviously affects, the crucial hermiticity test of the hamiltonian [\REq{Ham}] mentioned in connection with \REqs{moh}{bc.c1}; 
the second is related to the lack of orthogonality among the discontinuous eigenfunctions.
The severity of these warnings ensues from the realization that the normalization of the wavefunction [\REq{wfn}] may be endangered.
Well, finally a dim light at the end of the tunnel.
For the hamiltonian's hermiticity test, we must rewind to \REq{moh.i} and undo it to the unsimplified form
\begin{equation}\label{moh.i.g}
  \moHc - \moH = \intinf{[\cco{(\ham\wf)}\,\wf - \wfc\,\ham\wf]} = \hpib \intinf{\;\,\pd{}{}{x}\left(\wfc \pd{}{\wf}{x} - \wf \pd{}{\wfc}{x}\right)}
\end{equation}
The integral in \REq{ef.ortho.swp} appears after the substitution of the wavefunction's general solution [\REq{wgs.swp}] in \REq{moh.i.g}
\begin{equation}\label{ef.ortho.swp.hht}
  \intinf{\;\,\pd{}{}{x}\left(\wfc \pd{}{\wf}{x} - \wf \pd{}{\wfc}{x}\right)} =
  \sum_{n=1}^{N}\sum_{r=1}^{\infty}\sum_{m=1}^{N}\sum_{j=1}^{\infty} c_{nr} \cco{c}_{mj}\exp\left(i\frac{\cco{\epsilon}_{m}- \epsilon_{n}}{\hbar}t\right)
  \intinf{\;\,\pd{}{}{x} \left( \psi_{nr} \pd{}{\cco{\psi}_{mj}}{x} - \cco{\psi}_{mj} \pd{}{\psi_{nr}}{x}\right)}
\end{equation}
\REqb{ef.ortho.swp.nd.I.fin} taught us that the integral does not vanish if \itm{n\neq m} in the presence of discontinuous eigenfunctions; thus, in sequence, the hamiltonian's hermiticity test fails
\begin{equation}\label{moh.i.g.fail}
  \moHc - \moH \neq 0
\end{equation}
\REq{moh.i} breaks down, \REq{moe} prevails and the wavefunction's normalization becomes inhibited.
The latter occurrence can be seen more directly by substituting the wavefunction's general solution [\REq{wgs.swp}] into \REq{wfn}
\begin{equation}\label{wfn.fail}
   \intinf{\wfc(x,t)\cdot\wf(x,t)} =
   \sum_{n=1}^{N}\sum_{r=1}^{\infty}\sum_{m=1}^{N}\sum_{j=1}^{\infty} c_{nr} \cco{c}_{mj}\exp\left(i\frac{\cco{\epsilon}_{m}- \epsilon_{n}}{\hbar}t\right)
   \intinf{\cco{\psi}_{mj} \psi_{nr} }
\end{equation}
and by noticing that the lack of eigenfunctions' orthogonality prevents the disappearance of the temporal terms inside the series and kills the wavefunction's normalization
\begin{equation}\label{wfn.fail.fin}
   \intinf{\wfc(x,t)\cdot\wf(x,t)} \neq 1
\end{equation}
\REqdb{moh.i.g.fail}{wfn.fail.fin} end our quest: they constitute the lethal blow to discontinuous eigenfunctions and provide the physical justification for their rejection, consistently with Bohm's and Griffiths' teachings quoted between \REqd{moe}{moh}.
Both equations authorize us to set eigenfunctions' discontinuities to zero everywhere in this section and to retrieve comfortably all the familiar results presented in textbooks.

We reconnect with the reader's doubt expressed at the end of \Rse{ef.s}: is the detour via the \twp's study didactically justified and worth undertaking?
In the light of what we have discussed and the results achieved in this section, we believe the answer to be in the affirmative because the fact that, within the perspective of the \twp's study, eigenfunction's and first derivative's continuity are proven results pushes to question their role as initial assumptions of the \sta, to consider claims that they are based on unsatisfactory mathematical arguments, to believe that there must exist a more physically consistent manner to invoke their applicability within the \sta's context and to engage in the detective work to discover such a manner.
The process is certainly longer but definitely more enriching from both a teaching as well as a learning point of view.

\section{Conclusions}
Did we find any new results with respect to those already offered by the \sta? The honest answer is no; we reached the top of the mountain and enjoyed the same view.
But we climbed along an unexplored path: did we learn something new?
We feel confident to give a positive answer to such a question.
We had the opportunity to bring forth and discuss the importance of the boundary conditions in an insightful manner that puts in clear perspective how essential their impact on normalization condition [\REq{wfn}], hamiltonian's hermiticity [\REq{moh.i}]  and wavefunction's separability [\REq{vst} and \REqq{evp.s.bc}] is; and, if the crucial conceptual test represented by \REqq{evp.s.bc} is passed, how that impact propagates to eigenfunctions' uniqueness [\REq{w.b}].
We detoured the boundary conditions' effect on eigenfunctions' orthogonality [\REq{ef.ortho}] and eigenvalues' realness because we took for granted the details of the proofs provided in the literature; but we have learnt in a straightforward manner how those properties are in jeopardy if the boundary conditions do not make vanish the right-hand side of \REq{ef.ortho.swp}.
We have engaged in the didactic exercise to find analytical and numerical solutions of the eigenvalue problem for the \twp, a task we have accomplished by a thorough analysis and validated with the application of our findings to particular test cases taken from the literature.
An interesting product from the imposition of the initial wavefunction [\REq{wgs.0}] was the recognition of some important mathematical characteristics, or perhaps limitations hardly discussed in textbooks, of the wavefunction's general solution if the number of eigenvalues is finite.
We deduced the solution of the \swp's quantum mechanical problem as limit of that of the \twp\ when the slope of the potential's ramps becomes vertical and obtained as proven results the otherwise initial working assumptions of the \sta, particularly eigenfunction's and first derivative's continuity at the \swp's jump points.
That score encouraged and propelled us to attempt to fix their conceptual shakiness within the \sta's context and we were rewarded by discovering a sound physical fix: that discontinuous eigenfunctions, although mathematically admissible, must be rejected because they are incompatible with wavefunction's normalization and hamiltonian's hermiticity.

In conclusion, we trust we have provided a pedagogically worthy and sufficiently elaborated answer to the question ending the student's remark quoted in the beginning of \Rse{intro} by climbing along an unbeaten course.
We hope our contribution will help students on the educational part of their world line to acquire the equilibrium between how to take into proper consideration established knowledge but, at the same time, to gain strong confidence in their capabilities of free thinking and independent dexterity in undertaking scientific research.
In line with the mountaineering analogy introduced in the beginning of this section, it seems appropriate to exit by quoting an experienced mountain climber:\footnote{C. Anker; \url{www.conradanker.com}}
\begin{quote}
   \centering The summit is what drives us but the climb itself is what matters.
\end{quote}

\bibliographystyle{apsrev4-2}

%

\end{document}